\documentclass[%
reprint,
amsmath,amssymb,
aps,
pra,
]{revtex4-2}

\usepackage{graphicx}
\usepackage{dcolumn}
\usepackage{bm}

\begin{document}


\title{Spin of Photons: Nature of Polarisation}

\author{Shinichi Saito}
 \email{shinichi.saito.qt@hitachi.com}
\affiliation{Center for Exploratory Research Laboratory, Research \& Development Group, Hitachi, Ltd. Tokyo 185-8601, Japan.}

\date{\today}

\begin{abstract}
Stokes parameters (${\bf S}$) in Poincar\'e sphere are very useful values to describe the polarisation state of photons.
However, the fundamental principle of the nature of polarisation is not completely understood, yet, because we have no concrete consensus how to describe spin of photons, quantum-mechanically.
Here, we have considered a monochromatic coherent ray of photons, described by a many-body coherent state, and tried to establish a fundamental basis to describe the spin state of photons, in connection with a classical description based on Stokes parameters.
We show that a spinor description of the coherent state is equivalent to Jones vector for polarisation states, and obtain the spin operators (${\bf \hat{S}}$) of all components based on rotators in a $SU(2)$ group theory.
Polarisation controllers such as phase-shifters and rotators are also obtained as quantum-mechanical operators to change the phase of the wavefunction for polarisation states.
We show that the Stokes parameters are quantum-mechanical average of the obtained spin operators, ${\bf S} = \langle {\bf \hat{S}} \rangle $. 
\end{abstract}

\maketitle


\section{Introduction}
Stokes and Poincar\'e successfully established a systematic way to describe polarisation of lights by using several real value parameters, known as Stokes parameters, which are described as a vector in Poincar\'e sphere \cite{Stokes51,Poincare92,Max99,Jackson99,Yariv97}.
This is a spectacular achievement at the time, because it was before the discoveries of Plank and Einstein, that lights are composed of a quanta, named a photon, with both particle and wave characters to establish quantum mechanics \cite{Dirac30, Baym69,Sakurai14,Sakurai67}.
It is intriguing to learn from words of Einstein \cite{Lehner14}, {\it quote}, {\it All these fifty years of conscious brooding have brought me no nearer to the answer to the question, "What are light quanta?"}, {\it unquote}.

Here, we revisit a lemma of this grand challenge: {\it What is spin of a photon?}
Our answer to this question is {\it polarisation}.
One might think this is obvious and already well-established, but it is less obvious, because it is generally believed that the total orbital angular momentum of a photon is impossible to split \cite{Chen08} into spin and orbital angular momentum  \cite{Allen92,Enk94,Leader14,Barnett16,Yariv97,Jackson99,Grynberg10,Bliokh15} in a unique gauge invariant way \cite{Enk94,Leader14,Barnett16,Chen08,Ji10}.
It is beyond the scope of this paper to address this mystery, however, we will focus on understanding the spin of a photon.
We are interested in a monochromatic coherent ray of photons emitted from a laser source, such that we will investigate low-energy condensed-matter physics and we will not deal with the Lorentz invariance, required for high-energy physics.
The optical spin angular momentum was previously obtained by using many-body number operators, but it was shown that the operators are commutable  \cite{Enk94,Barnett16}.
Therefore, the quantum-mechanical nature of the spin of a photon is still not completely understood, yet.

We think some of these issues are coming from various ways to define the polarisation states of lights \cite{Goldstein11,Gil16,Pedrotti07,Hecht17}, spreading among literatures.
Unfortunately, there is no unique standard for the definitions, because the way to define rotation depends on whether we are evaluating the polarisation state seen from the light-source side or from the detector side.
It is also different among physicists and engineers whether we are going to use the phase evolution as ${\rm e}^{i(kz-\omega t)}$, which is common for physicists, or as ${\rm e}^{i(\omega t -kz)}$, which is more often used for engineers, where the parameters are time ($t$), the spatial axis along the direction of the propagation ($z$), the wavenumber ($k$), and the angular frequency ($\omega$), as usual.
Depending on this phase evolution over $t$ and $z$, the direction of the rotation of the polarisation state will be changed.
These differences impose unnecessary confusions among researchers for considering the polarisation states of lights.
Therefore, we have summarised our preferential definition in Appendixes.
Our convention is similar to the classical textbook of Jackson \cite{Jackson99}, but it is not necessarily common.

Spin is an intrinsic degree of freedom inherent to an elementary particle.
A photon has spin $1$ in the unit of Dirac constant ($\hbar$), and it is described by Bose statistics, because of this integer spin \cite{Dirac30, Baym69,Sakurai14,Sakurai67}.
For an elementary particle of spin $1$, in principle, there exists 3 major components to describe the polarisation state as fundamental basis states for Lie-algebra, however, one of the component with zero spin component is not observable  \cite{Sakurai67}.
This is coming from the fact that the lights are transverse waves, which is fundamentally coming from the theory of relativity, based on the principle that there is no rest frame for a photon, which is travelling at the speed of light ($c$) in a vacuum.
Consequently, the spin state of a photon can be described by Lie-algebra of spin $1/2$ \cite{Jones41,Payne52,Max99,Yariv97,Baym69,Sakurai14,Yariv97,Collett70,Luis02,Luis07,Bjork10,Castillo11,Sotto18,Sotto18b,Sotto19}.

The purpose of this work is to clarify the correlation between the classical description of polarisation states by using Stokes parameters in Poinca\'e sphere and a many-body description of spin.
We show that the {\it the vector described by Stokes parameters is actually the quantum-mechanical expectation value of spin operators.}
This means that the Stokes parameters are order parameters to describe a coherent state of a ray from a laser, which is essentially composed of a single mode with macroscopic number of photons degenerated due to the Bose-Einstein condensation of photons.
We also show the equivalence of Poinca\'e sphere with Bloch sphere, and explain how classical results for polarisation with various parameters such as orientation angle ($\Psi$), ellipticity angle ($\chi$), auxiliary angle ($\alpha$), and phase ($\delta$), are all derived from simple geometrical consideration in these spheres.
We also show that the change of the basis states are equivalent to the rotation in $SU(2)$ (Special Unitary) Hilbert space to describe the polarisation state.
Our results show that it is quite natural to believe that the spin operators are essentially equivalent to Stokes operators, which reasonably work as standard quantum-mechanical angular momentum operators, which are observable as polarisation state in Poincar\'e sphere, satisfying commutation relationship, working as generators of rotation, and describing the polarisation state of a coherent state of photons.

\section{Principles}

\subsection{Coherent state}
A photon is an elementary particle and it must follow the principle of quantum mechanics \cite{Dirac30, Baym69,Sakurai14,Sakurai67}.
A photon can be created in a laser source, or it can be annihilated in a detector.
The creation and annihilation are described by operators $\hat{a}_{\sigma}^{\dagger}$ and $\hat{a}_{\sigma}$, respectively, which satisfy the commutation relationships for Bose particles \cite{Sakurai67,Grynberg10,Fox06,Parker05} as $[\hat{a}_{\sigma},\hat{a}_{\sigma^{'}}] = 0 $, $ [\hat{a}_{\sigma}^{\dagger},\hat{a}_{\sigma^{'}}^{\dagger} ] = 0 $, and $[\hat{a}_{\sigma},\hat{a}_{\sigma^{'}}^{\dagger}] = \delta_{{\sigma},{\sigma}^{'}}$, where $\sigma$ stands for the polarisation states such as $\sigma={\rm H}$ for horizontally polarised state and $\sigma={\rm V}$ for vertically polarised state.
$\delta_{{\sigma},{\sigma}^{'}}$ is the Kronecker delta, which becomes 1 if the polarisation states of $\sigma$ and $\sigma^{'}$ coincide, and 0 if the polarisation states are orthogonal.
We can also choose another orthogonal base such as linearly polarised states along diagonal ($\sigma={\rm D}$) and anti-diagonal ($\sigma={\rm A}$) directions, or left ($\sigma={\rm L}$) and right ($\sigma={\rm R}$) circularly-polarised states (Appendixes).

We are interested in a monochromatic coherent ray of photons propagating in a material such as a waveguide or an optical fibre or in a vacuum, emitted from a laser source \cite{Yariv97}, because lasers are ubiquitously available.
The coherent states \cite{Grynberg10,Fox06,Parker05} are described as $|\alpha_{\rm H},\alpha_{\rm V}\rangle
=|\alpha_{\rm H}\rangle | \alpha_{\rm V}\rangle$, where 
\begin{eqnarray}
|\alpha_{\rm H} \rangle
&=&{\rm e}^{-\frac{|\alpha_{\rm H}|^2}{2}}
{\rm e}^{\alpha_{\rm H} \hat{a}_{\rm H}^{\dagger}}
|0\rangle \\
|\alpha_{\rm V} \rangle
&=&{\rm e}^{-\frac{|\alpha_{\rm V}|^2}{2}}
{\rm e}^{\alpha_{\rm V} \hat{a}_{\rm V}^{\dagger}}
|0\rangle, 
\end{eqnarray}
for which we assign $\alpha_{\rm H} =\sqrt{N_{\rm H}}=\sqrt{N} \cos \alpha$, $\alpha_{\rm V}=\sqrt{N_{\rm V}} {\rm e}^{i \delta} =\sqrt{N} \sin \alpha {\rm e}^{i \delta}$ with the average number of photons for each polarisation given by $N_{\rm H}$ and  $N_{\rm V}$, and the total number of photons in the system is $N=N_{\rm H}+N_{\rm V}$.
The auxiliary angle of $\alpha$ is the angle to split the electric field between horizontal and vertical directions (Supplementary Fig. 2 (a)) and the relative phase of $\delta=\delta_y-\delta_x$ is the amount of the phase shift for the vertical direction ($\delta_y$), measured from that for the horizontal direction ($\delta_x$). 
The coherent states have important characteristics, 
\begin{eqnarray}
\hat{a}_{\rm H}|\alpha_{\rm H} \rangle
&=&\alpha_{\rm H}|\alpha_{\rm H} \rangle \\
\hat{a}_{\rm V}|\alpha_{\rm V} \rangle
&=&\alpha_{\rm V}|\alpha_{\rm V} \rangle, 
\end{eqnarray}
which mean these are eigenstates of annihilation operators.

We consider the following complex electric field operator, defined as, 
\begin{eqnarray}
\bm{\hat{\mathcal{E}}}(z,t)=
\sqrt{
  \frac{2 \hbar \omega}{\epsilon V}
  }
{\rm e}^{i \beta}
\left(
  \hat{a}_{\rm H}
  \hat{\bf x}
  +\hat{a}_{\rm V}
  \hat{\bf y}
\right),
 \end{eqnarray}
where $\beta=kz-\omega t +\delta_x$, $\hat{\bf x}$ and $\hat{\bf y}$ are unit vectors along $x$ and $y$ directions, respectively, $\epsilon$ is the dielectric constant of the material, and $V$ is the volume of the system.
In a uniform material of the refractive index $n$, the dispersion is simply obtained as $\omega=vk=ck/n$ by solving the Maxwell equations \cite{Jackson99,Yariv97}.
By applying $\bm{\hat{\mathcal{E}}}(z,t)$ to $|\alpha_{\rm H},\alpha_{\rm V}\rangle
$ from the left, we obtain
\begin{eqnarray}
\bm{\hat{\mathcal{E}}}(z,t)
|\alpha_{\rm H},\alpha_{\rm V}\rangle
&=&
\bm{\mathcal{E}}(z,t)
|\alpha_{\rm H},\alpha_{\rm V}\rangle, 
\end{eqnarray}
which means the coherent state is an eigenstate of this operator and the operation did not change the state except for the factor, $\bm{\mathcal{E}}(z,t)$, which gives the complex amplitude of the electric field as 
\begin{eqnarray}
\bm{\mathcal{E}}(z,t)=
\left (
  \begin{array}{c}
    \mathcal{E}_{x} \\
    \mathcal{E}_{y}
  \end{array}
\right)
&=
E_{0}{\rm e}^{i\beta}
\left (
  \begin{array}{c}
    \cos \alpha \ \ \ \  \\
    \sin \alpha \ {\rm e}^{i\delta}
  \end{array}
\right),
 \end{eqnarray}
with the amplitude of $E_0=\sqrt{2 \hbar \omega N/(\epsilon V)}$.
Therefore, $\bm{\hat{\mathcal{E}}}(z,t)|\alpha_{\rm H},\alpha_{\rm V}\rangle$ also works as a wavefunction to describe the coherent state of photons.
Then, we recognise that $\bm{\mathcal{E}}(z,t)$ is actually a spinor representation of the wavefunction, and it is indeed rewritten as 
\begin{eqnarray}
\bm{\mathcal{E}}(z,t)
&=
E_{0}\Psi(z,t)
|{\rm Jones} \rangle,
 \end{eqnarray}
where $\Psi(z,t)={\rm e}^{i\beta}$ is the orbital part of the wavefunction, and $|{\rm Jones} \rangle$ it the Jones vector to describe the polarisation states (Appendixes).
Therefore, the Jones vector is actually the wavefunction itself to describe the spin state of the coherent photons, quantum mechanically.
It is interesting to note that the many-body coherent state is described simply by a single mode of $\Psi(z,t)$ with the spin state as inherent internal degrees of freedom.
This is coming from the nature of the Bose-Einstein condensation inherent to the coherent laser beam, in which macroscopic number of photons are degenerate to occupy the single mode.

In a real material with a specific geometrical structure, patterned into a form of a waveguide or a fibre, we must solve the Helmholtz equation 
\begin{eqnarray}
\nabla^2 
\Psi({\bf r})
=
\mu_0 \epsilon ({\bf r})
\frac{\partial^2}{\partial t^2}
\Psi({\bf r}),
\end{eqnarray}
because the dielectric constant has a profile in the form of $\epsilon=\epsilon ({\bf r})$, due to the spacial distribution of material compositions. 
We are aware that this is very important to take into account for the more realistic considerations.
By respecting the symmetry of the waveguide, we can also consider various forms of the orbital wavefunction, including the vortex nature of the beam with orbital angular momentum \cite{Allen92,Enk94,Leader14,Barnett16,Yariv97,Jackson99,Grynberg10,Bliokh15}.
Here, we consider only the plane wave solution of the Helmholtz equation as $\Psi(z,t)={\rm e}^{i\beta}$, which makes a lot of serious problems for separating the spin and orbital parts of the angular momentum  \cite{Enk94,Leader14,Barnett16,Chen08,Ji10}, as we shall see briefly below.
Nevertheless, the plane wave solution makes calculations easy, such that it is still useful for a theoretical perspective.

We also note that $\bm{\hat{\mathcal{E}}}(z,t)$ is not observable, since all physical observables must be real.
In order to observe the electric field, which is observable, we must define the {\it real} electric field operator given by 
\begin{eqnarray}
\bm{\hat{
{\bf E}
}}
=
\frac{1}{2}
\left (
\bm{\hat{\mathcal{E}}}
+
\bm{\hat{\mathcal{E}}}^{\dagger}
\right).
\end{eqnarray}
If we take the quantum-mechanical average over the coherent state, we obtain 
\begin{eqnarray}
{\bf E}(z,t)
&=&
\langle 
\bm{\hat{
{\bf E}
}}
\rangle \\
&=&
\langle \alpha_{\rm H},\alpha_{\rm V}|
\bm{\hat{
{\bf E}
}}
|\alpha_{\rm H},\alpha_{\rm V}\rangle \\
&=&
E_0
\left(
\cos \alpha
\cos \beta
  \hat{\bf x}
+
\sin \alpha
\cos (\beta+\delta)
  \hat{\bf y}
\right)\\
&=&
\Re
\left (
\mathcal{E}(z,t)
\right),
\end{eqnarray}
which is indeed real.
Please also note that the application of $\bm{\hat{{\bf E}}}$ to $|\alpha_{\rm H},\alpha_{\rm V}\rangle$ changes the original state, because the coherent state is {\it not} the eigenstate of the creation operator \cite{Grynberg10,Fox06,Parker05}.
Therefore, it is essential to treat the electric field by using a complex value rather than real value, so that the use of a complex value is not a mere mathematical convention.
The wavefunction is intrinsically described by a complex value for a photon, just like the other quantum-mechanical systems \cite{Dirac30, Baym69,Sakurai14,Sakurai67}.

We can also obtain the quantum-mechanical average of the energy for the photons 
\begin{eqnarray}
\overline{U}_{\rm QM}
&=&
\frac{V}{2}
\left \langle
\overline{
\hat{{\bf E}} \cdot \hat{{\bf D}}
}
+
\overline{
\hat{{\bf B}} \cdot \hat{{\bf H}}
}
\right \rangle
=
\epsilon V
\left \langle
\overline{
\hat{{\bf E}} \cdot \hat{{\bf E}}
}
\right \rangle\\
&=&
\frac{\epsilon V}{4}
\left \langle
\overline{
\hat{\mathcal{E}} \cdot \hat{\mathcal{E}^{\dagger}}
        }
+
\overline{
\hat{\mathcal{E}^{\dagger}} \cdot \hat{\mathcal{E}}
        }
+
\overline{
\hat{\mathcal{E}} \cdot \hat{\mathcal{E}}
        }
+
\overline{
\hat{\mathcal{E}^{\dagger}} \cdot \hat{\mathcal{E}^{\dagger}}
        }
\right \rangle\\
&=&
\frac{\epsilon V}{4}
\left \langle
\overline{
\hat{\mathcal{E}} \cdot \hat{\mathcal{E}^{\dagger}}
        }
+
\overline{
\hat{\mathcal{E}^{\dagger}} \cdot \hat{\mathcal{E}}
        }
\right \rangle\\
&=&
\frac{\epsilon V}{4}
\frac{2 \hbar \omega}{\epsilon V}
\left \langle
\hat{a}_{\rm H}^{\dagger}\hat{a}_{\rm H}
+\hat{a}_{\rm H}\hat{a}_{\rm H}^{\dagger}
+\hat{a}_{\rm V}^{\dagger}\hat{a}_{\rm V}
+\hat{a}_{\rm V}\hat{a}_{\rm V}^{\dagger}
\right \rangle\\
&=&
\hbar \omega
\left \langle
\hat{a}_{\rm H}^{\dagger}\hat{a}_{\rm H}
+\frac{1}{2}
+\hat{a}_{\rm V}^{\dagger}\hat{a}_{\rm V}
+\frac{1}{2}
\right \rangle\\
&=&
\hbar \omega 
\left(
N_{\rm H}+\frac{1}{2}
+
N_{\rm V}+\frac{1}{2}
\right) \\
&=&\hbar \omega 
\left( N+1 \right ),
\end{eqnarray}
where the bar stands for the $t$ average, and $1/2$ is coming from the zero-point oscillations.

\subsection{Electro-Magnetic field operators for lasers}
We will obtain various electro-magnetic field operators to describe a coherent ray of photons emitted from a laser.
We use a Coulomb gauge, which satisfy
\begin{eqnarray}
{\bf \nabla} \cdot \hat{\bf A}=0,
\end{eqnarray}
where $\hat{\bf A}$ is the vector potential operator.
The magnetic induction operator,  $\hat{\bf B}$, and  $\hat{\bf E}$ are obtained from  $\hat{\bf A}$, as 
\begin{eqnarray}
\hat{\bf B}
&=&
\nabla \times \hat{\bf A}
=\frac{{\bf k} \times \hat{\bf E}}{\omega} \\
\hat{\bf E}
&=& - \partial_{t} \hat{\bf A},
\end{eqnarray}
respectively.
Alternatively, we have already obtained $\hat{\bf E}$ as,
\begin{eqnarray}
\hat{\bf E}
= 
\sqrt{
      \frac
        {\hbar \omega}
        {2\epsilon V}
    }
\left(
  (\hat{a}_{\rm H} {\rm e}^{i \beta}
  +\hat{a}_{\rm H}^{\dagger} {\rm e}^{-i \beta}
  ) \hat{\bf x}
  +
  (\hat{a}_{\rm V} {\rm e}^{i \beta}
  +\hat{a}_{\rm V}^{\dagger} {\rm e}^{-i \beta}
  ) \hat{\bf y}
\right), \nonumber \\
\end{eqnarray}
we can obtain $\hat{\bf A}$, instead, as 
\begin{eqnarray}
\hat{\bf A}
= 
\frac{-i}{\omega}
\sqrt{
      \frac
        {\hbar \omega}
        {2\epsilon V}
    }
\left(
  (\hat{a}_{\rm H} {\rm e}^{i \beta}
  -\hat{a}_{\rm H}^{\dagger} {\rm e}^{-i \beta}
  ) \hat{\bf x}
  +
  (\hat{a}_{\rm V} {\rm e}^{i \beta}
  -\hat{a}_{\rm V}^{\dagger} {\rm e}^{-i \beta}
  ) \hat{\bf y}
\right). \nonumber \\
\end{eqnarray}
Consequently, we obtain
\begin{eqnarray}
\hat{\bf B}
&=& 
\frac{1}{v}
\sqrt{
      \frac
        {\hbar \omega}
        {2\epsilon V}
    }
\left(
  (\hat{a}_{\rm H} {\rm e}^{i \beta}
  +\hat{a}_{\rm H}^{\dagger} {\rm e}^{-i \beta}
  ) \hat{\bf y}
  -
  (\hat{a}_{\rm V} {\rm e}^{i \beta}
  +\hat{a}_{\rm V}^{\dagger} {\rm e}^{-i \beta}
  ) \hat{\bf x}
\right). \nonumber \\
\end{eqnarray}

Here, we assumed that the ray is described by a single mode, which is remarkably different from a standard description of the $\hat{\bf E}$ and $\hat{\bf B}$ in a Quantum Electro-Dynamics (QED) theory \cite{Dirac30,Sakurai67,Grynberg10,Fox06,Parker05}, for which the sum over all possible modes with different wavelengths are included.
For the application of a coherent ray from lasers, the single mode are dominated over other modes \cite{Yariv97}.
We can easily extend the theory to include a few modes for describing propagation of multiple modes just by summing up these contributions based on a superposition principle.
But, the main point is the Bose-Einstein condensation character of the coherent ray, and we do not have to consider infinite number of electromagnetic fields in a vacuum.
In that sense, our system, considering here for a standard optical laser lab, is remarkably different from situations in high energy physics, where the Lorentz invariance is inevitable \cite{Leader14,Chen08,Ji10}.
In a waveguide, the spatial symmetry is broken {\it a priori}, such that the orbital is not uniform, reflecting the shape and the profile of the material compositions.

The momentum operator of the electro-magnetic field, $\hat{\bf p}_{\rm field}$, is given from the Poynting vector operator, $\hat{\bf S}=\hat{\bf E}\times \hat{\bf H}$, with the magnetic field operator of $\hat{\bf H}$, as
\begin{eqnarray}
\hat{\bf p}_{\rm field}
=\epsilon (\hat{\bf E} \times \hat{\bf B})
=\epsilon \mu_0 (\hat{\bf E} \times \hat{\bf H})
=\frac{1}{v^2}  \hat{\bf S}, 
\end{eqnarray}
where $\mu_0$ is the permeability of a material, which is usually almost the same as that in a vacuum \cite{Yariv97}, for most of the optical materials, except for optical isolators.
By integrating over $V$, we obtain
\begin{eqnarray}
\hat{\bf P}_{\rm field}
&=
\hbar k
\hat{\bf z}
\left(
  (
  \hat{a}_{\rm H}^{\dagger} \hat{a}_{\rm H}
  +
  \frac{1}{2}
  )
  +
  (
  \hat{a}_{\rm V}^{\dagger} \hat{a}_{\rm V}
  +
  \frac{1}{2}
  )
\right),
\end{eqnarray}
for which we have used 
\begin{eqnarray}
\sqrt{
      \frac
        {\hbar \omega}
        {2\epsilon V}
    }
\frac{1}{v}
\sqrt{
      \frac
        {\hbar \omega}
        {2\epsilon V}
    }
\frac{1}{v^2 \mu_0}
V
=
\frac{1}{v}
\frac
        {\hbar \omega}
        {2}
\frac{1}{v^2 \epsilon \mu_0}
=
\frac{\hbar k}{2} 
\end{eqnarray}
for the factor, $\hat{\bf x}
\times 
\hat{\bf x}
=
\hat{\bf y}
\times 
\hat{\bf y}
=
0$, 
$
\hat{\bf x}
\times 
\hat{\bf y}
=
-
\hat{\bf y}
\times 
\hat{\bf x}
=
\hat{\bf z}
$, and for the boundary condition
\begin{eqnarray}
\int
\frac{dz}{L}
{\rm e}^{\pm 2i\beta}
=
0,
\end{eqnarray}
for the length of $L$ along $z$.
The contribution of $1/2$ for each polarisation states are coming from the zero-point fluctuations, which will cancel among contributions between modes propagating opposite directions ($\hbar k$ and $-\hbar k$).

Then, it is natural to expect that the total angular momentum operator for the ray should be given by \cite{Allen92,Enk94,Allen00,Leader14,Barnett16,
Yariv97,Jackson99,Grynberg10,Bliokh15,Enk94,Leader14,Barnett16,Chen08,Ji10}
\begin{eqnarray}
\hat{\bf J}_z=
\int d^3 {\bf r} \ {\bf r} \times \hat{\bf p}_{\rm field}
=
\epsilon
\int d^3 {\bf r} \ {\bf r} \times (\hat{\bf E} \times \hat{\bf B}).
\end{eqnarray}

By using the identities 
\begin{eqnarray}
&&{\bf r} \times (\hat{\bf E} \times (\nabla \times \hat{\bf A}))
=
\hat{\bf E} ({\bf r}  \cdot (\nabla \times \hat{\bf A}))
-
({\bf r} \cdot \hat{\bf E}) (\nabla \times \hat{\bf A}) \nonumber \\ \\
&&
{\bf r}  \cdot (\nabla \times \hat{\bf A})
=
r_{i} \epsilon_{ijk} \partial_{j} \hat{A}_k
=
\epsilon_{ijk} r_{i} \partial_{j} \hat{A}_k
=({\bf r} \times \nabla)\cdot \hat{\bf A}, \nonumber \\
\end{eqnarray}
$\hat{\bf J}_z=\hat{\bf L}_z+\hat{\bf S}_{z}$ is split into its orbital angular momentum part,
\begin{eqnarray}
\hat{\bf L}_z
=
\epsilon
\int d^3 {\bf r} \ 
\hat{\bf E} 
(({\bf r} \times \nabla) \cdot \hat{\bf A}),
\end{eqnarray}
and the spin angular momentum part,
\begin{eqnarray}
\hat{\bf S}_{z}
=
-
\epsilon
\int d^3 {\bf r} \ 
({\bf r} \cdot \hat{\bf E}) (\nabla \times \hat{\bf A}).
\end{eqnarray}
Furthermore, using the identity \cite{Jackson99,Yariv97}, 
\begin{eqnarray}
({\bf r} \cdot \hat{\bf E}) (\nabla \times \hat{\bf A})_{i}
=
(r_j \hat{E}_j)
\epsilon_{ilm}
\partial_l \hat{A}_m
=
\epsilon_{ilm}
(r_j \hat{E}_j)
(\partial_l \hat{A}_m), \nonumber \\
\end{eqnarray}
we obtain for the $i$-th component
\begin{eqnarray}
\hat{\bf S}_{i}
&=&
-
\epsilon
\int d^3 {\bf r} \ 
({\bf r} \cdot \hat{\bf E}) (\nabla \times \hat{\bf A})_{i} \\
&=&
-
\epsilon
\int d^3 {\bf r} \ 
\epsilon_{ilm}
(r_j \hat{E}_j)
(\partial_l \hat{A}_m) \\
&=&
-\epsilon
\left [
(r_j \hat{E}_j)
\hat{A}_m
\right ]_{-\infty}^{\infty}
+
\epsilon
\int d^3 {\bf r} \ 
\epsilon_{ilm}
\hat{A}_m
\partial_l(r_j \hat{E}_j),  \nonumber \\
\end{eqnarray}
whose first term vanishes \cite{Chen08} for the finite mode size in a waveguide.
After the integration only $i=z$ component survive, and we obtain
\begin{eqnarray}
\hat{\bf S}_{z}
&=&
\epsilon
\int d^3 {\bf r} \ 
(\hat{\bf E}\times \hat{\bf A}) \\
&=&
(-i)\hbar 
\hat{\bf z}
\left(
  \hat{a}_{\rm H}^{\dagger} \hat{a}_{\rm V}
  -
  \hat{a}_{\rm V}^{\dagger} \hat{a}_{\rm H}
\right), \\
\end{eqnarray}
for which we have used
\begin{eqnarray}
\epsilon \sqrt{
      \frac
        {\hbar \omega}
        {2\epsilon V}
    }
\frac{-i}{\omega}
\sqrt{
      \frac
        {\hbar \omega}
        {2\epsilon V}
    }
V
=
-i \frac{\hbar }{2}.
\end{eqnarray}

If we change the basis states for describing the polarisation states from horizontal/vertical linear polarised states to left/right circular polarised states by the transformations \cite{Yariv97,Goldstein11,Gil16}
\begin{eqnarray}
\hat{a}_{\rm L}^{\dagger}
&=&
\frac{1}{\sqrt{2}}
\left(
  \hat{a}_{\rm H}^{\dagger}
  +i
    \hat{a}_{\rm V}^{\dagger}
\right) \\
\hat{a}_{\rm R}^{\dagger}
&=&
\frac{1}{\sqrt{2}}
\left(
  \hat{a}_{\rm H}^{\dagger}
  -i
    \hat{a}_{\rm V}^{\dagger}
\right),
\end{eqnarray}
and their conjugate
\begin{eqnarray}
\hat{a}_{\rm L}
&=&
\frac{1}{\sqrt{2}}
\left(
  \hat{a}_{\rm H}
  -i
    \hat{a}_{\rm V}\right) \\
\hat{a}_{\rm R}
&=&
\frac{1}{\sqrt{2}}
\left(
  \hat{a}_{\rm H}
  +i
    \hat{a}_{\rm V}\right),
\end{eqnarray}
$\hat{\bf S}_{z}=\hat{S}_{z}\hat{\bf z}$ is further simplified to be
\begin{eqnarray}
\hat{S}_{z}
&=
\hbar 
\left(
  \hat{a}_{\rm L}^{\dagger} \hat{a}_{\rm L}
  -
  \hat{a}_{\rm R}^{\dagger} \hat{a}_{\rm R}
\right) .
\end{eqnarray}

We are aware that there are significant criticisms \cite{Allen92,Enk94,Leader14,Barnett16,
Yariv97,Jackson99,Grynberg10,Bliokh15,Enk94,Leader14,Barnett16,Chen08,Ji10} on this derivation such as the intentional choice of the Coulomb gauge, the artificial choice of the boundary condition, the apparent gauge dependent expression in the form of $\hat{\bf E}\times \hat{\bf A}$, the disappearance of the $x$ and $y$ components, and so on. 
We are not happy, either, and we will address some of these issues in a separate paper. 
Nevertheless, we think the last expression is quite intuitive, and we could diagonalise the spin component in the chiral bases, which implies the spin is deeply linked to the polarisation.
Moreover, this expression is not gauge dependent, since the number of photons should not depend on the choice of the gauge.
Here, we accept this form as an expected expression derived from the correspondence from classical expectation, although we do not know why only $z$ component appeared for spin operator.
In the next section, we will apply standard quantum-mechanical technique to consider the spin operators and their expectation values.
We naturally obtained Stokes operators simply from the $SU(2)$ consideration of the spin states, and established the expectation values of spin operators are Stokes parameters. 
Therefore, we show that the Poincar\'e sphere is essential the same as the Bloch sphere.

\section{Results and Discussions}
\subsection{Chiral representation}
Here, we describe the polarisation state of photons by a chiral representation using left  and right circular-polarised states,
\begin{eqnarray}
|{\rm L} \rangle
=|\circlearrowleft \ \rangle
&=&
\left (
  \begin{array}{c}
     1 \\
     0
  \end{array}
\right) \\
|{\rm R} \rangle
=|\circlearrowright \ \rangle
&=&
\left (
  \begin{array}{c}
     0 \\
     1
  \end{array}
\right).
 \end{eqnarray}
We will call this basis as LR-basis.
According to the result of the previous section, the spin operator along $z$ direction can be written as
\begin{eqnarray}
\hat{S}_{z}
&=&
\hbar 
\left(
  \begin{array}{cc}
     \hat{a}_{\rm L}^{\dagger}, & 
     \hat{a}_{\rm R}^{\dagger} 
  \end{array}
\right)
\left(
  \begin{array}{cc}
1 & 0 \\
0 & -1
  \end{array}
\right) 
\left(
  \begin{array}{c}
     \hat{a}_{\rm L} \\
     \hat{a}_{\rm R}
  \end{array}
\right) \\
&=&
\hbar 
     \bm{\hat{\psi}}_{\rm LR}^{\dagger}
\sigma_3
\bm{\hat{\psi}}_{\rm LR},
\end{eqnarray}
where $\bm{\hat{\psi}}_{\rm LR}^{\dagger}=( \hat{a}_{\rm L}^{\dagger}, \hat{a}_{\rm R}^{\dagger})$ and $\bm{\hat{\psi}}_{\rm LR}$ are the chiral representation of the creation and the annihilation operator, respectively, and the Pauli matrices are defined as  
\begin{eqnarray}
\sigma_1=
\left(
  \begin{array}{cc}
0 & 1 \\
1 & 0
  \end{array}
\right),
\sigma_2=
\left(
  \begin{array}{cc}
0 & -i \\
i & 0
  \end{array}
\right) , 
\sigma_3=
\left(
  \begin{array}{cc}
1 & 0 \\
0 & -1
  \end{array}
\right). \nonumber \\
\end{eqnarray}

Now, it is clear that the photon in the left-circular-polarised state has a spin of $\hbar$ along the direction of propagation ($\sigma=1$), and right-circular-polarised state has a spin of $-\hbar$ along the same direction ($\sigma=-1$).
Spin states pointing the other directions such as $x$ and $y$ would be realised by the superposition states of $\hat{a}_{\rm L}$ and $\hat{a}_{\rm R}$ in the chiral representation, because the 2 level systems are described by a $SU(2)$ theory \cite{Dirac30, Baym69,Sakurai14,Sakurai67}.
It is thus straightforward to expect the spin operators for $x$ and $y$ components as 
\begin{eqnarray}
\hat{S}_{x}
&=&
\hbar 
\left(
  \begin{array}{cc}
     \hat{a}_{\rm L}^{\dagger}, & 
     \hat{a}_{\rm R}^{\dagger} 
  \end{array}
\right)
\left(
  \begin{array}{cc}
0 & 1 \\
1 & 0
  \end{array}
\right) 
\left(
  \begin{array}{c}
     \hat{a}_{\rm L} \\
     \hat{a}_{\rm R}
  \end{array}
\right) \\
&=&
\hbar 
\bm{\hat{\psi}}_{\rm LR}^{\dagger}
\sigma_1
\bm{\hat{\psi}}_{\rm LR}, \\
\hat{S}_{y}
&=&
\hbar 
\left(
  \begin{array}{cc}
     \hat{a}_{\rm L}^{\dagger}, & 
     \hat{a}_{\rm R}^{\dagger} 
  \end{array}
\right)
\left(
  \begin{array}{cc}
0 & -i \\
i & 0
  \end{array}
\right) 
\left(
  \begin{array}{c}
     \hat{a}_{\rm L} \\
     \hat{a}_{\rm R}
  \end{array}
\right) \\
&=&
\hbar 
\bm{\hat{\psi}}_{\rm LR}^{\dagger}
\sigma_2
\bm{\hat{\psi}}_{\rm LR},
\end{eqnarray}
respectively.

The general spin state, pointing to the $(\theta,\phi)$ direction, is obtained by the Bloch state \cite{Dirac30, Baym69,Sakurai14,Sakurai67}
\begin{eqnarray}
| {\rm Bloch} \rangle 
&=&
 |\theta, \phi \rangle\\
&=&
\left (
  \begin{array}{c}
    {\rm e}^{-i\frac{\phi}{2}}  \cos \left( \frac{\theta}{2} \right)    \\
    {\rm e}^{+i\frac{\phi}{2}}\sin \left( \frac{\theta}{2} \right)  
  \end{array}
\right),
\end{eqnarray}
where $\theta$ is the polar angle and $\phi$ is the azimuthal angle (Fig. 1).

\begin{figure}[h]
\begin{center}
\includegraphics[width=4cm]{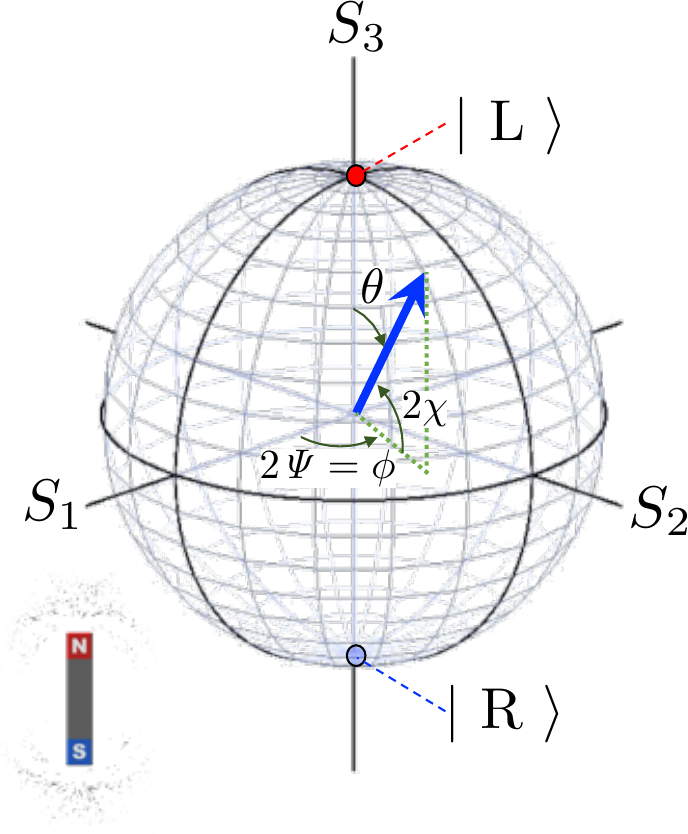}
\caption{
Bloch sphere for the polarisation states, described left and right circularly polarised states.
}
\end{center}
\end{figure}

Thus, the corresponding coherent state with the spin state $(\theta,\phi)$ is obtained as
$|\alpha_{\rm L}\alpha_{\rm R}\rangle=|\alpha_{\rm L}\rangle | \alpha_{\rm R}\rangle$, where 
\begin{eqnarray}
|\alpha_{\rm L} \rangle
&=&{\rm e}^{-\frac{|\alpha_{\rm L}|^2}{2}}
{\rm e}^{\alpha_{\rm L} \hat{a}_{\rm L}^{\dagger}}
|0\rangle \\
|\alpha_{\rm R} \rangle
&=&{\rm e}^{-\frac{|\alpha_{\rm R}|^2}{2}}
{\rm e}^{\alpha_{\rm R} \hat{a}_{\rm R}^{\dagger}}
|0\rangle, 
\end{eqnarray}
for which we assign $\alpha_{\rm L} =\sqrt{N}{\rm e}^{-i\frac{\phi}{2}}  \cos \left( \frac{\theta}{2} \right) $, $\alpha_{\rm R}=\sqrt{N}{\rm e}^{+i\frac{\phi}{2}}\sin \left( \frac{\theta}{2} \right)$.
The  complex electric field operator in the chiral representation is given by
\begin{eqnarray}
\bm{\hat{\mathcal{E}}}(z,t)=
\sqrt{
  \frac{2 \hbar \omega}{\epsilon V}
  }
{\rm e}^{i \beta}
\left(
  \hat{a}_{\rm L}
  \hat{\bf l}
  +\hat{a}_{\rm R}
  \hat{\bf r}
\right),
 \end{eqnarray}
where $\hat{\bf l}=(\hat{\bf x}+i\hat{\bf y})/\sqrt{2}$ and $\hat{\bf r}=(\hat{\bf x}-i\hat{\bf y})/\sqrt{2}$ are complex unit vectors to describe directions of left and right polarisation states with phases.
By applying this to the coherent state, we obtain the complex wavefunction in the chiral representation as

\begin{eqnarray}
\bm{\mathcal{E}}(z,t)
&=&
\left (
  \begin{array}{c}
    \mathcal{E}_{L} \\
    \mathcal{E}_{R}
  \end{array}
\right)\\
&=&
E_{0}{\rm e}^{i\beta}
\left (
  \begin{array}{c}
    {\rm e}^{-i\frac{\phi}{2}}  \cos \left( \frac{\theta}{2} \right)    \\
    {\rm e}^{+i\frac{\phi}{2}}\sin \left( \frac{\theta}{2} \right)  
  \end{array}
\right) \\
&=&
E_{0}\Psi(z,t)
|{\rm Bloch} \rangle,
 \end{eqnarray}

By calculating the expectation values of spin components by the coherent state, we obtain
\begin{eqnarray}
\langle \hat{\bf S} \rangle
&=&
\left (
  \begin{array}{c}
    \langle \hat{S}_x \rangle\\
    \langle \hat{S}_y \rangle\\
    \langle \hat{S}_z \rangle
  \end{array}
\right)\\
&=&
\hbar N
\left (
  \begin{array}{c}
    \sin \theta \cos \phi \\
    \sin \theta \sin \phi \\
    \cos \theta 
  \end{array}
\right) . \\
\end{eqnarray}
By realising the correspondences  between angles
\begin{eqnarray}
\theta &=&\frac{\pi}{2}-2\chi \\
\phi&=&2{\it \Psi},
\end{eqnarray}
we realised  
\begin{eqnarray}
\langle \hat{\bf S} \rangle
&=&
\hbar N
\left (
  \begin{array}{c}
     \cos (2 \chi) \cos (2{\it \Psi}) \\
    \cos (2 \chi) \sin (2{\it \Psi}) \\
    \sin (2 \chi) 
  \end{array}
\right) \\
&=&
\hbar N
\left (
  \begin{array}{c}
   S_1 \\
   S_2 \\
   S_3
  \end{array}
\right),
\end{eqnarray}
showing that the expectation values of the spin operators are essentially equivalent to the Stokes parameters.
Thus, the Poincar\'e sphere is equivalent to the Bloch sphere.

It is also useful to define the spin operator to represent the magnitude of the spin, 
\begin{eqnarray}
\hat{S}_{0}
&=
\hbar 
\left(
  \hat{a}_{\rm L}^{\dagger} \hat{a}_{\rm L}
  +
  \hat{a}_{\rm R}^{\dagger} \hat{a}_{\rm R}
\right), 
\end{eqnarray}
such that its expectation value is
\begin{eqnarray}
\langle
\hat{S}_{0}
\rangle
&=
\hbar 
N.
\end{eqnarray}
This actually shows the order parameter of the coherent states.
The onset of lasing is similar to the second order phase transition, which show the continuous increase of the macroscopic order parameter upon changing the control parameter such as temperature \cite{Ginzburg50,Bardeen57,Nambu59,Goldstone62,Schrieffer71,Nagaosa99,Wen04,Demler04}.
In the case for lasing, the control parameter is the pumping power, provided by injecting electrons and holes for a laser diode, or by optical populating of electrons to higher energy levels to realise a population inversion state.
Above the lasing threshold, the macroscopic number of photons are degenerate to occupy the single mode, such that $\langle \hat{S}_{0} \rangle$ can posses a non-zero value, and $\langle \hat{S}_{0} \rangle$ increases gradually upon the increase of the pumping power.

The theory of the order parameter description of the phase transition was first developed for the theory of superconductivity, as the Ginzburg-Landau theory \cite{Ginzburg50,Bardeen57,Nambu59,Goldstone62,Schrieffer71,
Nagaosa99,Wen04,Demler04}, for which the order parameter was the energy gap, 
$|\Delta|$, and the $U(1)$ gauge symmetry of the phase (${\rm e}^{i\phi}$) was broken.
Therefore, the order parameter described by a scalar.

In the case of lasing, two phases of the wave, such as $(\theta,\phi)$ in chiral representation and $(\alpha, \delta)$ in Jones representation, are fixed, and the order parameters are described by a vector, not by a scalar.
This is why $3D$ vectorial representation using the Poincar\'e sphere is so useful \cite{Yariv97,Goldstein11,Gil16}.

The $3D$ description of the order parameter similar to the Poincar\'e sphere is not restricted to the photonic systems, and actually they are ubiquitously available for describing various order parameters.
For example, magnetic Heisenberg model was used to describe the superfluid-solid phase transition for a liquid He \cite{Matsubara56}.
Another example is the $SO(5)$ theory, which was developed for describing antiferromagnetic-superconducting phase transition for high-critical-temperature superconducting cuprates \cite{Zhang97,Demler04}.

These spin operators are previously known as Stokes operators \cite{Payne52,Fano54,Collett70,Delbourgo77,Luis02,Luis07,Bjork10}, and their commutation relationships are
\begin{eqnarray}
\left[
\hat{S}_{x},
\hat{S}_{y}
\right]
=
2i
\hbar
\hat{S}_{z}, 
\left[
\hat{S}_{y},
\hat{S}_{z}
\right]
=
2i
\hbar
\hat{S}_{x}, 
\left[
\hat{S}_{z},
\hat{S}_{x}
\right]
=
2i
\hbar
\hat{S}_{y}, \nonumber \\
\end{eqnarray}
which are directly obtained by the commutation relationships of $\hat{a}_{\sigma}^{\dagger}$ and $\hat{a}_{\sigma}$.
The factor of $2$ is slightly unusual for orbital angular momentum.
Indeed, this is unusual, because we have just 2 degrees of freedom, regardless of the spin $1$ nature of a photon, which normally allow 3 states ($1,0,-1$) along the principle quantisation axis \cite{Dirac30, Baym69,Sakurai14,Sakurai67}.
This restriction is coming from the transverse nature of the ray of photons.

We also obtain
\begin{eqnarray}
\left[
\hat{S}_{0},
\hat{S}_{x}
\right]
=
\left[
\hat{S}_{0},
\hat{S}_{y}
\right]
=
\left[
\hat{S}_{0},
\hat{S}_{z}
\right]
=
0,
\end{eqnarray}
which are commutable, such that the magnitude can be a simultaneous eigenstate with the spin vector, 
${\bf{\hat{S}}}=
\left(
\hat{S}_x,
\hat{S}_y,
\hat{S}_z
\right)$.

It is also intuitive to evaluate the quantum fluctuations \cite{Luis02,Luis07,Bjork10} of  the spin of photons, by calculating 
\begin{eqnarray}
{\bf{\hat{S}}}
\cdot
{\bf{\hat{S}}}/\hbar^2
&=&
\left(
\hat{n}_{\rm L}
+
\hat{n}_{\rm R}
\right)
\left(
\hat{n}_{\rm L}
+
\hat{n}_{\rm R}
+2
\right) \\
&=&
\hat{n}
\left(
\hat{n}
+2
\right), 
\end{eqnarray}
where $\hat{n}=\hat{n}_{\rm L}+\hat{n}_{\rm R}$ is the total number operator, and then we obtain the expectation value of the quantum-mechanical fluctuation as \cite{Luis02,Luis07,Bjork10} 
\begin{eqnarray}
\delta S
&=&
\sqrt{
\frac{
\langle {\bf \hat{S}} \cdot {\bf \hat{S}}  \rangle
-
\langle
{\hat{S}_0}
\rangle^2
}
{\langle
{\hat{S}_0}
\rangle^2
}} \\
&=&
\sqrt{\frac{2}{N}},
\end{eqnarray}
which means that the quantum-mechanical fluctuation decreases significantly upon increasing the order parameter.
This is a quite typical behaviour similar to other macroscopic ordered quantum systems \cite{Ginzburg50,Schrieffer71,Nagaosa99,Wen04}.

\subsection{Jones vector representation}
Here, we will develop a similar description of spin of a photon, using Jones vector representation (Appendixes).
Our starting point is 
\begin{eqnarray}
\hat{S}_{z}
&=&
(-i)\hbar 
\left(
  \hat{a}_{\rm H}^{\dagger} \hat{a}_{\rm V}
  -
  \hat{a}_{\rm V}^{\dagger} \hat{a}_{\rm H}
\right), \\
&=&
\hbar 
\left(
  \begin{array}{cc}
     \hat{a}_{\rm H}^{\dagger}, & 
     \hat{a}_{\rm V}^{\dagger} 
  \end{array}
\right)
\left(
  \begin{array}{cc}
0 & -i \\
i & 0
  \end{array}
\right) 
\left(
  \begin{array}{c}
     \hat{a}_{\rm H} \\
     \hat{a}_{\rm V}
  \end{array}
\right) \\
&=&
\hbar 
\bm{\hat{\psi}}_{\rm HV}^{\dagger}
\sigma_2
\bm{\hat{\psi}}_{\rm HV},
\end{eqnarray}
where $\bm{\hat{\psi}}_{\rm HV}^{\dagger}=(\hat{a}_{\rm H}^{\dagger},\hat{a}_{\rm V}^{\dagger})$ and $\bm{\hat{\psi}}_{\rm HV}$ are the creation and annihilation operators in Jones vector representation.
We also call this basis as HV-basis.
In the HV-basis, it is expected that the spin state is diagonalised along the horizontal and vertical directions, such that we can expect the spin operator along $x$ as
\begin{eqnarray}
\hat{S}_{x}
&=&
\hbar 
\bm{\hat{\psi}}_{\rm HV}^{\dagger}
\sigma_3
\bm{\hat{\psi}}_{\rm HV},
\end{eqnarray}
and consequently,  
\begin{eqnarray}
\hat{S}_{y}
&=&
\hbar 
\bm{\hat{\psi}}_{\rm HV}^{\dagger}
\sigma_1
\bm{\hat{\psi}}_{\rm HV},
\end{eqnarray}
for the $y$ component, describing diagonal/anti-diagonal linear polarisation.

By taking the quantum-mechanical average over the coherent state, $|\alpha_{\rm H},\alpha_{\rm V}\rangle$, we obtain
\begin{eqnarray}
\langle {\bf \hat{S}} \rangle
&=
\hbar N
\left (
  \begin{array}{c}
    \cos (2 \alpha) \\
    \sin (2 \alpha) \cos \delta \\
    \sin (2 \alpha) \sin \delta 
  \end{array}
\right),
\end{eqnarray}
which is shown in a Poincar\'e sphere of Fig. 2.
\begin{figure}[h]
\begin{center}
\includegraphics[width=4cm]{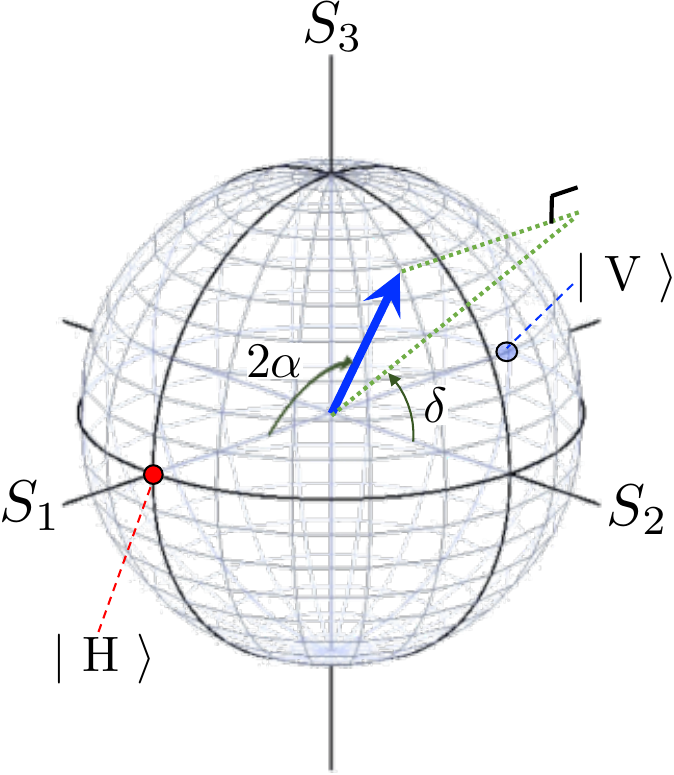}
\caption{
Poincar\'e sphere for the polarisation states, described by horizontally and vertically polarised states.
}
\end{center}
\end{figure}
The expectation value must be independent on the choice of the fundamental basis. 
By comparing $\langle {\bf \hat{S}} \rangle$, obtained for both LR- and HV-basis, we obtain the identities 
\begin{eqnarray}
\tan(2 {\it \Psi})
&=&
\tan(2 \alpha)
\cos\delta \nonumber \\
\sin (2 \chi)
&=&
\sin (2 \alpha) \sin \delta , 
\end{eqnarray}
for the transformations of angles.
These are exactly the same ones as those obtained classically, by rotating the horizontal axis to the principal axis of the polarisation ellipse (Appendix B).
Therefore, the rotation of the axes in the real space to change from $(\alpha,\delta)$ to $(\chi, \Psi)$ is equivalent to transforming from Jones vector representation to chiral representation.
The comparison between chiral and Jones representations are summarised in Table \ref{Table1} for Poincar\'e sphere.
\begin{table}[h]
\caption{\label{Table1}
Comparison between chiral and Jones representation for polarisation states.
}
\begin{ruledtabular}
\begin{tabular}{ccc}
\textrm{Representation}&
\textrm{Chiral }&
\textrm{Jones }\\
\colrule
Basis    &   $| {\rm L} \rangle$,  $| {\rm R} \rangle$ & $| {\rm H} \rangle$,  $| {\rm V}\rangle$\\
States   & Bloch vector & Jones vector\\
Sphere & Bloch  & Poincar\'e \\
Angles for $\bm{E}$ &$\chi$: Ellipticity   & $\alpha$: Auxiliary  \\
\textrm{ }&$\Psi$: Inclination & $\delta$: Phase \\
Angles for $\langle {\bf \hat{S}} \rangle$ &$\theta=\pi/2-2\chi$: Polar & 
$\gamma=2\alpha$:  Polar \\
\textrm{ }&$\phi=2\Psi$: Azimuthal & 
$\delta$: Azimuthal\\
${\bf {S}}=\langle {\bf \hat{S}} \rangle$ & 
$\hbar \langle (\sigma_0,\sigma_1,\sigma_2,\sigma_3)\rangle_{\rm LR}$ & 
$\hbar \langle (\sigma_0,\sigma_3,\sigma_1,\sigma_2)\rangle_{\rm HV}$ \\
$\bm{\mathcal{E}}$ & 
$
E_{0}{\rm e}^{i\beta}
\left (
  \begin{array}{c}
    {\rm e}^{-i\frac{\phi}{2}}  \cos \left( \frac{\theta}{2} \right)    \\
    {\rm e}^{+i\frac{\phi}{2}}\sin \left( \frac{\theta}{2} \right)  
  \end{array}
\right)
$ & 
$
E_{0}{\rm e}^{i\beta}
\left (
  \begin{array}{c}
    {\rm e}^{-i\delta/2}\cos \alpha \\
    {\rm e}^{+i\delta/2}\sin \alpha 
  \end{array}
\right)
$
\end{tabular}
\end{ruledtabular}
\end{table}
The reason why the factor of 2 appeared in front of angles such as $2\Psi$, $2\chi$, and $2\alpha$, is the quantum-mechanical average. 
By taking the complex conjugate and applying it to the original phase factor, we obtain this factor of 2, compared with the actual angle in the real space for $\bm{E}$.
This difference could be very important as for geometrical Pancharatnam-Berry's phase \cite{Pancharatnam56,Berry84}, since the adiabatic rotation in Bloch/Poincar\'e sphere would not change the expectation value, but nevertheless, it can change the sign of the electric field, which leads the non-trivial interference \cite{Tomita86}.

\subsection{Diagonal representation}
We can also consider another representation, using diagonal $|{\rm D} \rangle$ and anti-diagonal $|{\rm A} \rangle$ basis states.
In this basis, we will diagonalise the spin operator along $y$, and we obtain
\begin{eqnarray}
\hat{S}_{x}
&=&
\hbar 
     \bm{\hat{\psi}}_{\rm DA}^{\dagger}
\sigma_2
\bm{\hat{\psi}}_{\rm DA} \\
\hat{S}_{y}
&=&
\hbar 
     \bm{\hat{\psi}}_{\rm DA}^{\dagger}
\sigma_3
\bm{\hat{\psi}}_{\rm DA} \\
\hat{S}_{z}
&=&
\hbar 
     \bm{\hat{\psi}}_{\rm DA}^{\dagger}
\sigma_1
\bm{\hat{\psi}}_{\rm DA},
\end{eqnarray}
where $\bm{\hat{\psi}}_{\rm DA}^{\dagger}=( \hat{a}_{\rm D}^{\dagger}, \hat{a}_{\rm A}^{\dagger})$ and $\bm{\hat{\psi}}_{\rm DA}$ are the creation and the annihilation operator in the diagonal representation.
For this DA-representation, the coherent state and the average of the spin operators are best described by the polar angle $\theta^{\prime}$ measured from the $S_2$ axis and the azimuthal angle  $\phi^{\prime}$ measured from the $S_3$ axis (Fig. 3).
The expectation value is given by
\begin{eqnarray}
\langle {\bf \hat{S}} \rangle
&=
\hbar N
\left (
  \begin{array}{c}
    1 \\
    \sin \theta^{'} \sin \phi^{'}\\
    \cos \theta^{'}  \\
    \sin \theta^{'} \cos \phi^{'}
  \end{array}
\right).
\end{eqnarray}
As far as we are aware, this representation is barely used. 
\begin{figure}[h]
\begin{center}
\includegraphics[width=4cm]{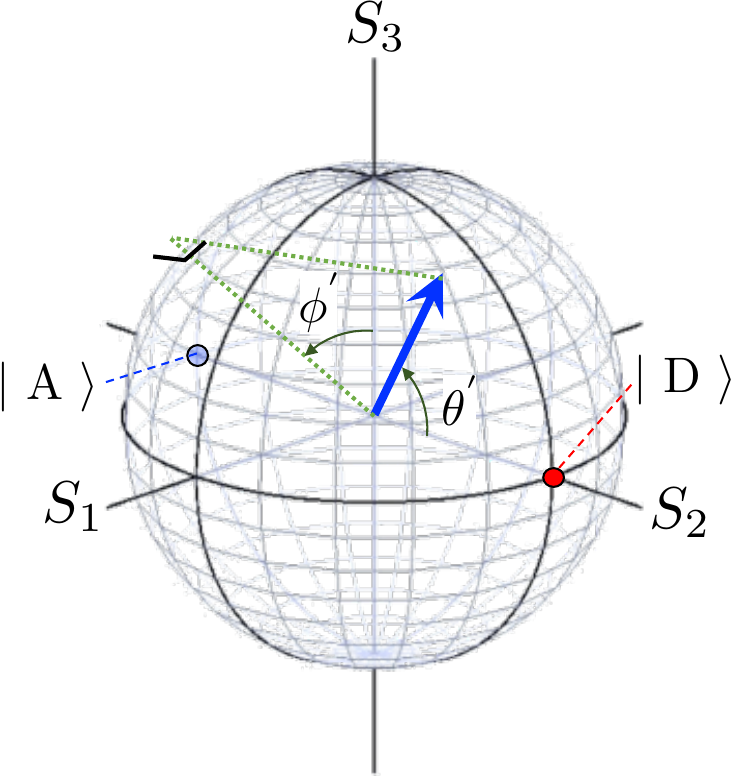}
\caption{
Poincar\'e sphere for the polarisation states, described by diagonally and anti-diagonally polarised states.
}
\end{center}
\end{figure}

\subsection{Unitary transformation from HV- to LR-bases}
Now, we realise the Jones vector treatments of the polarisation states are fully consistent with the quantum-mechanical treatment in $SU(2)$.
Here, we will double check this equivalence by transforming the Jones vector state to the corresponding representation in chiral state, which is made by the unitary transformation
\begin{eqnarray}
\left (
  \begin{array}{c}
    |{\rm H} \ \rangle \\
    |{\rm V} \ \rangle 
  \end{array}
\right)
=
\frac{{\rm e}^{i\gamma}}{\sqrt{2}}
\left (
  \begin{array}{cc}
    1 & 1 \\
   -i & i
  \end{array}
\right)
\left (
  \begin{array}{c}
    |{\rm L} \ \rangle \\
    |{\rm R} \ \rangle 
  \end{array}
\right),
\end{eqnarray}
where $\gamma$ is the uncertainty of the global $U(1)$ phase.

The original Jones vector is prepared as
\begin{eqnarray}
| \alpha, \delta \rangle
&=&
{\rm e}^{i\beta}
\left (
  \begin{array}{c}
    \cos \alpha \ \ \ \  \\
    \sin \alpha \ {\rm e}^{i\delta}
  \end{array}
\right) \\
&=&
  {\rm e}^{i\beta} \cos \alpha
  |{\rm H} \ \rangle
  +
  {\rm e}^{i\beta}\sin \alpha \ {\rm e}^{i\delta}
  |{\rm V} \ \rangle,
\end{eqnarray}
and after the unitary transformation, we obtain in the form of 
\begin{eqnarray}
| \alpha, \delta \rangle
=
C_{\rm L}|{\rm L} \ \rangle
+
C_{\rm R}|{\rm R} \ \rangle
=
| \theta, \phi \rangle.
\end{eqnarray}
Therefore, we need to determine $C_{\rm L}$ and $C_{\rm R}$, which are
\begin{eqnarray}
C_{\rm L}
&=&
{\rm e}^{i(\beta+\gamma)}
\frac{
      \cos \alpha 
       -i \ {\rm e}^{i\delta} \sin \alpha 
      }
      {
        \sqrt{2}
       }\\
C_{\rm R}
&=&
{\rm e}^{i(\beta+\gamma)}
\frac{
      \cos \alpha 
       +i \ {\rm e}^{i\delta} \sin \alpha 
      }
      {
        \sqrt{2}
       }.
\end{eqnarray}
We still need to express these as a function of $(\theta,\phi)$.

To this aid, we assume the expectation values of $\langle {\bf \hat{S}} \rangle$ are independent on the choice of the basis states.

The ratio of the coefficient becomes
\begin{eqnarray}
\frac{C_{\rm R}}{C_{\rm L}}
&=
\tan
\left ( 
\frac{\theta}{2}
\right)
{\rm e}^{i \phi}.
\end{eqnarray}
In addition, we can confirm that the wavefunction is normalised
\begin{eqnarray}
|C_{\rm L}|^2
+
|C_{\rm R}|^2
=1.
\end{eqnarray}
These equations for 2 complex values of $C_{\rm L}$ and $C_{\rm R}$ correspond to 3 equations for real values.
Therefore, we cannot determine the phase degree of freedom, ${\rm e}^{i \gamma}$.

Assuming $C_{\rm L} =l \in \ \Re$, we obtain $C_{\rm R}=l \tan \left (\theta/2\right){\rm e}^{i \phi}$. 
Inserting this into the normalisation condition, we obtain $l=\pm \cos \left ( \theta/2\right)$.
Thus, we obtain 2 states, 
\begin{eqnarray}
\left (
  \begin{array}{c}
    C_{\rm L} \\
    C_{\rm R}
  \end{array}
\right)
=
\pm
\left (
  \begin{array}{c}
\ \ \ \cos
\left ( 
\frac{\theta}{2}
\right)\\
{\rm e}^{i \phi}
\sin
\left ( 
\frac{\theta}{2}
\right)
  \end{array}
\right),
\end{eqnarray}
which yield the same expectation value but the overall sign is opposite each other.
This phase is different from the global phase, and this is nothing but a manifestation of Pancharatnam-Berry's phase \cite{Pancharatnam56,Berry84}. 
We know that the wavefunction of the polarisation state is the spinor representation of the complex electric field.
Therefore, the factor of $-1$ means that the change between $(\mathcal{E}_x,\mathcal{E}_y)$ and $(-\mathcal{E}_x,-\mathcal{E}_y)$, which cannot change the polarisation state, but the phase is observable in the interference experiments \cite{Pancharatnam56,Berry84,Tomita86}.
We can express these states together, by shifting the global phase, while keeping the relative phase, as 
\begin{eqnarray}
\left (
  \begin{array}{c}
    C_{\rm L} \\
    C_{\rm R}
  \end{array}
\right)
=
\left (
  \begin{array}{c}
{\rm e}^{-i \phi/2}
\cos
\left ( 
\frac{\theta}{2}
\right)\\
{\rm e}^{+i \phi/2}
\sin
\left ( 
\frac{\theta}{2}
\right)
  \end{array}
\right),
\end{eqnarray}
which is indeed the Bloch vector.
Here, we should consider the range of $\phi$ should be $(0,4\pi)$ to account for the change of the sign by the Pancharatnam-Berry's phase.
Thus, the Jones vector is equivalent to the Bloch vector.

It is less obvious of this Pancharatnam-Berry's phase in the above-defined Jones vector, if we describe the phase dependence as ${\rm e}^{i\delta}$.
This could be improved by shifting the global phase with the amount of ${\rm e}^{i\delta/2}$, and then Jones vector can be rewritten as
\begin{eqnarray}
| {\rm Jones} \rangle
&=&
{\rm e}^{i\beta}
\left (
  \begin{array}{c}
    {\rm e}^{-i\delta/2}\cos (\gamma/2)  \\
    {\rm e}^{i\delta/2}\sin (\gamma/2)
  \end{array}
\right) ,
\end{eqnarray}
where $\gamma=2\alpha$ is the azimuthal angle measured from $S_1$  in the Poincar\'e sphere (Fig. 2). 
In this spinor representation, it is clear that the state will change the sign after 1 rotation in Poincar\'e sphere, irrespective to whether the adiabatic rotation is the azimuthal rotation along the equator or the polar rotation along the meridian.
This is exactly the same form of the Bloch state in chiral state, such that the change of basis from LR to HV simply corresponds to change from $(\theta,\phi)$ to $(\gamma,\delta)$ in polar coordinates.
This corresponds to the cyclic exchange of Pauli matrices  from $(\sigma_1,\sigma_2,\sigma_3)$ to $(\sigma_3,\sigma_1,\sigma_2)$ (Table \ref{Table2}).
\begin{table}[h]
\caption{\label{Table2}
Summary of spin operators for each representation.
}
\begin{ruledtabular}
\begin{tabular}{cccc}
\textrm{Representation}&
\textrm{Chiral }&
\textrm{Jones }&
\textrm{Diagonal }\\
\colrule
Basis    &   $| {\rm L} \rangle$,  $| {\rm R} \rangle$ &
$| {\rm H} \rangle$,  $| {\rm V}\rangle$ &
$| {\rm D} \rangle$,  $| {\rm A}\rangle$ \\
$\hat{S}_1=\hat{S}_x$ & $\sigma_1$ &  $\sigma_3$ &  $\sigma_2$ \\
$\hat{S}_2=\hat{S}_y$ & $\sigma_2$ &  $\sigma_1$ &  $\sigma_3$ \\
$\hat{S}_3=\hat{S}_z$ & $\sigma_3$ &  $\sigma_2$ &  $\sigma_1$ 
\end{tabular}
\end{ruledtabular}
\end{table}

\subsection{Spin rotation by a $SU(2)$ group theory}
Now, we understand the spin state of a photon is described by a $SU(2)$ group theory \cite{Dirac30, Baym69,Sakurai14,Sakurai67}.
By using a standard Lie algebra, using Pauli matrices, we can obtain the many-body spin operators for the coherent monochromatic ray for photons, in a more elegant way.
The general rotation operator \cite{Dirac30, Baym69,Sakurai14,Sakurai67} along the direction ${\bf \hat{n}}$ with the amount of $\delta \phi$ is defined as
\begin{eqnarray}
\hat{\mathcal{D}}({\bf \hat{n}},\delta \phi)
&=\exp 
\left (
-i 
{\bm \sigma}\cdot {\bf \hat{n}}
\left (
\frac{\delta \phi}{2}
\right)
\right),
\end{eqnarray}
where $|\hat{\bf \hat{n}}|=1$ and ${\bm \sigma}=(\sigma_1,\sigma_2,\sigma_3)$.
We have chosen the direction of rotation in a standard mathematical way.
Specifically, the positive rotation along $z$ is equivalent to the left-hand rotation (anti-clock-wise) rotation in $xy$-plane, seen from the top of the $z$ axis,  which is equivalent to see from the observer in the detector side.
By expanding the exponential and using the formulas, 
$\left \{\sigma_{i},\sigma_{j}\right \}=2\delta_{ij}{\bf 1}$, and 
$\left [ \sigma_{i},\sigma_{j} \right ]=2 i \epsilon_{ijk}\sigma_{k}$,  
we obtain  \cite{Dirac30, Baym69,Sakurai14,Sakurai67}
\begin{eqnarray}
\hat{\mathcal{D}} ({\bf \hat{n}},\delta \phi)
=& {\bf 1} \cos \left( \frac{\delta \phi}{2} \right) 
-i {\bm \sigma}\cdot {\bf \hat{n}} \sin \left( \frac{\delta \phi}{2} \right).
\end{eqnarray}
In particular, we describe the rotation around $x$, $y$, and $z$, axes as $\hat{\mathcal{D}}_1 (\delta \phi)=\hat{\mathcal{D}}_x (\delta \phi)=\hat{\mathcal{D}}(\hat{\bf x},\delta \phi)$,  $\hat{\mathcal{D}}_2 (\delta \phi)=\hat{\mathcal{D}}_y (\delta \phi)=\hat{\mathcal{D}}(\hat{\bf y},\delta \phi)$, and  $\hat{\mathcal{D}}_3 (\delta \phi)=\hat{\mathcal{D}}_z (\delta \phi)=\hat{\mathcal{D}}(\hat{\bf z},\delta \phi)$, for simplicity.

Previously, as outlined above, the spin operator along the direction of propagation ($\hat{S}_z$) was obtained by using the Poynting vector and considerations of orbital angular momentum \cite{Allen92,Enk94,Leader14,Barnett16,
Yariv97,Jackson99,Grynberg10,Bliokh15,Enk94,Leader14,Barnett16,Chen08,Ji10}.
Then, we can obtain the spin operator along x, $\hat{S}_x$, by rotating $\hat{S}_z$ with the amount of $\pi/2$ along $y$, and therefore, 
\begin{eqnarray}
\hat{S}_x
&=&
\hat{\mathcal{D}}_y  
\left(
  \frac{\pi}{2}
\right)
\hat{S}_z
\hat{\mathcal{D}}_y ^{\dagger} 
\left(
  \frac{\pi}{2}
\right)\\
&=&
\hat{\mathcal{D}}_y  
\left(
  \frac{\pi}{2}
\right)
{\bm \psi}_{\rm LR}^{\dagger}
\hbar 
\sigma_3
{\bm \psi}_{\rm LR}
\hat{\mathcal{D}}_y ^{\dagger} 
\left(
  \frac{\pi}{2}
\right)
\\
&=&
{\bm \psi}_{\rm LR}^{\dagger}
\hat{\mathcal{D}}_y  
\left(
  \frac{\pi}{2}
\right)
\hbar 
\sigma_3
\hat{\mathcal{D}}_y ^{\dagger} 
\left(
  \frac{\pi}{2}
\right)
{\bm \psi}_{\rm LR}
\\
&=&
\hbar 
{\bm \psi}_{\rm LR}^{\dagger}
\sigma_1
{\bm \psi}_{\rm LR}.
\end{eqnarray}

Similarly, we obtain $\hat{S}_y$ from $\hat{S}_z$ by rotating along x with the amount of $-\pi/2$ as
\begin{eqnarray}
\hat{S}_y
&=&
\hat{\mathcal{D}}_x  
\left(
  -\frac{\pi}{2}
\right)
\hat{S}_z
\hat{\mathcal{D}}_x ^{\dagger} 
\left(
  -\frac{\pi}{2}
\right)\\
&=&
\hat{\mathcal{D}}_x  
\left(
  - \frac{\pi}{2}
\right)
{\bm \psi}_{\rm LR}^{\dagger}
\hbar 
\sigma_3
{\bm \psi}_{\rm LR}
\hat{\mathcal{D}}_x ^{\dagger} 
\left(
 - \frac{\pi}{2}
\right)
\\
&=&
{\bm \psi}_{\rm LR}^{\dagger}
\hat{\mathcal{D}}_x  
\left(
  - \frac{\pi}{2}
\right)
\hbar 
\sigma_3
\hat{\mathcal{D}}_x ^{\dagger} 
\left(
  - \frac{\pi}{2}
\right)
{\bm \psi}_{\rm LR}
\\
&=&
\hbar 
{\bm \psi}_{\rm LR}^{\dagger}
\sigma_2
{\bm \psi}_{\rm LR}.
\end{eqnarray}

Alternatively, we can also rotate $2\pi/3$ along $(1,1,1)/\sqrt{3}$ direction, for cyclic permutation of axes.
The rotation operator becomes
\begin{eqnarray}
\hat{\mathcal{D}}
\left(
\frac{(1,1,1)}{\sqrt{3}},\frac{2 \pi}{3}
\right)
&=
\frac{1}{2}
-
\frac{i}{2}
(
\sigma_1 + \sigma_2 + \sigma_3
),
\end{eqnarray}
which yields
\begin{eqnarray}
\hat{\mathcal{D}}
\left(
\frac{(1,1,1)}{\sqrt{3}},\frac{2 \pi}{3}
\right)
\sigma_3
\hat{\mathcal{D}}
\left(
\frac{(1,1,1)}{\sqrt{3}},\frac{2 \pi}{3}
\right) ^{\dagger} 
=
\sigma_1.
\end{eqnarray}
Therefore, we successfully obtain $\hat{S}_x$, and the opposite rotation yield $\hat{S}_y$.

For the expressions using HV-basis, we can use a unitary transformation
\begin{eqnarray}
\left (
  \begin{array}{c}
    \hat{a}_{\rm H}^{\dagger} \\
    \hat{a}_{\rm V}^{\dagger}
  \end{array}
\right)
=
\frac{1}{\sqrt{2}}
\left (
  \begin{array}{cc}
    1 & 1\\
    -i & i
  \end{array}
\right)
\left (
  \begin{array}{c}
    \hat{a}_{\rm L}^{\dagger} \\
    \hat{a}_{\rm R}^{\dagger}
  \end{array}
\right),
\end{eqnarray}
and its conjugate
\begin{eqnarray}
\left (
  \begin{array}{c}
    \hat{a}_{\rm H} \\
    \hat{a}_{\rm V}
  \end{array}
\right)
=
\frac{1}{\sqrt{2}}
\left (
  \begin{array}{cc}
    1 & 1\\
    i & -i
  \end{array}
\right)
\left (
  \begin{array}{c}
    \hat{a}_{\rm L} \\
    \hat{a}_{\rm R}
  \end{array}
\right).
\end{eqnarray}
We can, of course, come back to LR-basis from HV-basis by the inverse unitary transformation.
The transfer to the DA-basis is also straightforward by using the unitary transformation
\begin{eqnarray}
\left (
  \begin{array}{c}
    \hat{a}_{\rm D}^{\dagger} \\
    \hat{a}_{\rm A}^{\dagger}
  \end{array}
\right)
=
\frac{1}{\sqrt{2}}
\left (
  \begin{array}{cc}
    1 & 1\\
    1 & -1
  \end{array}
\right)
\left (
  \begin{array}{c}
    \hat{a}_{\rm H}^{\dagger} \\
    \hat{a}_{\rm V}^{\dagger}
  \end{array},
\right)
\end{eqnarray}
and its conjugate
\begin{eqnarray}
\left (
  \begin{array}{c}
    \hat{a}_{\rm D} \\
    \hat{a}_{\rm A}
  \end{array}
\right)
=
\frac{1}{\sqrt{2}}
\left (
  \begin{array}{cc}
    1 & 1\\
    1 & -1
  \end{array}
\right)
\left (
  \begin{array}{c}
    \hat{a}_{\rm H} \\
    \hat{a}_{\rm V}
  \end{array}
\right).
\end{eqnarray}

The summary of the assignments of Pauli matrices to spin operator components for each representation is given by Table \ref{Table2}.

\subsection{Rotation in real space}
Here, we consider the rotation in real space rather than Hilbert  space for spin.
We define the rotation operators  for the amount of the rotation of $\delta \phi$ along $x$, $y$, and $z$ axes as $\mathcal{R}_x(\delta \phi)$, $\mathcal{R}_y(\delta \phi)$ , $\mathcal{R}_z(\delta \phi)$, respectively.
These are rotations in a $SO(3)$ (Special Orthogonal) group theory.
By applying 2 successive rotations along $y$  and $x$, 
\begin{eqnarray}
\mathcal{R}_x
\left(
\frac{\pi}{2}
\right) 
\mathcal{R}_y
\left(
\frac{\pi}{2}
\right)  
\hat{\bf x}
&=&
\hat{\bf y}
\\
\mathcal{R}_x
\left(
\frac{\pi}{2}
\right) 
\mathcal{R}_y
\left(
\frac{\pi}{2}
\right)  
\hat{\bf y}
&=&
\hat{\bf z}
\\
\mathcal{R}_x
\left(
\frac{\pi}{2}
\right) 
\mathcal{R}_y
\left(
\frac{\pi}{2}
\right)  
\hat{\bf z}
&=&
\hat{\bf x}
\end{eqnarray}
we can perform cyclic exchange of axes from $(x,y,z)$ to $(y,z,x)$ as
\begin{eqnarray}
\mathcal{R}_x
\left(
\frac{\pi}{2}
\right) 
\mathcal{R}_y
\left(
\frac{\pi}{2}
\right)  
=
\left (
  \begin{array}{ccc}
    0 & 0 & 1 \\
    1 & 0 & 0 \\
    0 & 1 & 0 
  \end{array}
\right).
\end{eqnarray}

If we apply this rotations to $\hat{{\bf E}}$, we obtain the corresponding electric field operator after 2 successive rotations as
\begin{eqnarray}
\hat{
{\bf E}}^{''}=&&
\frac{E_0}{2}
\left(
  \hat{a}_{\rm H}
  {\rm e}^{i \beta^{''}}
  +
  \hat{a}_{\rm H}^{\dagger}
  {\rm e}^{-i \beta^{''}}
\right)
  \hat{\bf y}\nonumber \\
&&+
\frac{E_0}{2}
\left(
  \hat{a}_{\rm V}
  {\rm e}^{i \beta^{''}}
  +
  \hat{a}_{\rm V}^{\dagger}
  {\rm e}^{-i \beta^{''}}
\right)
  \hat{\bf z}, \nonumber \\
\end{eqnarray}
where $\beta^{''}=kx-\omega t + \delta_x$.

Then, by applying the same argument using the Poynting vector, we obtain the momentum operator and the total angular momentum after the rotations as
\begin{eqnarray}
\hat{\bf P}^{''}
=&
\hbar k_{n_0}
\left(
\hat{n}_{\rm H}
+
\hat{n}_{\rm V}
+1
\right)
\hat{\bf x},
\end{eqnarray}

\begin{eqnarray}
\hat{\bf J}^{''}_z
=
\hat{\bf S}^{''}_z
=
\hat{S}_z
\hat{\bf x},
\end{eqnarray}
where 
\begin{eqnarray}
\hat{S}_z
=&
\hbar
\left(
\hat{n}_{\rm L}
-
\hat{n}_{\rm R}
\right).
\end{eqnarray}
This means that the spatial rotations simply change the direction of propagation, but the polarisation state is not changed.
We could also confirm that the polarisation state has not been changed by the opposite rotation for the left cyclic exchange from $(x,y,z)$ to  $(z,y,x)$ by using the $SO(3)$ rotation
\begin{eqnarray}
\mathcal{R}_x
\left(
-
\frac{\pi}{2}
\right) 
\mathcal{R}_y
\left(
-
\frac{\pi}{2}
\right)  
=
\left (
  \begin{array}{ccc}
    0 & 1 & 0 \\
    0 & 0 & 1 \\
    1 & 0 & 0 
  \end{array}
\right).
\end{eqnarray}

Of course, the polarisation should not depend on the choice of the spatial coordinate, since the polarisation is a measure to evaluate the relative phase between orthogonal polarisation states, which cannot be changed by the rotation for the direction of propagation, which is responsible to the overall phase of both polarisation states, equally.

$\hat{S}_z$ and thus $S_3=\langle \hat{S}_z \rangle$ are inherently linked to the direction of the propagation.
Therefore, it is natural to regard $\hat{S}_z$ as the helicity operator \cite{Barnett12},
\begin{eqnarray}
\hat{h}_z
=&
\hbar
\left(
\hat{n}_{\rm L}
-
\hat{n}_{\rm R}
\right).
\end{eqnarray}
The helicity is usually defined  as the projection of  the spin onto the direction of the propagation, and in fact $\hat{h}_z={\bf \hat{S}}\cdot \hat{\bf z}$, for the light propagating along $z$. 
We have defined the polarisation states from the electro-magnetic field oscillations seen from the detector side, such that the helicity operator becomes 
\begin{eqnarray}
\hat{h}
=&
\hbar
\left(
\hat{n}_{\rm L}
-
\hat{n}_{\rm R}
\right),
\end{eqnarray}
independent on the direction of propagation.

We could also define our spin operators as 
\begin{eqnarray}
\hat{S}_{1}
&=&
\hbar 
     \bm{\hat{\psi}}_{\rm LR}^{\dagger}
\sigma_1
\bm{\hat{\psi}}_{\rm LR} \\
\hat{S}_{2}
&=&
\hbar 
     \bm{\hat{\psi}}_{\rm LR}^{\dagger}
\sigma_2
\bm{\hat{\psi}}_{\rm LR} \\
\hat{S}_{3}
&=&
\hbar 
     \bm{\hat{\psi}}_{\rm LR}^{\dagger}
\sigma_3
\bm{\hat{\psi}}_{\rm LR},
\end{eqnarray}
to emphasise the direct relevance to the Stokes parameters as $\langle \hat{S}_{i} \rangle = S_i$ for  $^{\forall}i=0,1,2,3$.
Even in this notation, we still need to clarify the  direction of the propagation, otherwise the spin state and the polarisation state cannot be properly specified. 
The way to define the rotation, whether the phase front is rotating to the left (anti-clock-wise) or to the right (clock-wise), depends crucially on the definition of the direction of the propagation.
The definition of the rotation is also important and we have assumed the polarisation is seen from the observer (detector) in this paper.
We have discussed, by assuming the direction of the propagation is mostly along $z$, and defined spin operators, accordingly.
Here, we have shown that the polarisation state, thus defined, should not depend on the direction of the propagation, such that the spin operators and associated expectation values as Stokes parameters should not depend on the choice of the coordinate.
The direction of the propagation of photons naturally set the quantisation axis for their inherent spin states, as confirmed by the spatial  integration of the the outer product between ${\bf r}$ and the Poynting vector.
We confirmed $\hat{S}_{z}=\hat{S}_{3}=\hat{h}_z=\hat{h}$ is always aligned to the direction of propagation.

One might attempt to align the direction of spin operators to a specific axis in an arbitrary chosen coordinate.
However, in this case, the artificially fixed spin operator is not always aligned to the direction of the propagation, such that $\hat{S}_{z}$ may not be aligned to the direction of the propagation.
In such a coordinate, it is very difficult to discuss the polarisation state, even if it is possible.

The choice of the coordinate should be arbitrary, according to Einstein's theory of relativity.
The quantisation axis of the spin operator for describing the amount of the circular polarised state is naturally aligned to the direction of the propagation.
We do not know why the spin quantisation axis is locked to the direction of the propagation, but if we accept this as a principle, we could construct spin operators for other components, just by following a standard quantum-mechanical prescription and a $SU(2)$ group theory.

\section{Applications}
As applications of our formalism, we consider several typical optical components to control the polarisation states \cite{Jones41,Stokes51,Poincare92,Yariv97,Goldstein11,Gil16}. 
Practically, this is nothing new compared with well-established Jones matrix formulation, but the purpose of this consideration is to establish a fundamental basis to justify the calculation of  polarisation states using Jones matrices based on a many-body quantum physics and a $SU(2)$ group theory.

\subsection{Phase-shifter}
\subsubsection{Phase-shifter in HV-basis}
A phase-shifter is an optical component, which control the phase of $\delta$ by injecting a coherent laser beam with a specific polarisation state into it and changing the polarisation state of the output beam \cite{Yariv97,Goldstein11,Gil16}. 
It is also called as a retarder, but we prefer to call it as a phase-shifter, because we can allow both retardation and advancement of the phase, just by changing the angle of the optical component.
It is also called as a wave-plate.
It is best-described by HV-basis, so that we will discuss in HV-basis, first and then, transform the formulas to those in LR-basis.

The working principle of the phase-shifter is quite simple. 
It is based on a birefringence of a transparent single crystal such as quartz, LiNbO$_3$, and other  transparent single crystals \cite{Yariv97,Goldstein11,Gil16}.
In these birefringent materials, the values of the refractive index depend on the direction of the propagation against their crystal axis.
The axis for the large refractive index ($n_{s}$) is called as a slow axis, and  the axis for the small refractive index ($n_{f}$) is called as a fast axis, because the phase velocity of the slow axis ($v_{\rm s}$) is slower than the phase velocity of the fast axis ($v_{\rm f}$).
We abbreviate slow axis as SA and fast axis as FA.
This induces the polarisation dependent phase-shift, through the factor of ${\rm e}^{ikx}$.

\begin{figure}[h]
\begin{center}
\includegraphics[width=6cm]{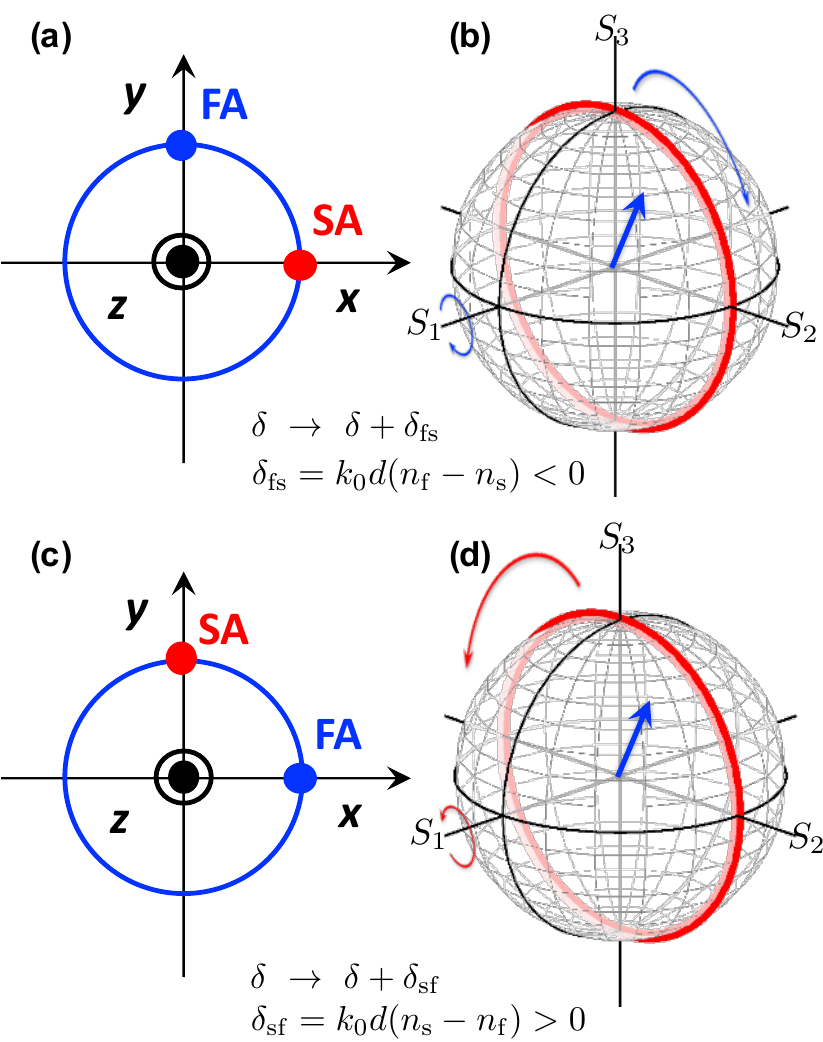}
\caption{
Phase-shifter and its impact on the polarisation state.
(a) Retarder configuration. Slow axis (SA) is aligned horizontally.
(b) Clock-wise rotation of the polarisation state by a retarder.
(c) Phase-shifter (phase-advancement) configuration. Fast axis (FA) is aligned horizontally.
(d) Anti-clock-wise rotation of the polarisation state by a phase-shifter.
This rotation is  described by $\Delta_{\rm HV}(\delta_{\rm sf})$ or $\Delta_{\rm LR}(\delta_{\rm sf})$.
}
\end{center}
\end{figure}

Specifically, first, we consider the retarder configuration, (Fig. 4 (a) and (b)), which means that the SA is aligned horizontally, and we expect the phase delay given by $\delta_{\rm fs}=k_0 (n_{\rm f}-n_{\rm s}) d<0$, where $k_0=2\pi/\lambda$ is the wavenumber in the vacuum, $\lambda$ is  the wavelength in the vacuum, and $d$ is the thickness of the wave plate.
The wavenumbers for SA and FA are given by $k_{\rm s}=k_0 n_{\rm s}$ and $k_{\rm f}=k_0 n_{\rm f}$, respectively.
We also define  the average wavenumber as $\bar{k}=(k_{\rm s}+k_{\rm f})/2$.

The many-body operator to describe this change is given by the following phase-shifter operator
\begin{eqnarray}
\hat{\Delta}_{\rm HV}
(\delta_{fs})
=
{\rm e}^{ik_{\rm s}d}
\frac{{\rm e}^{i\beta}}{\sqrt{N}}
\hat{a}_{\rm H}\hat{\bf x}
+
{\rm e}^{ik_{\rm f}d}
\frac{{\rm e}^{i\beta}}{\sqrt{N}}
\hat{a}_{\rm V}\hat{\bf y}.
\end{eqnarray}
By applying this to the coherent state, we obtain
\begin{eqnarray}
&&\hat{\Delta}_{\rm HV}
(\delta_{fs})
|\alpha_{\rm H},\alpha_{\rm V}\rangle
\nonumber \\ 
&&=
{\rm e}^{i\beta}
{\rm e}^{i\overline{k}d}
\left (
  \begin{array}{cc}
    {\rm e}^{-i\frac{\delta_{\rm fs}}{2}} & 0 \\
    0 & {\rm e}^{+i\frac{\delta_{\rm fs}}{2}} 
  \end{array}
\right)
\left (
  \begin{array}{c}
    {\rm e}^{-i\frac{\delta}{2}} \cos \alpha \\
    {\rm e}^{+i\frac{\delta}{2}}\sin \alpha \   
\end{array}
\right)
|\alpha_{\rm H},\alpha_{\rm V}\rangle \nonumber \\
\end{eqnarray}
which  means that $\hat{\Delta}_{\rm HV}$ does not change the number of photons and just affect the polarisation state.
If we multiply the ket vector of the coherent state, $\langle \alpha_{\rm H},\alpha_{\rm V}|$, from the left, we obtain the expectation value of the phase-shifter operator in the retarder configuration as 
\begin{eqnarray}
&&
\langle \alpha_{\rm H},\alpha_{\rm V}|
\hat{\Delta}_{\rm HV}
(\delta_{fs})
|\alpha_{\rm H},\alpha_{\rm V}\rangle
\nonumber \\ 
&&=
{\rm e}^{i\beta}
{\rm e}^{i\overline{k}d}
\left (
  \begin{array}{cc}
    {\rm e}^{-i\frac{\delta_{\rm fs}}{2}} & 0 \\
    0 & {\rm e}^{+i\frac{\delta_{\rm fs}}{2}} 
  \end{array}
\right)
\left (
  \begin{array}{c}
    {\rm e}^{-i\frac{\delta}{2}} \cos \alpha \\
    {\rm e}^{+i\frac{\delta}{2}}\sin \alpha \   
\end{array}
\right) \\
&&=
\Delta_{\rm HV} (\delta_{\rm fs}) | {\rm Jones}\rangle,
\end{eqnarray}
where we have defined the Jones matrix for the phase-shifter as
\begin{eqnarray}
\Delta_{\rm HV} (\delta_{\rm fs})
=
{\rm e}^{i\beta}
{\rm e}^{i\overline{k}d}
\left (
  \begin{array}{cc}
    {\rm e}^{-i\frac{\delta_{\rm fs}}{2}} & 0 \\
    0 & {\rm e}^{+i\frac{\delta_{\rm fs}}{2}} 
  \end{array}
\right)
\end{eqnarray}
Therefore, we can calculate the polarisation state of the ray after the propagation of the retarder by using the Jones matrix \cite{Jones41,Stokes51,Poincare92,Yariv97,Goldstein11,Gil16}.
In the Poincar\'e sphere, this corresponds to rotate $\bf{S}$ along $S_1$ with the amount of $\delta_{\rm fs}$ (Fig. 4 (b)).
This corresponds to the clock-wise rotation, since $\delta_{\rm fs}<0$.

Next, we consider the phase-shifter configuration, which corresponds to increase the phase by aligning FA horizontally (Figs. 4(c) and (d)).
In this case, the phase-shifter operator is given by
\begin{eqnarray}
\hat{\Delta}_{\rm HV}
(\delta_{sf})
=
{\rm e}^{ik_{\rm f}d}
\frac{{\rm e}^{i\beta}}{\sqrt{N}}
\hat{a}_{\rm H}\hat{\bf x}
+
{\rm e}^{ik_{\rm s}d}
\frac{{\rm e}^{i\beta}}{\sqrt{N}}
\hat{a}_{\rm V}\hat{\bf y},
\end{eqnarray}
 which just corresponds to exchanging SA and FA, so that the above formulas are valid just by replacing $\delta_{\rm fs}$ to $\delta_{\rm sf}=k_0 (n_{\rm s}-n_{\rm f}) d>0$.
Therefore, we obtain the Jones matrix
\begin{eqnarray}
\Delta_{\rm HV} (\delta_{\rm sf})
=
{\rm e}^{i\beta}
{\rm e}^{i\overline{k}d}
\left (
  \begin{array}{cc}
    {\rm e}^{-i\frac{\delta_{\rm sf}}{2}} & 0 \\
    0 & {\rm e}^{+i\frac{\delta_{\rm sf}}{2}} 
  \end{array}
\right).
\end{eqnarray}
In this case, the operation of the phase-shifter is simply the phase-shift of $\delta \rightarrow \delta+\delta_{\rm sf}$.
Alternatively, we can regard the retarder as a special case of the phase-shifter with opposite rotation.
The phase-shifter configuration (horizontal FA) is our preferable configuration to think about the rotation in Poincar\'e sphere, since we can consider positive rotation, but of course, we can use both configurations depending on the applications.

It is now clear that the phase-shifter operator will change the polarisation state of the coherent state as an out put beam,
\begin{eqnarray}
&&|{\rm output} \rangle 
=\hat{\Delta}_{\rm HV}
(\delta_{fs})
|{\rm input}\rangle
\nonumber \\ 
&&=
{\rm e}^{i\beta}
{\rm e}^{i\overline{k}d}
\left (
  \begin{array}{c}
    {\rm e}^{-i\frac{\delta+\delta_{\rm fs}}{2}} \cos \alpha \\
    {\rm e}^{+i\frac{\delta+\delta_{\rm fs}}{2}}\sin \alpha \   
\end{array}
\right)
|{\rm input}\rangle, \\
\end{eqnarray}
where the input beam is $|{\rm input}\rangle = |\alpha_{\rm H},\alpha_{\rm V}\rangle$.
Therefore, the phase-shifter changes the relative phase to describe the spin states, while the coherency of the monochromatic ray of photons, described by coherent states is preserved. 
This aspect can be more clearly confirmed by calculating the average quantum-mechanical expectation value of the spin of photons, using the output state, 
\begin{eqnarray}
\langle {\rm output}| {\bf \hat{S}} |{\rm output} \rangle
&=
\hbar N
\left (
  \begin{array}{c}
    \cos (\gamma) \\
    \sin (\gamma) \cos \left( \delta +\delta_{\rm fs}  \right)\\
    \sin (\gamma) \sin \left( \delta +\delta_{\rm fs}  \right)
  \end{array}
\right). \nonumber \\
\end{eqnarray}
Thus, the spin is rotated along $S_1$ with the amount of $\delta_{\rm fs}$  by the phase-shifter (Fig. 4 (d)).

\subsubsection{Phase-shifter as a rotator in $SU(2)$ Hilbert space}
Aside from the  overall phase factor, the phase shifter can be described by a rotation in $SU(2)$ Hilbert space for spin of a photon.
The phase-shifter corresponds to the rotation along $S_1$, such that the phase-shifter matrix in HV-basis is described as
\begin{eqnarray}
\mathcal{D}_1^{\rm HV}(\delta_{\rm sf})
&=&
\exp 
\left (
  -\frac{i \sigma_3 \delta_{\rm sf}}{2}
\right) \\
&=&
{\bf 1}
\cos 
\left(
  \frac{\delta_{\rm sf}}{2}
\right)
-
i 
\sigma_3
\sin 
\left(
  \frac{\delta_{\rm sf}}{2}
\right)\\
&=&
\left (
  \begin{array}{cc}
    \exp \left( -i\frac{\delta_{\rm sf}}{2} \right) & 
   0  \\
    0  & 
    \exp \left(+ i\frac{\delta_{\rm sf}}{2} \right)
  \end{array}
\right).
\end{eqnarray}
Combined with the overall phase factor, coming from the average global phase of the orbital wavefunction, we obtain 
\begin{eqnarray}
{\Delta}_{\rm HV}(\delta_{\rm sf})
&=&
{\rm e}^{i\beta}
{\rm e}^{i\overline{k}d}
\mathcal{D}_1^{\rm HV}(\delta_{\rm sf}) \\
&=&
{\rm e}^{i\beta}
{\rm e}^{i\overline{k}d}
\left (
  \begin{array}{cc}
    {\rm e}^{-i\frac{\delta_{\rm sf}}{2}} & 
   0  \\
    0  & 
    {\rm e}^{+ i\frac{\delta_{\rm sf}}{2} }
  \end{array}
\right),
\end{eqnarray}
which  is exactly the same as that obtained previously.
Therefore, the $SU(2)$ group theory is a powerful method to consider the impact of the phase-shifter in Poincar\'e sphere.

The retarder configuration (horizontal SA) is obviously corresponds to the opposite rotation (Figs. 4 (b) and  (d)), whose operator form is obtained by the change of sign, due to $\delta_{\rm fs}=-\delta_{\rm sf}$.

\subsubsection{Phase-shifter in LR-basis}
Here, we obtain the phase-shifter operator in chiral LR-basis.
In the LR-basis, the rotation along $S_1$ is described by $\sigma_1$ (Table \ref{Table2}).
Therefore, the Jones matrix for the phase-shifter in LR-basis is 
\begin{eqnarray}
{\Delta}_{\rm LR}(\delta_{\rm sf})
&=&
{\rm e}^{i\beta}
{\rm e}^{i\overline{k}d}
\mathcal{D}_1^{\rm LR}(\delta_{\rm sf}) \\
&=&
{\rm e}^{i\beta}
{\rm e}^{i\overline{k}d}
\exp 
\left (
  -\frac{i \sigma_1 \delta_{\rm sf}}{2}
\right)\\
&=&
{\rm e}^{i\beta}
{\rm e}^{i\overline{k}d}
\left(
{\bf 1}
\cos 
\left(
  \frac{\delta_{\rm sf}}{2}
\right)
-
i 
\sigma_1
\sin 
\left(
  \frac{\delta_{\rm sf}}{2}
\right)
\right) \nonumber \\
&=&
{\rm e}^{i\beta}
{\rm e}^{i\overline{k}d}
\left (
  \begin{array}{cc}
    \cos \left( \frac{\delta_{\rm sf}}{2} \right) & 
   -i\sin \left( \frac{\delta_{\rm sf}}{2} \right) \\
   -i\sin \left( \frac{\delta_{\rm sf}}{2} \right) & 
    \cos \left( \frac{\delta_{\rm sf}}{2} \right)
  \end{array}
\right).
\end{eqnarray}

\subsubsection{Unitary operation}
LR-basis can be transferred to HV-basis by a unitary transformation, 
\begin{eqnarray}
U_{\rm HV}
=
\frac{1}{\sqrt{2}}
\left (
  \begin{array}{cc}
    1 &     1  \\
   i  &     -i 
  \end{array}
\right),
\end{eqnarray}
and {\it vice versa} by the inverse
\begin{eqnarray}
U^{\dagger}_{\rm HV}
=
\frac{1}{\sqrt{2}}
\left (
  \begin{array}{cc}
    1 &     -i  \\
   1  &     i 
  \end{array}
\right).
\end{eqnarray}
Therefore, any operator in HV-basis, $O_{\rm HV}$, can be transferred to that in LR-basis, $O_{\rm HV}$, by the unitary transformation 
\begin{eqnarray}
O_{\rm LR}
=U^{-1}_{\rm HV}O_{\rm HV}U_{\rm HV},
\end{eqnarray}
which means that the state in LR-basis is first transformed to HV-basis by $U_{\rm HV}$, operated in HV-space by $O_{\rm HV}$ , and finally brought back to LR-basis by $U^{-1}_{\rm HV}$.
We confirm this operation for the above obtained phase-shifter.
In order to confirm, we directly calculated 
\begin{eqnarray}
&&\Delta_{\rm LR}(\delta_{\rm sf})
=U^{-1}_{\rm HV} \Delta_{\rm HV}U_{\rm HV}  \\
&&=
\frac{1}{\sqrt{2}}
\left (
  \begin{array}{cc}
    1 & -i \\
    1 &  i
  \end{array}
\right)
{\rm e}^{i\beta}
{\rm e}^{i\overline{k}d}
\left (
  \begin{array}{cc}
    {\rm e}^{-i\frac{\delta_{\rm sf}}{2}} & 
   0  \\
    0  & 
    {\rm e}^{+ i\frac{\delta_{\rm sf}}{2} }
  \end{array}
\right)
\frac{1}{\sqrt{2}}
\left (
  \begin{array}{cc}
    1 & 1 \\
    i & -i
  \end{array}
\right) \nonumber \\
&&=
{\rm e}^{i\beta}
{\rm e}^{i\overline{k}d}
\left (
  \begin{array}{cc}
    \cos \left( \frac{\delta_{\rm sf}}{2} \right) & 
   -i\sin \left( \frac{\delta_{\rm sf}}{2} \right) \\
   -i\sin \left( \frac{\delta_{\rm sf}}{2} \right) & 
    \cos \left( \frac{\delta_{\rm sf}}{2} \right)
  \end{array}
\right),
\end{eqnarray}
which is indeed successfully transferred.
Therefore, the unitary transformation is useful to change the basis states.

\subsection{Rotator}
\subsubsection{Rotator in LR-basis}
The idea of the phase-shifter is to utilise the orbital degree of the wavefunction to tune the relative phase between two orthogonal polarisation states by using a polarisation dependent material for changing the polarisation state.
The directional dependence of the refractive indexes was the key ingredient for enabling the phase-shift.

Here, we show very similar formulation is applicable to a rotator. 
In a rotator, the chiral dependence of the refractive indexes are used to control the relative phase between left and right circular polarised states.
Consequently, it is straightforward to construct an operator in the chiral LR-basis.

The key ingredient for enabling the chirality control is the refractive index dependence on chirality in materials like quartz and liquid crystal \cite{Hecht17,Barger87}.
One of the most important application of the control of chirality of photons is the use for a Liquid-Crystal-Display (LCD).
By applying voltage to the transparent capacitor, organic molecule sandwiched between two parallel electrodes of capacitors can align towards the electric field, which changes the refractive index to switch the pixel on and off.
Left-handed quartz and right-handed quartz are also known to be optically active materials due to their chiral atomic arrangements of the network of silicon-oxide bonds \cite{Hecht17}.
A material with a chiral dependence should have such atomic or molecular structures, which are optically active dependent on the polarisation state of the chiral $S_3$ component.

We assume a rotator made of an optically active material with the thickness of $d$ and the refractive indexes for left and right circular-polarised states are $n_{\rm L}$ and $n_{\rm R}$, whose wavenumbers are $k_{\rm L}=k_{0}n_{\rm L}$ and $k_{\rm R}=k_{0}n_{\rm R}$,  respectively.
Then, the many-body rotator operator is given by
\begin{eqnarray}
\hat{\mathcal{R}}_{\rm LR}
(\Delta \phi)
=
{\rm e}^{ik_{\rm L}d}
\frac{{\rm e}^{i\beta}}{\sqrt{N}}
\hat{a}_{\rm L}\hat{\bf l}
+
{\rm e}^{ik_{\rm R}d}
\frac{{\rm e}^{i\beta}}{\sqrt{N}}
\hat{a}_{\rm R}\hat{\bf r},
\end{eqnarray}
where $\Delta \phi$ is the amount of the rotation, which we shall obtain next.
By applying this to the coherent state, we obtain the output state
\begin{eqnarray}
&&|{\rm output} \rangle
=\hat{\mathcal{R}}_{\rm LR}
(\Delta \phi)
|\alpha_{\rm L},\alpha_{\rm R}\rangle
\nonumber \\ 
&&=
{\rm e}^{i\beta}
{\rm e}^{i\overline{k}d}
\left (
  \begin{array}{cc}
    {\rm e}^{-i\frac{\Delta \phi}{2}} & 0 \\
    0 & {\rm e}^{+i\frac{\Delta \phi}{2}} 
  \end{array}
\right)
\left (
  \begin{array}{c}
    {\rm e}^{-i\frac{\phi}{2}} \cos (\theta/2) \\
    {\rm e}^{+i\frac{\phi}{2}}\sin (\theta/2) \   
\end{array}
\right)
|\alpha_{\rm L},\alpha_{\rm R}\rangle, \nonumber \\
\end{eqnarray}
where $\Delta \phi=2 \rho d$ is the rotation angle of the azimuthal direction in Poincar\'e sphere, $\Delta \Psi=\rho d$ it the rotation angle of the inclination angle for the electric field of the principal axis in the polarisation ellipse, and $\rho=(k_{\rm R}-k_{\rm L})d/2=\pi (n_{\rm R}-n_{\rm L})/\lambda$.
By applying $\langle \alpha_{\rm L},\alpha_{\rm R}|$ from the left, we obtain
\begin{eqnarray}
&&\langle \alpha_{\rm L},\alpha_{\rm R}| 
\hat{\mathcal{R}}_{\rm LR}
(\Delta \phi)
|\alpha_{\rm L},\alpha_{\rm R}\rangle
\nonumber \\ 
&&=
{\rm e}^{i\beta}
{\rm e}^{i\overline{k}d}
\left (
  \begin{array}{cc}
    {\rm e}^{-i\frac{\Delta \phi}{2}} & 0 \\
    0 & {\rm e}^{+i\frac{\Delta \phi}{2}} 
  \end{array}
\right)
\left (
  \begin{array}{c}
    {\rm e}^{-i\frac{\phi}{2}} \cos (\theta/2) \\
    {\rm e}^{+i\frac{\phi}{2}}\sin (\theta/2) \   
\end{array}
\right) \\
&&=
\mathcal{R}_{\rm LR}
(\Delta \phi)
| {\rm Bloch} \rangle , 
\end{eqnarray}
where we have obtained the Jones matrix for a rotator as
\begin{eqnarray}
\mathcal{R}_{\rm LR}
(\Delta \phi) 
=
{\rm e}^{i\beta}
{\rm e}^{i\overline{k}d}
\left (
  \begin{array}{cc}
    {\rm e}^{-i\frac{\Delta \phi}{2}} & 0 \\
    0 & {\rm e}^{+i\frac{\Delta \phi}{2}} 
  \end{array}
\right).
\end{eqnarray}

After the propagation in the rotator, the output beam state becomes
\begin{eqnarray}
&&|{\rm output} \rangle 
=\hat{\mathcal{R}}_{\rm LR}
(\Delta \phi) 
|{\rm input}\rangle
\nonumber \\ 
&&=
{\rm e}^{i\beta}
{\rm e}^{i\overline{k}d}
\left (
  \begin{array}{c}
    {\rm e}^{-i\frac{\phi+\Delta \phi}{2}} \cos (\theta/2) \\
    {\rm e}^{+i\frac{\phi+\Delta \phi}{2}}\sin (\theta/2)\   
\end{array}
\right)
|{\rm input}\rangle. \\
\end{eqnarray}
By taking the quantum-mechanical expectation values of the output state, we obtain
\begin{eqnarray}
\langle {\rm output}| {\bf \hat{S}} |{\rm output} \rangle
&=
\hbar N
\left (
  \begin{array}{c}
    \sin \theta \cos(\phi+\Delta \phi)\\
    \sin \theta \sin(\phi+\Delta \phi)\\
    \cos \theta
  \end{array}
\right), \nonumber \\
\end{eqnarray}
which means that the rotator successfully rotate the polarisation state as the expectation value of the spin vector in Poincar\'e sphere with the amount of $\Delta \phi$ along the $S_3$ axis (Fig. 5).
\begin{figure}[h]
\begin{center}
\includegraphics[width=5.5cm]{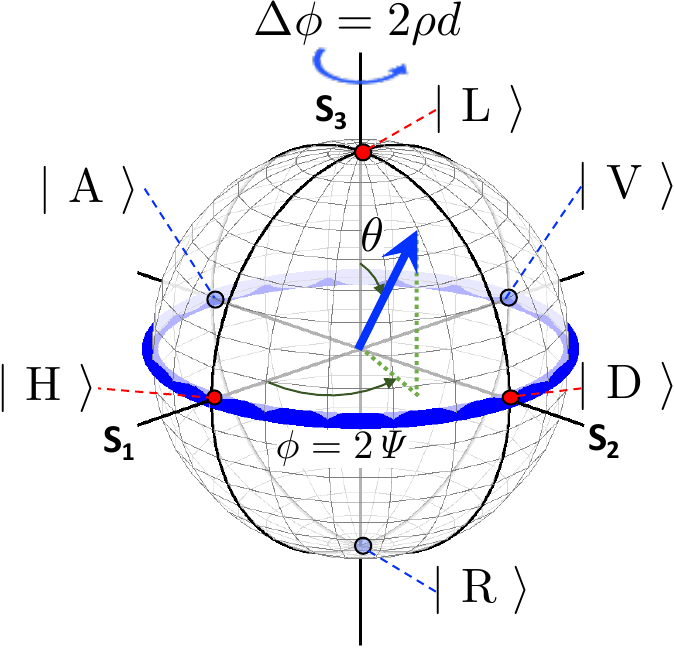}
\caption{
Rotator and its impact on the polarisation state.
}
\end{center}
\end{figure}

\subsubsection{Rotator as a \underline{rotator} in $SU(2)$ Hilbert space}
We understand the chiral phase-control corresponds to the rotation of around $S_3$ in Poincar\'e sphere.
$\hat{S}_3$ corresponds to $\sigma_3$ in the chiral LR-basis (Table \ref{Table2}).
Then, we can construct a rotator based on a $SU(2)$ group theory by
\begin{eqnarray}
&&\mathcal{D}_3^{\rm LR}({\it \Delta \phi})
=\exp 
\left (
  -\frac{i \sigma_3 {\it \Delta \phi}}{2}
\right) \\
&&=
{\bf 1}
\cos 
\left(
  \frac{{\it \Delta \phi}}{2}
\right)
-
i 
\sigma_3
\sin 
\left(
  \frac{{\it \Delta \phi}}{2}
\right)\\
&&=
\left (
  \begin{array}{cc}
    \exp \left( -i\frac{{\it \Delta \phi}}{2} \right) & 
   0  \\
    0  & 
    \exp \left(+ i\frac{{\it \Delta \phi}}{2} \right)
  \end{array}
\right).
\end{eqnarray}
With the inclusion of the overall phase for the global orbital contribution, the Jones matrix becomes
\begin{eqnarray}
\mathcal{R}_{\rm LR}({\it \Delta \phi})
=
{\rm e}^{i\beta}
{\rm e}^{i\overline{k}d}
\mathcal{D}_{3}^{\rm LR}({\it \Delta \phi}),
\end{eqnarray}
which is in agreement with the previous many-body operator based calculation.

\subsubsection{Rotator in HV-basis}
It is also straightforward to obtain the rotator in HV-basis by a $SU(2)$ group theory.
$\hat{S}_3$ corresponds to $\sigma_2$ in the HV-basis (Table \ref{Table2}), such that, we can construct a rotator by the operator
\begin{eqnarray}
&&\mathcal{D}_3^{\rm HV}({\it \Delta \phi})
=\exp 
\left (
  -\frac{i \sigma_2 {\it \Delta \phi}}{2}
\right)\\
&&=
{\bf 1}
\cos 
\left(
  \frac{{\it \Delta \phi}}{2}
\right)
-
i 
\sigma_2
\sin 
\left(
  \frac{{\it \Delta \phi}}{2}
\right)\\
&&=
\left (
  \begin{array}{cc}
    1 & 0 \\
    0 & 1
  \end{array}
\right)
\cos 
\left(
  \frac{{\it \Delta \phi}}{2}
\right)
+
\left (
  \begin{array}{cc}
    0 & -1 \\
    1 & 0
  \end{array}
\right)
\sin
\left(
  \frac{{\it \Delta \phi}}{2}
\right)\\
&&=
\left (
  \begin{array}{cc}
    \cos \left( \frac{{\it \Delta \phi}}{2} \right) & 
   -\sin \left( \frac{{\it \Delta \phi}}{2} \right) \\
    \sin \left( \frac{{\it \Delta \phi}}{2} \right) & 
    \cos \left( \frac{{\it \Delta \phi}}{2} \right)
  \end{array}
\right)\\
&&=
\left (
  \begin{array}{cc}
    \cos \left( {\it \Delta \Psi} \right) & 
   -\sin \left( {\it \Delta \Psi} \right) \\
    \sin \left( {\it \Delta \Psi} \right) & 
    \cos \left( {\it \Delta \Psi} \right)
  \end{array}
\right).
\end{eqnarray}
With the phase factor, Jones matrix for a rotator in HV-basis becomes
\begin{eqnarray}
\mathcal{R}_{\rm HV}({\it \Delta \phi})
=
{\rm e}^{i\beta}
{\rm e}^{i\overline{k}d}
\mathcal{D}_{3}^{\rm HV}({\it \Delta \phi}).
\end{eqnarray}
We realise that this corresponds to a standard $SO(3)$ rotation 
\begin{eqnarray}
R_{z}({\it \Delta \Psi})
=
\left (
  \begin{array}{cc}
    \cos ({\it \Delta \Psi}) & -\sin ({\it \Delta \Psi}) \\
    \sin ({\it \Delta \Psi}) & \cos ({\it \Delta \Psi}) 
  \end{array}
\right)
\end{eqnarray}
of $\bm{\mathcal{E}}$ along $z$.
In fact, the complex electric field is simply rotated upon the operation of the rotator as
\begin{eqnarray}
\left (
  \begin{array}{c}
    \mathcal{E}^{'}_{x} \\
    \mathcal{E}^{'}_{y}
  \end{array}
\right)
=
{\rm e}^{i\beta}
{\rm e}^{i\overline{k}d}
R_{z}({\it \Delta \Psi})
\left (
  \begin{array}{c}
    \mathcal{E}_x \\
    \mathcal{E}_y
  \end{array}
\right),
\end{eqnarray}
including the global phase factor, where $\bm{\mathcal{E}^{\prime}}=( \mathcal{E}^{'}_{x}, \mathcal{E}^{'}_{y})$ is the complex output electric field.
Therefore, the impact of a rotator as an passive optical component is equivalent to a physical anti-clockwise rotation of the electric field, which is also identical to the rotation of $(x,y)$-axes along $z$ in the opposite direction (clock-wise).
During this rotation, the angle of $\chi$ is preserved, such that the shape of the polarisation ellipse is not affected.
The rotation of the principal axis in the real space of ${\it \Delta \Psi}$ corresponds to the rotation of $\langle {\bf \hat{S}} \rangle$ with the amount of ${\it \Delta \phi}=2{\it \Delta \Psi}$.
The factor of 2 is again coming from the quantum-mechanical expectation value.
It is very useful to remember that the rotation in real space affects twice the rotation of $\langle {\bf \hat{S}} \rangle$ in Poincar\'e sphere.
We should always be aware of this difference between real space and Poincar\'e sphere, because this difference will lead the Pancharatnam-Berry's phase \cite{Pancharatnam56,Berry84}.

It is also interesting to note that the physical rotation of a rotator cannot change the operation.
This can be checked simply by calculating
\begin{eqnarray}
\mathcal{D}_3^{\dagger}({\it \Delta \phi^{'}})
\mathcal{D}_3({\it \Delta \phi})
\mathcal{D}_3({\it \Delta \phi^{'}})
=\mathcal{D}_3({\it \Delta \phi}),
\end{eqnarray}
which is valid for both LR- and HV-bases.
Therefore, the physical rotation of a rotator will not change the polarisation state of the output beam.

\subsubsection{Unitary transformation}
We have obtained Jones matrix for a rotator both in LR- and HV-bases.
As is expected for other quantum systems, these descriptions must be identical and transferable to one from the other by unitary transformation.
In order to transfer from $\mathcal{R}_{\rm HV}({\it \Delta \phi}) $ to $\mathcal{R}_{\rm LR}({\it \Delta \phi})$, we use $U_{\rm HV}$ and we confirmed
\begin{eqnarray}
&&\mathcal{R}_{\rm LR}({\it \Delta \phi})
=U^{-1}_{\rm HV} \mathcal{R}_{\rm HV}({\it \Delta \phi}) U_{\rm HV}\\
&&=
\frac{{\rm e}^{i\beta}
{\rm e}^{i\overline{k}d}
}{2}
\left (
  \begin{array}{cc}
    1 & -i \\
    1 &  i
  \end{array}
\right)
\left (
  \begin{array}{cc}
    \cos \left( \frac{{\it \Delta \phi}}{2} \right) & 
   -\sin \left( \frac{{\it \Delta \phi}}{2} \right) \\
    \sin \left( \frac{{\it \Delta \phi}}{2} \right) & 
    \cos \left( \frac{{\it \Delta \phi}}{2} \right)
  \end{array}
\right)
\left (
  \begin{array}{cc}
    1 & 1 \\
    i & -i
  \end{array}
\right) \nonumber \\
&&=
{\rm e}^{i\beta}
{\rm e}^{i\overline{k}d}
\left (
  \begin{array}{cc}
    \exp \left( -i\frac{{\it \Delta \phi}}{2} \right) & 
   0  \\
    0  & 
    \exp \left(+ i\frac{{\it \Delta \phi}}{2} \right)
  \end{array}
\right),
\end{eqnarray}
which is in agreement with our result.

For the opposite transformation, from $\mathcal{R}_{\rm LR}({\it \Delta \phi})$ to $\mathcal{R}_{\rm HV}({\it \Delta \phi}) $, we use $U_{\rm LR}=U_{\rm HV}^{-1}$ and we confirmed
\begin{eqnarray}
&&\mathcal{R}_{\rm HV}({\it \Delta \phi})
=U^{-1}_{\rm LR} \mathcal{R}_{\rm LR} U_{\rm LR}\\
&&=
\frac{{\rm e}^{i\beta}{\rm e}^{i\overline{k}d}}{2}
\left (
  \begin{array}{cc}
    1 & 1 \\
    i &  -i
  \end{array}
\right)
\left (
  \begin{array}{cc}
    {\rm e}^{-i\frac{{\it \Delta \phi}}{2}} & 
   0  \\
    0  & 
    {\rm e}^{+ i\frac{{\it \Delta \phi}}{2}}
  \end{array}
\right)
\left (
  \begin{array}{cc}
    1 & -i \\
    1 &  i
  \end{array}
\right) \nonumber \\
&&=
{\rm e}^{i\beta}
{\rm e}^{i\overline{k}d}
\left (
  \begin{array}{cc}
    \cos \left( \frac{{\it \Delta \phi}}{2} \right) & 
   -\sin \left( \frac{{\it \Delta \phi}}{2} \right) \\
    \sin \left( \frac{{\it \Delta \phi}}{2} \right) & 
    \cos \left( \frac{{\it \Delta \phi}}{2} \right)
  \end{array}
\right).
\end{eqnarray}
Therefore, a standard unitary transformation and its inverse are applicable to the polarisation operators.

\subsection{Rotated phase-shifter}
\subsubsection{Rotated phase-shifter in HV-basis}
As an application of the phase-shifter and the rotator, discussed above, we consider a rotated phase-shifter with the amount of ${\it \Delta \Psi}$, whose FA was originally aligned horizontally (Fig. 6).
\begin{figure}[h]
\begin{center}
\includegraphics[width=7cm]{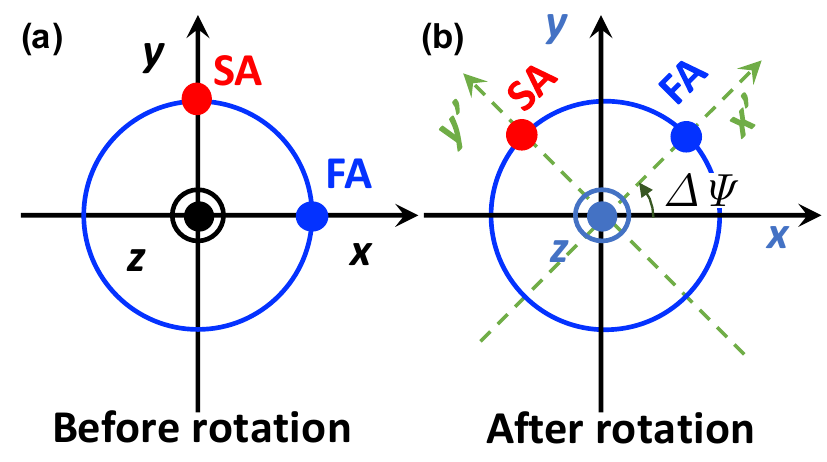}
\caption{
Rotated phase-shifter. 
(a) Phase-shifter arrangement before rotation. The fast axis (FA) is horizontally aligned.
(b) After rotation of ${\it \Delta \Psi}$ in the anti-clock-wise direction (left rotation), seen from the top of the $z$-axis.
}
\end{center}
\end{figure}

It is straightforward to obtain the Jones matrix of the rotated phase-shifter, by following the several steps.
First, we consider the rotation of the coordinate from $(x,y)$ to $(x^{\prime},y^{\prime})$-axes, which is described by a standard rotation around $z$-axis as
\begin{eqnarray}
\left (
  \begin{array}{c}
    x^{'} \\
    y^{'} 
  \end{array}
\right)
=R_{z}({\it \Delta \Psi})
\left (
  \begin{array}{c}
    x \\
    y 
  \end{array}
\right).
\end{eqnarray}
The rotation of the coordinate is equivalent to the rotation of the physical vector (in this case, complex electric field) in the opposite direction 
\begin{eqnarray}
\left (
  \begin{array}{c}
    \mathcal{E}^{'}_{x} \\
    \mathcal{E}^{'}_{y}
  \end{array}
\right)
=R_{z}(-{\it \Delta \Psi})
\left (
  \begin{array}{c}
    \mathcal{E}_x \\
    \mathcal{E}_y
  \end{array}
\right).
\end{eqnarray}
Next, we will apply the phase-shifter operation in the rotated frame as
\begin{eqnarray}
\left (
  \begin{array}{c}
    \mathcal{E}_{x}^{''} \\
    \mathcal{E}_{y}^{''} 
  \end{array}
\right)
=
{\Delta}_{\rm HV}(\delta_{\rm sf})R_{z}(-{\it \Delta \Psi})
\left (
  \begin{array}{c}
    \mathcal{E}_{x} \\
    \mathcal{E}_{y} 
  \end{array}
\right).
\end{eqnarray}
Finally, we will bring back to the original frame as
\begin{eqnarray}
\left (
  \begin{array}{c}
    \mathcal{E}_{x}^{'''} \\
    \mathcal{E}_{y}^{'''} 
  \end{array}
\right)
=
R_{z}({\it \Delta \Psi})
{\Delta}_{\rm HV}(\delta_{\rm sf})
R_{z}(-{\it \Delta \Psi})
\left (
  \begin{array}{c}
    \mathcal{E}_{x} \\
    \mathcal{E}_{y} 
  \end{array}
\right).
\end{eqnarray}
Then, we obtain the Jones matrix for the rotated phase-shifter as
\begin{eqnarray}
&&{\Delta}_{\rm HV}({\it \Delta \phi},\delta_{\rm sf}) \nonumber \\
&&=
{\rm e}^{i \beta}
{\rm e}^{i\overline{k}d}
\left (
\cos (\frac{\delta_{\rm sf}}{2}) 
{\bf 1}
-i
\sin (\frac{\delta_{\rm sf}}{2}) 
\left (
  \begin{array}{cc}
    \cos ({\it \Delta \phi}) &  \sin ({\it \Delta \phi})\\
    \sin ({\it \Delta \phi}) &   -\cos ({\it \Delta \phi}) 
  \end{array}
\right)
\right), \nonumber \\
\end{eqnarray}
where $\Delta \phi=2\Delta \Psi$ as usual.

It is especially important at $\Delta \phi=\pi/2$ ($\Delta \Psi=\pi/4$), which corresponds to the case of $45^{\circ}$ rotated phase-shifter,  given by
\begin{eqnarray}
{\Delta}_{\rm HV}({\it \Delta \phi}=\pi/2,\delta_{\rm sf}) 
&=&
{\rm e}^{i \beta}
{\rm e}^{i\overline{k}d}
\left (
\cos (\frac{\delta_{\rm sf}}{2}) 
{\bf 1}
-i
\sin (\frac{\delta_{\rm sf}}{2}) 
\sigma_1
\right), \nonumber \\
&=&
{\rm e}^{i \beta}
{\rm e}^{i\overline{k}d}
\exp 
\left (
  -\frac{i \sigma_1 {\delta_{\rm sf}}}{2}
\right) \\
&=&
{\rm e}^{i \beta}
{\rm e}^{i\overline{k}d}
\mathcal{D}_2^{\rm HV}(\delta_{\rm sf}).
\end{eqnarray}
because the rotation along $S_2$ corresponds to $\sigma_1$ in the HV-basis (Table \ref{Table2}).
This is useful to use, when we want to convert from LR-states to HV-states and {\it vice versa} (Fig. 7).
\begin{figure}[h]
\begin{center}
\includegraphics[width=4cm]{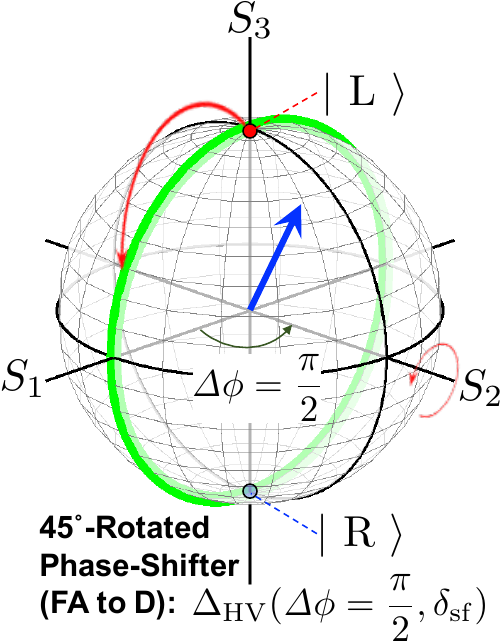}
\caption{
$45^{\circ}$ rotated phase-shifter. 
The operation corresponds to the rotation of polarisation state along $S_2$, which is described bby ${\Delta}_{\rm HV}({\it \Delta \phi}=\pi/2,\delta_{\rm sf}) $.
}
\end{center}
\end{figure}

If we use this $45^{\circ}$-rotated phase-shifter to the coherent state, described by diagonal basis (Fig. 3), the spin expectation value of the output state, 
\begin{eqnarray}
|{\rm output} \rangle&=& 
{\Delta}_{\rm HV}({\it \Delta \phi}=\pi/2,\delta_{\rm sf}) |{\rm input} \rangle,
\end{eqnarray}
is rotated $\delta_{\rm sf}$ along $S_2$, and therefore, we obtain
\begin{eqnarray}
\langle {\rm output}| {\bf \hat{S}} |{\rm output} \rangle
&=
\hbar N
\left (
  \begin{array}{c}
    \sin \theta^{'} \sin (\phi^{'}+\delta_{\rm sf})\\
    \cos \theta^{'}  \\
    \sin \theta^{'} \cos (\phi^{'}+\delta_{\rm sf})
  \end{array}
\right). \nonumber \\
\end{eqnarray}

\subsubsection{Rotated phase-shifter in LR-basis}
Similarly, it is straightforward to obtain the general rotated phase-shifter in LR-basis as
\begin{eqnarray}
&&{\Delta}_{\rm LR}({\it \Delta \phi},\delta_{\rm sf})
=
\mathcal{R}_{\rm LR}({\it \Delta \phi})
{\Delta}_{\rm LR}(\delta_{\rm sf})
\mathcal{R}_{\rm LR}(-{\it \Delta \phi}) \nonumber \\
&&=
{\rm e}^{i\beta}
{\rm e}^{i\overline{k}d}
\left (
  \begin{array}{cc}
   \cos \left( \frac{\delta_{\rm sf}}{2} \right) & 
   -i {\rm e}^{-i\Delta \phi}  \sin \left( \frac{\delta_{\rm sf}}{2} \right) \\
   -i {\rm e}^{+i\Delta \phi}   \sin \left( \frac{\delta_{\rm sf}}{2} \right) & 
    \cos \left( \frac{\delta_{\rm sf}}{2} \right)
  \end{array}
\right), \nonumber \\
\end{eqnarray}
which yields at $\Delta \phi=\pi/2$
\begin{eqnarray}
{\Delta}_{\rm LR}({\it \Delta \phi}=\pi/2,\delta_{\rm sf}) 
&=&
{\rm e}^{i\beta}
{\rm e}^{i\overline{k}d}
\left (
  \begin{array}{cc}
   \cos \left( \frac{\delta_{\rm sf}}{2} \right) & 
   - \sin \left( \frac{\delta_{\rm sf}}{2} \right) \\
    \sin \left( \frac{\delta_{\rm sf}}{2} \right) & 
    \cos \left( \frac{\delta_{\rm sf}}{2} \right)
  \end{array}
\right),  \nonumber \\
&=&
{\rm e}^{i \beta}
{\rm e}^{i\overline{k}d}
\exp 
\left (
  -\frac{i \sigma_2 {\delta_{\rm sf}}}{2}
\right) \\
&=&
{\rm e}^{i \beta}
{\rm e}^{i\overline{k}d}
\mathcal{D}_2^{\rm LR}(\delta_{\rm sf}),
\end{eqnarray}
because the rotation along $S_2$ corresponds to $\sigma_2$ in  LR-basis (Table \ref{Table2}).

\subsubsection{Rotated phase-shifter in $SU(2)$ Hilbert space}
We can easily obtain above formulas in consideration of a quantum-mechanical $SU(2)$ theory.
The phase-shifter is described by a {\it rotator} of a $SU(2)$ group with the rotation axis in the $S_1$-$S_2$ plane.
If we rotate the phase-shifter with the amount of ${\it \Delta \Psi}$ in the real space (Fig. 6 (b)), the rotation axis in Poincar\'e sphere corresponds to the direction ${\bf n}=(\cos ({\it \Delta \phi}),\sin ({\it \Delta \phi}),0)$, where ${\it \Delta \phi}=2{\it \Delta \Psi}$.
Therefore, the rotated phase-shifter corresponds to the {\it rotator} of spin states in $SU(2)$ Hilbert space along ${\bf n}$ with the amount of ${\it \Delta \phi}$.
Away from the global phase factor, this rotation is described by the operator, for HV-basis, as
\begin{eqnarray}
&&\mathcal{D}_{xy}^{\rm HV}(\delta_{\rm sf}) \nonumber \\
&&=
{\bf 1}
\cos 
\left(
  \frac{\delta_{\rm sf}}{2}
\right)
-
i 
\sigma_3
\cos 
\left(
  {\it \Delta \phi}
\right)
\sin 
\left(
  \frac{\delta_{\rm sf}}{2}
\right) \nonumber \\
&&
-
i 
\sigma_1
\sin 
\left(
  {\it \Delta \phi}
\right)
\sin 
\left(
  \frac{\delta_{\rm sf}}{2}
\right) \nonumber \\
&&=
\cos (\frac{\delta_{\rm sf}}{2}) 
{\bf 1}
-i
\sin (\frac{\delta_{\rm sf}}{2}) 
\left (
  \begin{array}{cc}
    \cos ({\it \Delta \phi}) &  \sin ({\it \Delta \phi})\\
    \sin ({\it \Delta \phi}) &   -\cos ({\it \Delta \phi}) 
  \end{array}
\right) \nonumber \\
\end{eqnarray}
By including the global phase, we obtain
\begin{eqnarray}
{\Delta}_{\rm HV}({\it \Delta \phi},\delta_{\rm sf})
=
{\rm e}^{i\beta}
{\rm e}^{i\overline{k}d}
\mathcal{D}_{xy}^{\rm HV}(\delta_{\rm sf}),
\end{eqnarray}
which agreed with the previous result.

For LR-basis, we define a similar operator
\begin{eqnarray}
&&\mathcal{D}_{xy}(\delta_{\rm sf}) \nonumber \\
&&=
{\bf 1}
\cos 
\left(
  \frac{\delta_{\rm sf}}{2}
\right)
-
i 
\sigma_1
\cos 
\left(
  {\it \Delta \phi}
\right)
\sin 
\left(
  \frac{\delta_{\rm sf}}{2}
\right)  \nonumber \\
&&
-
i 
\sigma_2
\sin 
\left(
  {\it \Delta \phi}
\right)
\sin 
\left(
  \frac{\delta_{\rm sf}}{2}
\right) \nonumber \\
&&=
\cos 
\left(
  \frac{\delta_{\rm sf}}{2}
\right) 
{\bf 1}
-i
\sin 
\left(
  \frac{\delta_{\rm sf}}{2}
\right) 
\left (
  \begin{array}{cc}
    0 &  \exp (-i{\it \Delta \phi})\\
    \exp (+i{\it \Delta \phi}) &   0
  \end{array}
\right), \nonumber \\
\end{eqnarray}
which yields
\begin{eqnarray}
{\Delta}_{\rm LR}({\it \Delta \phi},\delta_{\rm sf})
=
{\rm e}^{i\beta}
{\rm e}^{i\overline{k}d}
\mathcal{D}_{xy}^{\rm LR}(\delta_{\rm sf}),
\end{eqnarray}
which also agreed with the previous result.

Therefore, a $SU(2)$ group theory is a powerful tool to describe the polarisation control by a phase-shifter, a rotator, and a combination of these operations.

\subsection{Half- and quarter-wavelenth phase-shifters and rotators}
\subsubsection{Rotated half-wavelength phase-shifters}
One of the most frequently used phase-shifters is the half-wavelength phase-shifter, which $\delta_{\rm sf}=2\pi(n_{\rm s} - n_{\rm f})d/\lambda=\pi$.
This corresponds to the difference of the half-wavelength ($\lambda/2$) for the effective optical path distances in the phase-shifter for the SA ($n_{\rm s} d$) and the FA ($n_{\rm f} d$).
In the arrangement of the FA aligned horizontally, the phase of SA is advanced due to the phase factor, coming from the orbital ${\rm e}^{ik_0(n_{\rm s}-n_{\rm f})d}={\rm e}^{i\pi}$.
This means that the half-wavelength phase-shifter rotate $\langle \bf{\hat{S}} \rangle$ along $S_1$ with the amount of $\pi$.
Therefore, $|{\rm L}\rangle$ is transformed to $|{\rm R}\rangle$, $|{\rm D}\rangle$ is transformed to $|{\rm A}\rangle$, and {\it vice versa} (Fig. 8).

\begin{table}[h]
\caption{\label{Table3}
Summary of the rotated half-wavelength phase-shifter.
${\it \Delta \Psi}$ and ${\it \Delta \phi}$ correspond to rotations in real space and in Poincar\'e sphere, respectively.
The operators away from the phase factor of ${\rm e}^{i\beta}{\rm e}^{i\overline{k}d}$ are listed.
}
\begin{ruledtabular}
\begin{tabular}{cccccc}
${\it \Delta \Psi}$&
${\it \Delta \phi}$&
$\Delta_{\rm HV}/({\rm e}^{i\beta}{\rm e}^{i\overline{k}d})$ &
${\it \Delta \Psi}$&
${\it \Delta \phi}$&
$\Delta_{\rm LR}/({\rm e}^{i\beta}{\rm e}^{i\overline{k}d})$ \\
\colrule
$0$ & $0$ &  $-i\sigma_3$ & $0$ & $0$ &  $-i\sigma_1$  \\
$\pi/4$ & $\pi/2$ &  $-i\sigma_1$ & $\pi/4$ & $\pi/2$ &  $-i\sigma_2$  \\
$\pi/2$ & $\pi$ &  $+i\sigma_3$ & $\pi/2$ & $\pi$ &  $+i\sigma_1$  \\
$-\pi/4$ & $-\pi/2$ &  $+i\sigma_3$ & $-\pi/4$ & $-\pi/2$ &  $+i\sigma_2$  \\
$\pi$ & $2\pi$ &  $-i\sigma_3$ & $\pi$ & $2\pi$ &  $-i\sigma_1$  
\end{tabular}
\end{ruledtabular}
\end{table}
\begin{eqnarray}
\end{eqnarray}

The operators of the rotated half-wavelength phase-shifter at major angles are summarised in Table \ref{Table3}.
Away from the global phase factor of  ${\rm e}^{i\beta}{\rm e}^{i\overline{k}d}$, the operations are very simple.
For example, without the rotation, the operation in HV-basis becomes, 
\begin{eqnarray}
\mathcal{D}_1^{\rm HV}(\delta_{\rm sf}=\pi)
=-i\sigma_{3},
\end{eqnarray}
which represents the $\pi$ rotation in Poincar\'e sphere around $S_1$.
The $45^{\circ}$ rotated half-wavelength phase-shifter becomes, 
\begin{eqnarray}
\mathcal{D}_2^{\rm HV}(\delta_{\rm sf}=\pi)
=-i\sigma_{1}.
\end{eqnarray}
Similarly, the half-wavelength rotator is given by 
\begin{eqnarray}
\mathcal{D}_3^{\rm HV}({\it \Delta \phi}=\pi)
=-i\sigma_{2},
\end{eqnarray}
which is independent on the physical rotation, as we confirmed.
These are corresponding to the original Pauli matrices, responsible for these rotations (Table \ref{Table2}).

Another important quantum-mechanical aspect of these rotations are Pancharatnam-Berry's phase \cite{Pancharatnam56,Berry84}.
If we consider $2\pi$ rotations by successive application of these operators, we confirm the phase change of $-1$ as
\begin{eqnarray}
\mathcal{D}_1^{\rm HV}(\delta_{\rm sf}=\pi)
\mathcal{D}_1^{\rm HV}(\delta_{\rm sf}=\pi)
&=&
-\sigma_{3}\sigma_{3}
=
-{\bf 1} \\
\mathcal{D}_2^{\rm HV}(\delta_{\rm sf}=\pi)
\mathcal{D}_2^{\rm HV}(\delta_{\rm sf}=\pi)
&=&
-\sigma_{1}\sigma_{1}
=
-{\bf 1} \\
\mathcal{D}_3^{\rm HV}(\delta_{\rm sf}=\pi)
\mathcal{D}_3^{\rm HV}({\it \Delta \phi}=\pi)
&=&
-\sigma_{2}\sigma_{2}
=
-{\bf 1}.
\end{eqnarray}
This means that the polarisation control is not classical at all, but fully quantum mechanical.
The non-trivial Pancharatnam-Berry's phase is successfully incorporated in the spin rotation operator of a $SU(2)$ group theory.

We have obtained the same relationship in LR-basis for the half-wavelength phase-shift, its rotated one, and the half-wavelength rotator, as
\begin{eqnarray}
\mathcal{D}_1^{\rm LR}(\delta_{\rm sf}=\pi)=-i\sigma_{1}, \\
\mathcal{D}_2^{\rm LR}(\delta_{\rm sf}=\pi)=-i\sigma_{2}, \\
\mathcal{D}_3^{\rm LR}({\it \Delta \phi}=\pi)=-i\sigma_{3},
\end{eqnarray}
respectively, which are consistent with Table \ref{Table2}.
It is also obvious that Pancharatnam-Berry's phase is properly included for $2\pi$ rotations, since $\sigma_i^2=1$ for $^{\forall}i=1,2,3$.

\begin{figure}[h]
\begin{center}
\includegraphics[width=7cm]{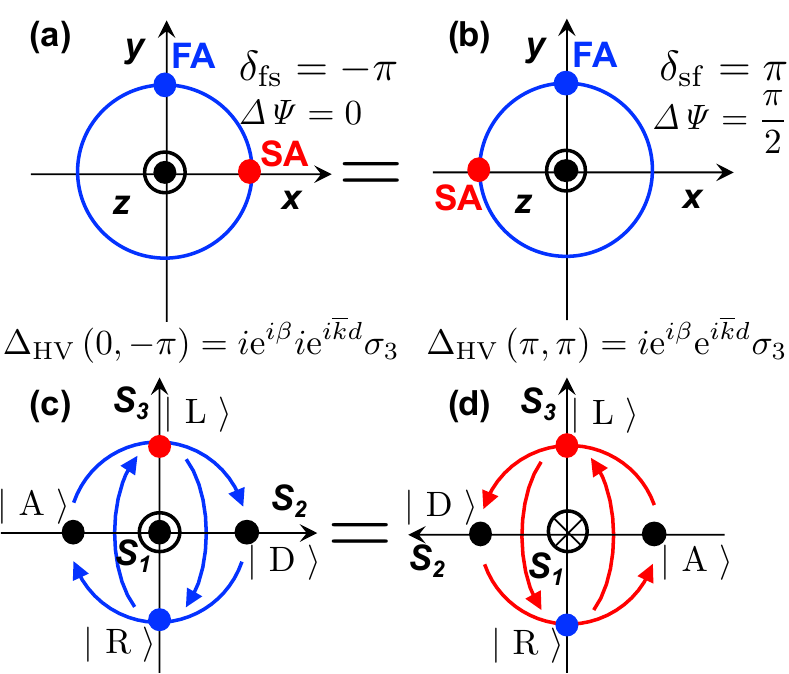}
\caption{
Impact of a flip-flop exchange.
(a) An example of the half-wavelength phase-shifter, whose slow axis (SA) is aligned horizontally.
(b) The flip-flop exchanged configuration.
(c) Operation of the phase-shifter before the exchange in Poincar\'e sphere.
(d) Operation of the flip-flop exchanged phase-shifter.
}
\end{center}
\end{figure}

It is also interesting to note that a phase-shifter, which aligns its optical axis (SA or FA) horizontally, is not affected by a flip-flop exchange (Fig. 8).
This must be true, because the crystal has a mirror symmetry against both SA and FA.
Therefore, there is no difference between the front side and the back side with regard to the amount of the polarisation rotation, achieved by the phase-shifter. 
Nevertheless, the alignment of the SA or FA to the designated direction is important, such that if the phase-shifter is rotated, to align its FA to the diagonal direction, the flip-flop by the mirror symmetric exchange against $y$-axis will make the FA align to the anti-diagonal direction.
This corresponds to rotate in the opposite direction.
Still, there is no difference for the half-wavelength phase-shifter, but it does make a difference for a quarter-wavelength phase-shifter.

\begin{figure}[h]
\begin{center}
\includegraphics[width=4cm]{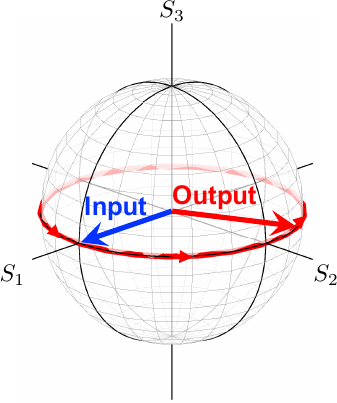}
\caption{
The trajectory of the output Stokes vector by the rotated phase-shifter with the input of the horizontally polarised state.
}
\end{center}
\end{figure}

As an example of the application of the rotated half-wavelength phase-shifter, we consider the input state of $|{\rm H}\rangle$.
In this case, the expectation value of spin by the output state, becomes
\begin{eqnarray}
\langle {\rm output} | {\bf \hat{S}} |  {\rm output} \rangle
&=&
\hbar N
\left (
  \begin{array}{c}
     \cos (2\Delta \phi) \\
     \sin (2\Delta \phi) \\
     0 \\
  \end{array}
\right) \nonumber \\
&=&
\hbar N
\left (
  \begin{array}{c}
     \cos (4\Delta {\it \Psi}) \\
     \sin (4\Delta {\it \Psi}) \\
     0 \\
  \end{array}
\right),
\end{eqnarray}
which means that the average spin vector (Stokes vector) will rotate 4-times in Poincar\'e sphere, while rotating the phase-shifter 1-time in real space ($\Delta {\it \Psi}$ changing from $0$ to $2\pi$).
The trajectory of the Stokes vector is shown in Fig. 9.
In this case, the Stokes vector is always located in the $S_1-S_2$ plane, and the output state will rotate anti-clock-wise, seen from the top of $S_3$ upon the rotation of the phase-shifter towards the anti-clock-wise, seen from the detector side.

\subsubsection{Rotated quarter-wavelength phase-shifters}
Another frequently used phase-shifters is a quarter-wavelength phase-shifter at $\delta_{\rm sf}=2\pi(n_{\rm s} - n_{\rm f})d/\lambda=\pi/2$, which corresponds to the deference in path lengths of the quarter wavelength between SA and FA.
Operators of the quarter-wavelength phase-shifters at major angles are summarised in Table \ref{Table4}.

\begin{table}[h]
\caption{\label{Table4}
Summary of the rotated quarter-wavelength phase-shifter.
${\it \Delta \Psi}$ and ${\it \Delta \phi}$ correspond to rotations in real space and in Poincar\'e sphere, respectively.
The operators away from the phase factor of ${\rm e}^{i\beta}{\rm e}^{i\overline{k}d}$ are listed.
}
\begin{ruledtabular}
\begin{tabular}{cccccc}
${\it \Delta \Psi}$&
${\it \Delta \phi}$&
$\Delta_{\rm HV}/({\rm e}^{i\beta}{\rm e}^{i\overline{k}d})$ &
${\it \Delta \Psi}$&
${\it \Delta \phi}$&
$\Delta_{\rm LR}/({\rm e}^{i\beta}{\rm e}^{i\overline{k}d})$ \\
\colrule
$0$ & $0$ &  $\frac{1}{\sqrt{2}}\left({\bf 1}-i\sigma_3\right)$ 
& $0$ & $0$ &  $\frac{1}{\sqrt{2}}\left({\bf 1}-i\sigma_1\right)$   \\
$\pi/4$ & $\pi/2$ &  $\frac{1}{\sqrt{2}}\left({\bf 1}-i\sigma_1\right)$  & 
$\pi/4$ & $\pi/2$ &  $\frac{1}{\sqrt{2}}\left({\bf 1}-i\sigma_2\right)$   \\
$\pi/2$ & $\pi$ &  $\frac{1}{\sqrt{2}}\left({\bf 1}+i\sigma_3\right)$  & 
$\pi/2$ & $\pi$ &  $\frac{1}{\sqrt{2}}\left({\bf 1}+i\sigma_1\right)$   \\
$-\pi/4$ & $-\pi/2$ &  $\frac{1}{\sqrt{2}}\left({\bf 1}+i\sigma_1\right)$  & 
$-\pi/4$ & $-\pi/2$ &  $\frac{1}{\sqrt{2}}\left({\bf 1}+i\sigma_2\right)$   \\
$\pi$ & $2\pi$ &  $\frac{1}{\sqrt{2}}\left({\bf 1}-i\sigma_3\right)$ 
& $\pi$ & $2\pi$ &  $\frac{1}{\sqrt{2}}\left({\bf 1}-i\sigma_1\right)$   
\end{tabular}
\end{ruledtabular}
\end{table}

\begin{figure}[h]
\begin{center}
\includegraphics[width=7cm]{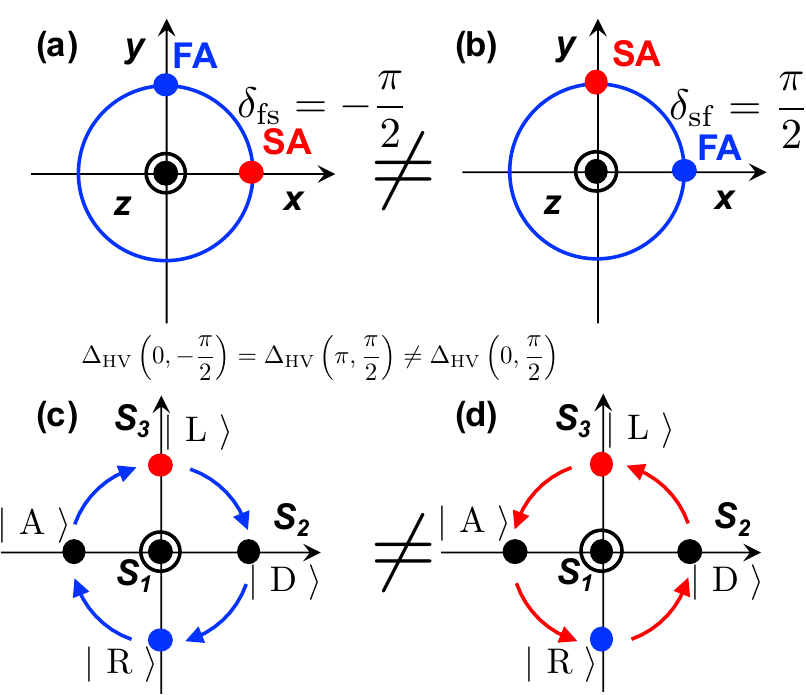}
\caption{
The difference of the alignment of the optical axis for the quarter-wavelength phase-shifter.
(a) Slow axis (SA) aligned horizontally.
(b) Fast axis (FA) aligned horizontally.
(c) Rotation of polarisation state in Poincar\'e sphere, when SA is aligned horizontally.
(d) Rotation of polarisation state in Poincar\'e sphere, when FA is aligned horizontally.
}
\end{center}
\end{figure}

An example of the operation using the quarter wavelength phase-shifter is shown in Fig. 10.
It is crucial to align SA or FA properly for the desired operation, because it determines the direction of rotation whether right (clock-wise) or left (anti-clock-wise) rotations.

We can recognise that 2 successive applications of the quarter-wavelength phase-shifters correspond to 1 application of the half-wavelength phase-shifter, except for the global phase, 
\begin{eqnarray}
\left(
\frac{\Delta_{\rm HV}(\delta_{\rm sf}=\pi/2)}{{\rm e}^{i\beta}{\rm e}^{i\overline{k}d}}
\right)^2
&=&
\left(
\frac{1}{\sqrt{2}}\left({\bf 1}-i\sigma_3\right)
\right)^2 \\
&=&
-i\sigma_3
\\
&=&
\frac{\Delta_{\rm HV}(\delta_{\rm sf}=\pi)}{{\rm e}^{i\beta}{\rm e}^{i\overline{k}d}},
\end{eqnarray}
because of the additive nature of the rotation, $\mathcal{D}_1^{\rm HV}(\pi/2)\mathcal{D}_1^{\rm HV}(\pi/2)=\mathcal{D}_1^{\rm HV}(\pi)$, which simply means that 2 quarters-rotations are equivalent to 1 half-rotation.
Consequently, 4 quarters-rotations are equivalent to 1-whole-rotation in Poincar\'e sphere, which yields the Pancharatnam-Berry's phase \cite{Pancharatnam56,Berry84} of $\left( \mathcal{D}_1^{\rm HV}(\pi/2) \right)^4=-1$.

\begin{figure}[h]
\begin{center}
\includegraphics[width=7cm]{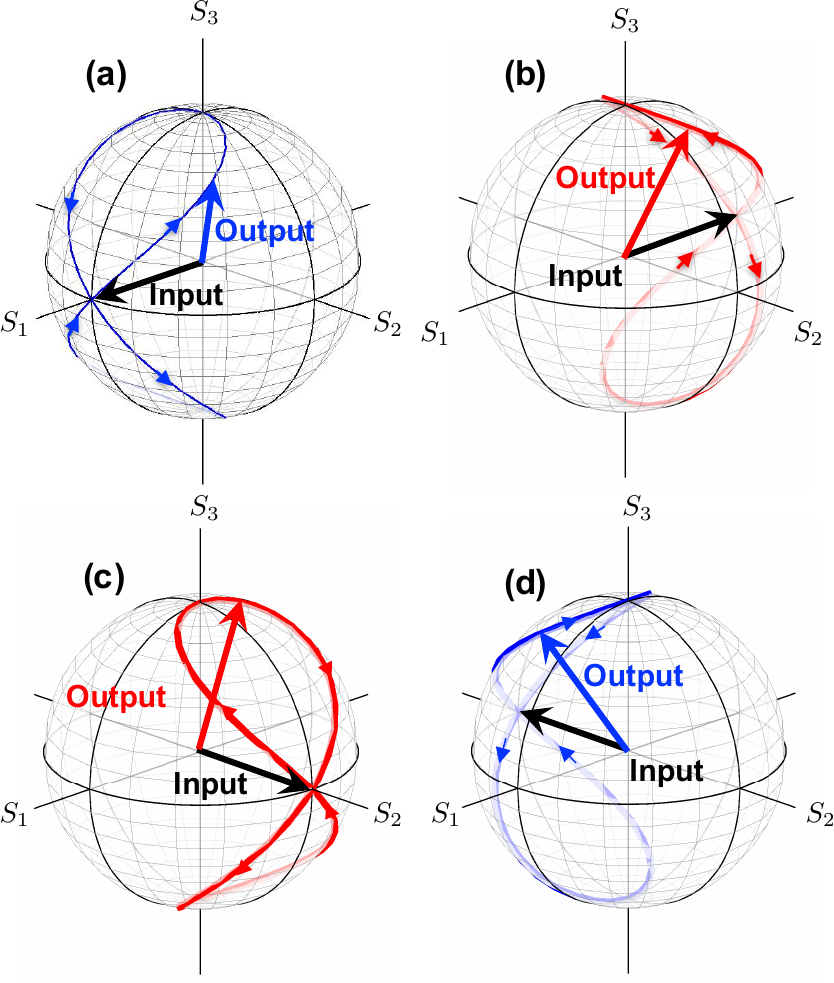}
\caption{
Trajectories of output polarisation states by the rotated quarter-wavelength phase-shifters.
The input is linearly polarised state.
(a) The input is a horizontally polarised state, $|{\rm H}\rangle$ .
(b) The input is a vertically polarised state, $|{\rm V}\rangle$ .
(c) The input is a diagonally polarised state, $|{\rm D}\rangle$ .
(d) The input is an anti-diagonally polarised state, $|{\rm A}\rangle$ .
}
\end{center}
\end{figure}

In Fig. 11, we show the trajectories of the output polarisation state, $\langle {\rm output} | {\bf \hat{S}} |  {\rm output} \rangle$, when the linearly polarised input states are rotated by the rotated quarter-wavelength phase-shifters.
Upon physically-rotating the phase-shifter, the linearly polarised state changes its ellipticity, arriving to the poles, which correspond to left- and right-circularly polarised states, and coming back to the original state. 
For example, if the input is the horizontally polarised state, the output spin becomes
\begin{eqnarray}
\langle {\rm output} | {\bf \hat{S}} |  {\rm output} \rangle
&=&
\hbar N
\left (
  \begin{array}{c}
     \cos^2 (\Delta \phi) \\
     \sin (\Delta \phi) \cos (\Delta \phi) \\
     - \sin (\Delta \phi) \\
  \end{array}
\right).
\end{eqnarray}
While physically-rotating the quarter-wavelength phase-shifter 1-time in real space, the Stokes vector will rotate 2-times, because $\Delta \phi=2\Delta {\it \Psi}$.
The horizontally polarised state can never arrive to be the vertically polarised state by the quarter-wavelength phase-shifter, because the horizontally polarised state does not contain any contribution of the orthogonal vertically polarised state and the $\pi/4$ change of the phase-shift is not large enough.

\subsection{Polariser}

So far, we have discussed phase-shifters and rotators, which are described energy-conserving unitary operators.
The advantages of Jones vector treatments are the capabilities to extend the operator analysis to the situations with energy-dissipations.
One of the most important polarisation optical components in that aspect is a polariser \cite{Jones41,Yariv97,Goldstein11,Gil16}.
Polarisers are made, for example, by patterning thin metallic layer into arrays of metallic wires with sub-wavelength widths \cite{Yamazaki16}.
This allows to reflect and absorb the most of the polarisation component along the direction of the long axis of the wires, while the polarisation component perpendicular to the wires can transmit the polariser.
There are many other types of polarisers by using polymers with polarisation-dependent absorption coefficients or reflectors using birefringent prisms or reflections at Brewster's angle.

\subsubsection{Polariser in HV-basis}
It is straightforward to consider a polariser operator in HV-basis, 
\begin{eqnarray}
\hat{\mathcal P}_{x}^{\rm HV}
=
{\rm e}^{ik_0 n_x d - \alpha_x d}
\frac{{\rm e}^{i\beta}}{\sqrt{N}}
\hat{a}_{\rm H}\hat{\bf x}
+
{\rm e}^{ik_0 n_y d - \alpha_y d}
\frac{{\rm e}^{i\beta}}{\sqrt{N}}
\hat{a}_{\rm V}\hat{\bf y}, \nonumber \\
\end{eqnarray}
where $n_x$ and $n_y$ are refractive indices, and $\alpha_x$ and $\alpha_y$ are absorption coefficients per unit propagation length for polarisation components along $x$ and $y$, respectively, and $d$ is the thickness of the polariser.
For the ideal horizontal polariser, we take the limits of $\alpha_x d \rightarrow 1$ and $\alpha_y d \rightarrow \infty$, and we obtain
\begin{eqnarray}
\hat{\mathcal P}_{x}^{\rm HV}
=
{\rm e}^{i\overline{k} d}
\frac{{\rm e}^{i\beta}}{\sqrt{N}}
\hat{a}_{\rm H}\hat{\bf x}
, 
\end{eqnarray}
where $\overline{k}=k_0 n_x$.
By considering the spinor representation of the output state, we obtain
\begin{eqnarray}
|{\rm output} \rangle
&&=\hat{\mathcal P}_{x}^{\rm HV}
|\alpha_{\rm H},\alpha_{\rm V}\rangle
\nonumber \\ 
&&=
{\rm e}^{i\beta}
{\rm e}^{i\overline{k}d}
\left (
  \begin{array}{cc}
    1 & 0 \\
    0 & 0 
  \end{array}
\right)
\left (
  \begin{array}{c}
    {\rm e}^{-i\frac{\delta}{2}} \cos \alpha \\
    {\rm e}^{+i\frac{\delta}{2}}\sin \alpha \   
\end{array}
\right)
|{\rm input} \rangle, \nonumber \\
\end{eqnarray}
Therefore, we obtain 
\begin{eqnarray}
&&
\langle \alpha_{\rm H},\alpha_{\rm V}|
\hat{\mathcal P}_{x}^{\rm HV}
|\alpha_{\rm H},\alpha_{\rm V}\rangle
\nonumber \\ 
&&=
{\rm e}^{i\beta}
{\rm e}^{i\overline{k}d}
\left (
  \begin{array}{cc}
    1 & 0 \\
    0 & 0 
  \end{array}
\right)
\left (
  \begin{array}{c}
    {\rm e}^{-i\frac{\delta}{2}} \cos \alpha \\
    {\rm e}^{+i\frac{\delta}{2}}\sin \alpha \   
\end{array}
\right) \\
&&=
{\mathcal P}_{x}^{\rm HV}
 | {\rm Jones}\rangle,
\end{eqnarray}
where we have defined the Jones matrix for the horizontal polariser as
\begin{eqnarray}
{\mathcal P}_{x}^{\rm HV}
&=&
{\rm e}^{i\beta}
{\rm e}^{i\overline{k}d}
\left (
  \begin{array}{cc}
    1 & 0 \\
    0 & 0 
  \end{array}
\right) \\
&=&
\frac{
{\rm e}^{i\beta}
  {\rm e}^{i\overline{k}d}
      }{2}
({\bf 1}+\sigma_{z}).
\end{eqnarray}

Similarly, we obtain the Jones matrix for vertical polariser as 
\begin{eqnarray}
{\mathcal P}_{y}^{\rm HV}
&=&
{\rm e}^{i\beta}
{\rm e}^{i\overline{k}d}
\left (
  \begin{array}{cc}
    0 & 0 \\
    0 & 1 
  \end{array}
\right) \\
&=&
\frac{
{\rm e}^{i\beta}
  {\rm e}^{i\overline{k}d}
      }{2}
({\bf 1}-\sigma_{z}).
\end{eqnarray}

It is important to recognise that polarisers are projectors to remove one of the orthogonal components, and the intensity of the ray will be reduced, accordingly.
Consequently, it is dissipative irreversible operation, and thus the inverse of the polariser matrix does not exist.
It is also interesting to note that the transmitted output state is not altered except for the global phase of ${\rm e}^{i\beta}{\rm e}^{i\overline{k}d}$.

\subsubsection{Polariser in LR-basis}
After obtaining the polariser operators in HV-basis, it is straightforward to obtain the corresponding operators in LR-basis simply by unitary transformations.
We obtain
\begin{eqnarray}
\mathcal{P}_{x}^{\rm LR}
&=&U^{-1}_{\rm HV} \mathcal{P}_{x}^{\rm HV} U_{\rm HV} \\
&=&
\frac{{\rm e}^{i\overline{k}d}}{2}
\left (
  \begin{array}{cc}
    1 & -i \\
    1 & i
  \end{array}
\right)
\left (
  \begin{array}{cc}
    1 & 0 \\
    0 & 0
  \end{array}
\right)
\left (
  \begin{array}{cc}
    1 & 1 \\
    i & -i
  \end{array}
\right)\\
&=&
\frac{
  {\rm e}^{i\overline{k}d}
    }{2}
\left (
  \begin{array}{cc}
    1 &  1  \\
    1  &  1
  \end{array}
\right) \\
&=&
\frac{
  {\rm e}^{i\overline{k}d}
      }{2}
({\bf 1}+\sigma_{x}),
\end{eqnarray}
and
\begin{eqnarray}
\mathcal{P}_{y}^{\rm LR}
&=&
U^{-1}_{\rm HV} \mathcal{P}_{y}^{\rm HV} U_{\rm HV} \\
&=&
\frac{{\rm e}^{i\overline{k}d}}{2}
\left (
  \begin{array}{cc}
    1 & -i \\
    1 & i
  \end{array}
\right)
\left (
  \begin{array}{cc}
    0 & 0 \\
    0 & 1
  \end{array}
\right)
\left (
  \begin{array}{cc}
    1 & 1 \\
    i & -i
  \end{array}
\right) \\
&=&
{\rm e}^{i\overline{k}d}
\left (
  \begin{array}{cc}
    1 &  -1  \\
    -1  &  1
  \end{array}
\right) \\
&=&
\frac{
  {\rm e}^{i\overline{k}d}
      }{2}
({\bf 1}-\sigma_{x}).
\end{eqnarray}

\subsubsection{Rotated polarisers}
It is also straightforward to obtain the rotated polarisers with the angle of $\Delta \Psi = \Delta \phi /2$. 
In HV-basis, it becomes
\begin{eqnarray}
\mathcal{P}_{x}^{\rm HV}({\it \Delta \phi})
&=&
R_{\rm HV}({\it \Delta \Psi})
\mathcal{P}_{x}^{\rm HV}
R_{\rm HV}(-{\it \Delta \Psi}) \\
&=&
\frac{
  {\rm e}^{i\overline{k}d}
  }{2}
\left (
  \begin{array}{cc}
    1+\cos ({\it \Delta \phi}) & \sin ({\it \Delta \phi}) \\
     \sin ({\it \Delta \phi}) & 1- \cos ({\it \Delta \phi}) 
  \end{array}
\right).
 \nonumber \\
\end{eqnarray}
In LR-basis, it becomes
\begin{eqnarray}
\mathcal{P}_{x}^{\rm LR}({\it \Delta \phi})
&=&
R_{\rm LR}({\it \Delta \Psi})
\mathcal{P}_{x}^{\rm LR}
R_{\rm LR}(-{\it \Delta \Psi})
\\
&=&
\frac{
  {\rm e}^{i\overline{k}d}
  }{2}
\left (
  \begin{array}{cc}
    1 & {\rm e}^{- i \it \Delta \phi} \\
     {\rm e}^{ i \it \Delta \phi} & 1
  \end{array}
\right).
\end{eqnarray}
The rotated polarisers at typical angles are summarised in Table \ref{Table5}.
\begin{table}[h]
\caption{\label{Table5}
Summary of the rotated polarisers.
${\it \Delta \Psi}$ and ${\it \Delta \phi}$ correspond to rotations in real space and in Poincar\'e sphere, respectively.
The operators away from the phase factor of ${\rm e}^{i\beta}{\rm e}^{i\overline{k}d}$ are listed.
}
\begin{ruledtabular}
\begin{tabular}{cccccc}
${\it \Delta \Psi}$&
${\it \Delta \phi}$&
${\mathcal P}_x^{\rm HV}/({\rm e}^{i\beta}{\rm e}^{i\overline{k}d})$ &
${\it \Delta \Psi}$&
${\it \Delta \phi}$&
${\mathcal P}_x^{\rm LR}/({\rm e}^{i\beta}{\rm e}^{i\overline{k}d})$ \\
\colrule
$0$ & $0$ &  $\frac{1}{2}\left({\bf 1}+\sigma_3\right)$ 
& $0$ & $0$ &  $\frac{1}{2}\left({\bf 1}+\sigma_1\right)$   \\
$\pi/4$ & $\pi/2$ &  $\frac{1}{2}\left({\bf 1}+\sigma_1\right)$  & 
$\pi/4$ & $\pi/2$ &  $\frac{1}{2}\left({\bf 1}+\sigma_2\right)$   \\
$\pi/2$ & $\pi$ &  $\frac{1}{2}\left({\bf 1}-\sigma_3\right)$  & 
$\pi/2$ & $\pi$ &  $\frac{1}{2}\left({\bf 1}-\sigma_1\right)$   \\
$-\pi/4$ & $-\pi/2$ &  $\frac{1}{2}\left({\bf 1}-\sigma_1\right)$  & 
$-\pi/4$ & $-\pi/2$ &  $\frac{1}{2}\left({\bf 1}-\sigma_2\right)$   \\
$\pi$ & $2\pi$ &  $\frac{1}{2}\left({\bf 1}+\sigma_3\right)$ 
& $\pi$ & $2\pi$ &  $\frac{1}{2}\left({\bf 1}+\sigma_1\right)$   
\end{tabular}
\end{ruledtabular}
\end{table}

\subsection{Pancharatnam-Berry's phase}

\begin{figure}[h]
\begin{center}
\includegraphics[width=7cm]{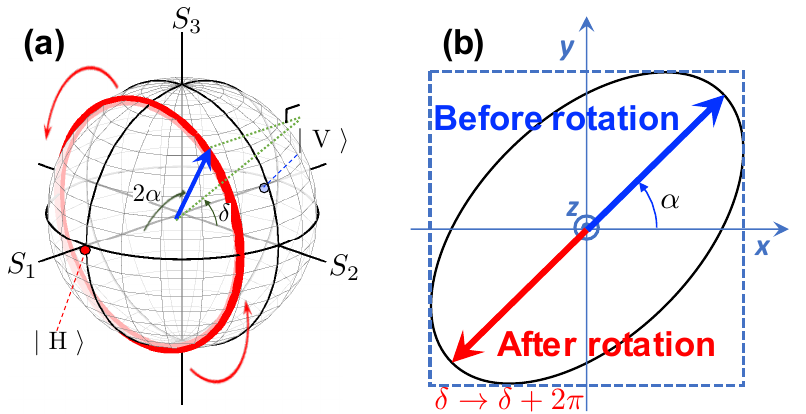}
\caption{
The impact of Pancharatnam-Berry's phase.
(a) 1-time rotation of the Stokes vector in Poincare\'e sphere. 
The polarisation state, seen from the average of the expectation value, is not changed at all upon the rotation.
(b) The corresponding polarisation ellipse, in the real space. 
The sign of the electric field changes both for $x$ and $y$ directions.
This change of the phase is different from the global phase coming from the orbital, and thus, the phase change can be observed by the interference with the original wave before the rotation. 
}
\end{center}
\end{figure}

We think it is worth for summarising our simple picture on the Pancharatnam-Berry's phase \cite{Pancharatnam56,Berry84} for polarisation states \cite{Tomita86}.
We consider a situation, where the Stokes vector rotate 1-time in Poincar\'e sphere (Fig. 12).
Suppose this could be achieved by phase-shifters, e.g. by 2 successive operations of the half-wavelength phase-shifters.
The spin expectation value, $\langle {\rm output} | {\bf \hat{S}} |  {\rm output} \rangle$, would not be changed upon this rotation, such that the polarisation state is equivalent in terms of the expectation value.
However, the electric field has changed its sign, such that the destructive interference can be observable upon the coupling to the original wave before the rotation.
We could successfully include this Pancharatnam-Berry's phase in our $SU(2)$ description of the spin state of photons, because $\mathcal{D}_1^{\rm HV}(\delta_{\rm sf}=2\pi)=-1$.

The acquisition of this Pancharatnam-Berry's phase upon the 1-time rotation does not depend on how we rotate the polarisation state nor the bases which we are going to use.
Let us consider to use the chiral LR-basis and rotate the polarisation state $(\theta,\phi)$ 1-time upon rotating $\theta$ to $\theta-2\pi$ (Fig. 13 (a)).
This corresponds to use the rotation axis, which is rotated to the clock-wise with the amount of $\Delta \phi^{\prime}=\pi/2-\phi$ from $S_1$.
Therefore, the rotation axis is pointing to ${\bf n}=(\cos(-\Delta \phi^{\prime}),\sin(-\Delta \phi^{\prime}),0)$, which corresponds to rotate the phase-shifter with the amount of $\Delta {\it \Psi}^{\prime}=\pi/4-\phi/2$ in the clock-wise direction (Fig. 13 (b)).
Upon this $2\pi$ rotation in Poincar\'e sphere, the spin expectation value will not be changed, since the spin is coming back to point to the original direction.
However, the electric field in the polarisation ellipse changes its sign in the polarisation ellipse (Fig. 13 (d)), because the rotation corresponds to rotate $\chi$ with the amount of $\pi$.
This difference of the factor of $2$ among $\theta$ in the Poincar\'e sphere and $\chi$ in the polarisation ellipse is responsible for the emergence of the geometrical phase.
In a $SU(2)$ theory, this simply corresponds to  $\mathcal{D}^{\rm LR}({\bf n},\delta_{\rm sf}=2\pi)=-1$.

We can also confirm the impact of the polarisation rotation by a rotator, which is equivalent to the adiabatic change of the coordinate (Fig. 13 (c)).
The rotation corresponds to change $\phi$ to be $\phi + 2 \pi$ around $S_3$.
This corresponds to rotate the inclination angle ${\it \Psi}$ with the amount of $\pi$, because $\phi=2{\it \Psi}$ (Fig. 13 (e)).
In a $SU(2)$ theory, it is guaranteed by $\mathcal{D}_3^{\rm LR}(\delta_{\rm sf}=2\pi)=-1$.

\begin{figure}[h]
\begin{center}
\includegraphics[width=8cm]{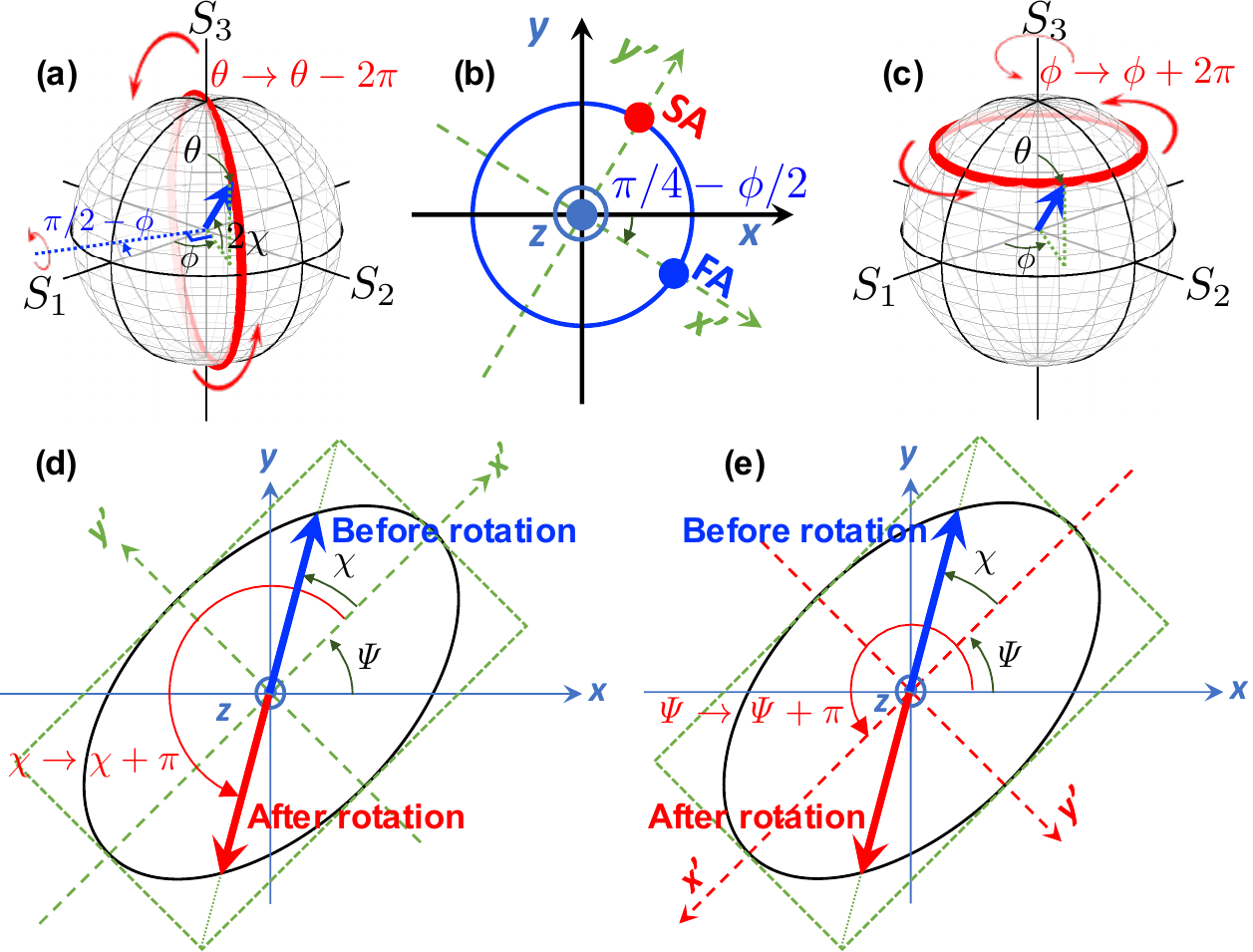}
\caption{
The Pancharatnam-Berry's phase, seen from the chiral bases.
(a) 1-time rotation of the Stokes vector by the phase-shifter in Poincare\'e sphere. 
The polarisation state is not changed upon the rotation.
(b) Corresponding phase-shifter arrangement.
(c) 1-time rotation of the Stokes vector by a rotator in Poincare\'e sphere.
(d) Corresponding polarisation ellipse after the rotation by the phase-shifter of (a) and (b).
(d) Corresponding polarisation ellipse after the rotation by the rotator of (c).
}
\end{center}
\end{figure}

Away from the global $U(1)$ phase factor of ${\rm e}^{i\beta}{\rm e}^{i\overline{k}d}$, the sign of the rotator operation depends on the direction of rotation, because $\mathcal{D}_3^{\rm LR}( \pi )=-i\sigma_{3}$ and $\mathcal{D}_3^{\rm LR}(- \pi )=+i\sigma_{3}$.
This is equivalent to the difference of the $2\pi$ due to the Pancharatnam-Berry's phase.
The difference of the sign of the operators are indispensable to guarantee the identity, $\mathcal{D}_3^{\rm LR}( \pi )\mathcal{D}_3^{\rm LR}(- \pi )=\mathcal{D}_3^{\rm LR}(0)=\sigma_{3}^2={\bf 1}$.
We can also confirm this in HR-basis as $\mathcal{D}_3^{\rm HR}( \pi )=-i\sigma_{2}$ and $\mathcal{D}_3^{\rm HR}( - \pi )=+i\sigma_{2}$, and consequently $\mathcal{D}_3^{\rm HR}( \pi )\mathcal{D}_3^{\rm HR}(- \pi )=\mathcal{D}_3^{\rm HR}(0)=\sigma_{2}^2={\bf 1}$.

\begin{figure}[h]
\begin{center}
\includegraphics[width=8.5cm]{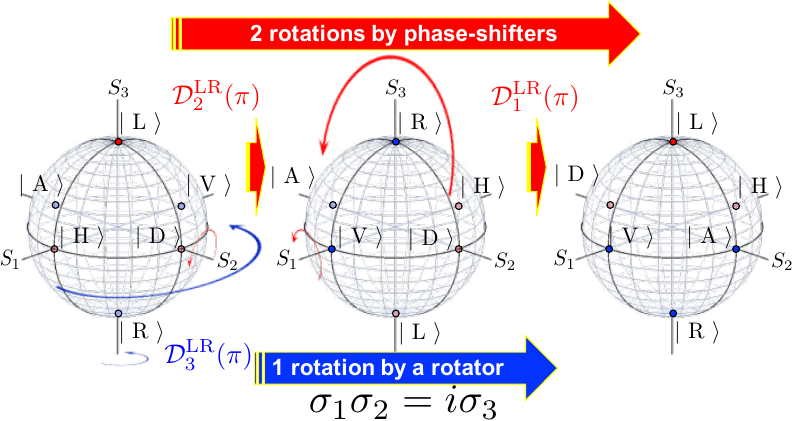}
\caption{
Equivalence of 2 half-wavelength phase-shifter operators and 1 half-wavelength rotator operation.
We considered the phase-shifter rotation along $S_2$ and the successive rotation along $S_1$ in chiral LR-basis.
This is equivalent to the 1-rotation along $S_3$ by a rotator.
It is important to consider the overall phase by this operation, since this corresponds to $\sigma_1 \sigma_2 = - \sigma_2 \sigma_1  = i \sigma_3$, which depends on the order of operations. 
In HR-basis, this corresponds to $\sigma_3 \sigma_1 = - \sigma_1 \sigma_3  = i \sigma_2$.
}
\end{center}
\end{figure}

Another quantum-mechanical aspect of the spin of photons is found by successive application of rotation operations in Poincar\'e sphere (Fig. 14).
We have confirmed that the half-wavelength phase-shifters and the rotator are equivalent to Pauli matrices, 
\begin{eqnarray}
i \mathcal{D}_i^{\rm LR}(\pi)=\sigma_{i}, \\
\end{eqnarray}
where $i=1,2,3$, which obey commutation and anti-commutation relationships.
The commutation relationship is a fundamental basis of quantum-mechanics and thus it is also essential for spin of photons.
For polarisation, we can easily manipulate spin of photons by phase-shifter and rotators, but it is important to aware the order is important, because the sign can be changed as
\begin{eqnarray}
\sigma_1 \sigma_2 = - \sigma_2 \sigma_1  = i \sigma_3.
\end{eqnarray}
This shows that 2 successive rotations by phase-shifters, one aligned its FA to the diagonal direction and the other aligned its FA to the horizontal direction, are equivalent to 1 rotation by a rotator.
We can change the order of operations, without changing the final polarisation state as an expectation value of the spin operators.
However, the phase is different depending on the order of applications of the phase-shifters.
This difference of the sign should also be observable in the interference experiments.
Essentially, this is equivalent to the difference of 1-time rotation by a rotator, because $i\sigma_3$ and $-i\sigma_3$ correspond to the rotation of $\pi$ and $-\pi$, respectively, such that the difference is $2\pi$-rotation, as we have explained above.
This is nothing but a Pancharatnam-Berry's phase.

\subsection{Jones vector and Bloch vector in $SU(2)$ Hilbert space}
As we have seen above, a $SU(2)$ group theory is a powerful tool to understand various rotations of $\langle {\bf \hat{S}}\rangle$ in Poincar\'e sphere.
Here, we apply a $SU(2)$ group theory to confirm some of concepts for polarisation states.

First, we obtain the unitary transformation from HV-basis to LR-basis by {\it rotators} in $SU(2)$ Hilbert space.
The choice of the bases is based on our preference of the quantisation axis.
The HV-basis is based on the alignment of the quantisation axis $\sigma_3$ to the $S_1$ axis (Table \ref{Table2}), while the LR-basis is based on the alignment of $\sigma_3$ to the $S_3$ axis (Figs. 15 (a) and (b)).
The expectation values should not depend on the choice of the basis, such that we should be able to transfer from HV- to LR-basis by a unitary transformation, which is described by the following 2 steps of rotations.
First, we start from HV-basis and apply the rotation along $S_2$ for the amount of $-\pi/2$ as
\begin{eqnarray}
\mathcal{D}_2^{\rm HV}
  \left( 
    -\frac{\pi}{2}
  \right)
=
\frac{1}{\sqrt{2}}
\left (
  {\bf 1}+i\sigma_{x}
\right)
=
\frac{1}{\sqrt{2}}
\left (
  \begin{array}{cc}
    1 & i \\
    i & 1
  \end{array}
\right).
\end{eqnarray}
Next, we rotate along the $S_3$ for the amount of $-\pi/2$ as
\begin{eqnarray}
\mathcal{D}_3^{\rm HV}
  \left( 
    -\frac{\pi}{2}
  \right)
=
\frac{1}{\sqrt{2}}
\left (
  {\bf 1}+i\sigma_{y}
\right)
=
\frac{1}{\sqrt{2}}
\left (
  \begin{array}{cc}
    1 & 1 \\
    -1 & 1
  \end{array}
\right).
\end{eqnarray}
These 2 successive rotations in $SU(2)$ space give the unitary 

\begin{eqnarray}
U_{\rm HV}
&=&\mathcal{D}_3^{\rm HV}
  \left( 
    -\frac{\pi}{2}
  \right)
\mathcal{D}_2^{\rm HV}
  \left( 
    -\frac{\pi}{2}
  \right)
\\
&=&
\frac{1}{2}
\left (
  \begin{array}{cc}
    1 & 1 \\
    -1 & 1
  \end{array}
\right)
\left (
  \begin{array}{cc}
    1 & i \\
    i & 1
  \end{array}
\right) \\
&=&
\frac{{\rm e}^{i\pi/4}}{\sqrt{2}}
\left (
  \begin{array}{cc}
    1 & 1 \\
    i & -i
  \end{array}
\right),
\end{eqnarray}
which is in agreement with the previous result away from the irrelevant overall $U(1)$ phase factor of ${\rm e}^{i\pi/4}$.

\begin{figure}[h]
\begin{center}
\includegraphics[width=7cm]{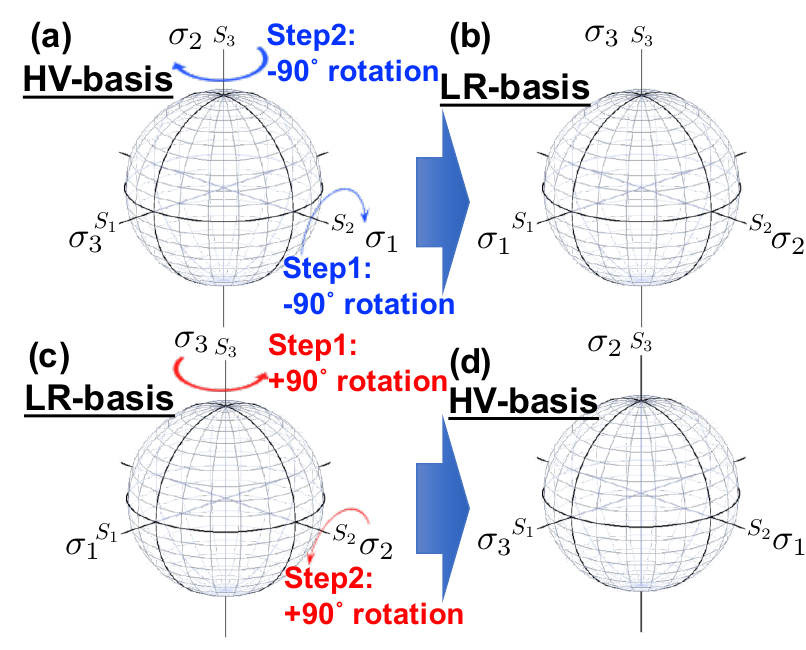}
\caption{
Unitary transformation between linear horizontal/vertical (HV)-basis and chiral left/right (LR)-basis.
(a) Original HV-basis.
(b) Rotated LR-basis.
(c) Original LR-basis.
(d) Rotated HV-basis.
}
\end{center}
\end{figure}

Similarly, we also confirmed the inverse transformation (Figs 15. (c) and (d)) as
\begin{eqnarray}
U_{\rm LR}
&=&
\mathcal{D}_2^{\rm LR}
  \left( 
    \frac{\pi}{2}
  \right)
\mathcal{D}_3^{\rm LR}
  \left( 
    \frac{\pi}{2}
  \right)
\\
&=&
\frac{1}{2}
\left (
  \begin{array}{cc}
    1 & -1 \\
    1 & 1
  \end{array}
\right)
\left (
  \begin{array}{cc}
    {\rm e}^{-i\pi/4} & 0 \\
    0 & {\rm e}^{i\pi/4}
  \end{array}
\right) \\
&=&
\frac{{\rm e}^{-i\pi/4}}{\sqrt{2}}
\left (
  \begin{array}{cc}
    1 & -i \\
    1 & i
  \end{array}
\right).
\end{eqnarray}

Finally, we obtain the Jones vector and the Bloch vector by the {\it rotations} in a $SU(2)$ Hilbert space.
For Jones vector, we use linear HV-basis, and start from $| {\rm H}\rangle$.
Then, we will rotate along $S_3$ with the amount of $\gamma$, and then subsequently rotate along $S_1$ with the amount of $\delta$.
Consequently, we obtain
\begin{eqnarray}
|\gamma,\delta \rangle 
&=&
\mathcal{D}_1^{\rm HV}(\delta)
\mathcal{D}_3^{\rm HV}(\gamma)
| {\rm H}\rangle \\
&=&
\left (
  \begin{array}{cc}
    {\rm e}^{-i\frac{\delta}{2}} & 
   0  \\
    0  & 
    {\rm e}^{+i\frac{\delta}{2}} 
 \end{array}
\right)
\left (
  \begin{array}{cc}
    \cos (\frac{\gamma}{2}) & 
   -\sin (\frac{\gamma}{2}) \\
    \sin (\frac{\gamma}{2}) & 
   \  \ \cos (\frac{\gamma}{2})
  \end{array}
\right)
\left (
  \begin{array}{c}
    1\\
    0  
 \end{array}
\right) \nonumber \\
&=&
\left (
  \begin{array}{c}
    {\rm e}^{-i\frac{\delta}{2}}\cos (\frac{\gamma}{2})\\
    {\rm e}^{+i\frac{\delta}{2}}\sin (\frac{\gamma}{2})  
 \end{array}
\right) \\
&=&
|{\rm Jones}\rangle,
\end{eqnarray}
which is indeed the Jones vector.

For Bloch state, we start from chiral LR-basis, and start from $| {\rm L}\rangle$
Then, we rotate the state along $S_2$ with the amount of $\theta$ and rotate it along $S_3$ with the amount of $\phi$.
Then, we obtain
\begin{eqnarray}
|\theta,\phi \rangle 
&=&
\mathcal{D}_3^{\rm LR}(\phi)
\mathcal{D}_2^{\rm LR}(\theta)
| {\rm L}\rangle \\
&=&
\left (
  \begin{array}{cc}
    {\rm e}^{-i\frac{\phi}{2}} & 
   0  \\
    0  & 
    {\rm e}^{+i\frac{\phi}{2}} 
 \end{array}
\right)
\left (
  \begin{array}{cc}
    \cos (\frac{\theta}{2}) & 
   -\sin (\frac{\theta}{2}) \\
    \sin (\frac{\theta}{2}) & 
   \  \ \cos (\frac{\theta}{2})
  \end{array}
\right)
\left (
  \begin{array}{c}
    1\\
    0  
 \end{array}
\right) \nonumber \\
&=&
\left (
  \begin{array}{c}
    {\rm e}^{-i\frac{\phi}{2}}\cos (\frac{\theta}{2})\\
    {\rm e}^{+i\frac{\phi}{2}}\sin (\frac{\theta}{2})  
 \end{array}
\right) \\
&=&
|{\rm Bloch}\rangle,
\end{eqnarray}
which is the Bloch state.
Therefore, polarisation states are described by spin states and operators based on a $SU(2)$ group theory.

\section{Conclusions}
We have discussed {\it what is spin of a photon?}
Our hypothesis is that spin of a photon is an intrinsic {\it quantum-mechanical} degree of freedom inherent to a photon, which was suggested by the classical description of the angular momentum expression together with the Poynting vector.
While the chiral spin component was obtained by this analogy, it was not completely understood whether other components exist or not, and if any, whether the spin operators of photons are well-defined quantum observables or not.
We have accepted as a principle, that the chiral spin operator of a photon is aligned to the direction of propagation, and applied a standard quantum-mechanical prescription and a $SU(2)$ group theory, assuming the spin state of a photon is described by a 2-level quantum-mechanical system.
Then, by a rotation in the $SU(2)$ Hilbert space, we obtained all 3 spatial components of the spin operator, and established that the quantum-mechanical expectation values of the spin of photons are Stokes parameters in Poincar\'e sphere.
Based on this analogy and the comparison with coherent state of a monochromatic ray of photons from a laser, we identified that the  zero-th component of Stokes parameter, $S_0$, is the order parameter of the coherent ray, which becomes zero after the time average below the lasing threshold, while it becomes finite after the on-set of lasing.
The reason why the laser beam is described by a single mode is deeply rooted to the Bose-Einstein condensation nature of photons, which allow macroscopic number of photons occupy the same level with the phase coherence including polarisation states.
Based on this identification, the description of Stokes parameters in Poincar\'e shpere corresponds to the visualisation of the spin expectation values for all 3 spatial components of the vectorial order parameters.
There is no obvious contradiction or a difficulty, as far as we accept the spin operators of photons exist as quantum-mechanical many-body operators, and evaluate their expectation values by a coherent state.
This does not necessarily mean that the principle to define the spin operators is true, however, with the definition of the spin operators and a standard quantum many-body theory, we can practically deal and understand polarisation states as standard quantum-mechanical states in a 2-level system.

We admit that the most of our equations presented in this paper were already appeared in research papers and textbooks on photonics.
Nevertheless, it was not completely clear for us to explain the nature of spin of a photon.
The situation might have some similarity with the development of the special theory of relativity by Einstein \cite{Eintein1905}, for which the Lorentz transformation\cite{Lorentz1899} was known at that time, while its actual implication on the principle of relativity was not established, yet.
Therefore, the equations were not enough to understand the principle behind them.
The theory of relativity established the principle that the speed of light, thus, the momentum of a photon in a vacuum is invariant under the Lorentz transformation.
{\it
The principle behind our $SU(2)$ theory of a photon is the rotational symmetry of the angular momentum of a photon in a vacuum.
}
In other words, there is no particular preferential polarisation state for a photon to be realised in a vacuum, no matter which direction the photon is propagating along with.
The rotational symmetry is broken for the coherent light from a laser source, considered in this paper, in the sense that some fixed polarisation state is realised when the Bose-Einstein condensation of photons occurred upon exceeding the threshold of pumping for lasing.
In a material with a broken directional symmetry (phase-shifter) or a broken rotational symmetry (rotator), the polarisation state can be rotated, because of the difference of the phases acquired during the transmission of the material between the orthogonal components of the polarisation state.
A remarkable difference from the time of Einstein was that we have almost everything we need to consider the spin of a photon, such as quantum many-body theories, coherent state descriptions, spinor representation, Jones vectors, and so on.
We have just applied the existing framework of the quantum field theory to a coherent state of photons to understand the spin state of the photons in a straightforward way.

Now, we have a clear view in confidence that the spin of a photon is well-defined quantum-mechanical observable and the Jones vector is equivalent to the Bloch vector to describe the quantum-mechanical state of polarisation.
Thus, the Poincar\'e sphere is essentially equivalent to the Bloch sphere.
The only possible difference between the Poincar\'e sphere and the Bloch sphere is the statistics of the particles, which we are dealing with; 
Stokes parameters are used for photons, which are Bose particles, while the Bloch sphere is usually used for the spin state of an electron, which is a Fermi particle. 
Because of the Bose-Einstein nature of a coherent ray of photons, we obtained the magnitude of the order parameter as $S_0=\hbar N$, which is macroscopically large compared with $\hbar/2$ for an electron.
What is intriguing is that the quantum-mechanical feature of a spin state of photons is controllable by conventional optical components such as phase-shifters and rotators.
It is surprising that the macroscopic coherent state of a laser beam is so easily controlled, and the standard quantum-mechanical operations for a 2-level system work successfully for the ray. 
We hope that we have justified the treatment of Stokes parameters, based on a standard many-body quantum theory and a $SU(2)$ group theory.
The commutation relationship of these spin operators are obtained, with the additional factor of 2 due to the peculiar situation, that a photon has the spin of $1$ but only 2 orthogonal states are allowed due to the transverse condition of the ray.
Linear and chiral bases are mutually transferred by unitary transformations, and Jones vector is obtained by the rotation of states in $SU(2)$ Hilbert space.

We have also found that the Pancharatnam-Berry's phase is essential to consider the commutation relationship of spin state of a photon beyond the expectation value.
The $2\pi$ rotation in the Poincar\'e sphere resulted in the $\pi$ rotation of the electric field in the real space, such that the Pancharatnam-Berry's phase can destructively interfere with the original wave before the rotation, regardless of the same expectation values of the spin.
It is required to rotate $4\pi$ in the Poincar\'e sphere to come back to the original polarisation state with the same phase, which is indeed coming from the quantum-mechanical character governed by the commutation relationship.
We can also regard that the Pancharatnam-Berry's phase guarantees non-Abelian relationship of the spin operators for photons.
We believe that our interpretation is new.

It is remarkable that Stokes and Poincar\'e could arrive to the correct formulas, before the establishments of quantum-mechanics, a quantum many-body theory, Maxwell equations, a Ginzburg-Landau theory, Jones vector calculus, and so on.
They have captured all important aspects of polarisation states, and certainly, acquired quantum-mechanical nature of polarisation states.
In the modern perspective, the remaining challenge for us is to justify the splitting of spin and orbital angular momentum at least for photons in a coherent monochromatic state from a laser source.
This is not a trivial task at all, and we will revisit this issue, separately.

In conclusion, we believe that spin of a photon is an intrinsic quantum mechanical degree of freedom for polarisation. 
We accepted the principle that the chiral spin state of a photon is aligned to the direction of the propagation, and applied a many-body quantum field theory to obtain the other spin operators.
We cannot derive the spin commutation relationship from a correspondence from classical mechanics. 
Instead, we accepted the validity of the quantum commutation relationship as a principle for a polarisation state of a photon.
It is important to recognise that photons can take any polarisation state, described by a superposition state of the arbitrary chosen 2 orthogonal polarisation states (e.g., left/right circularly polarised states or horizontal/vertical linearly polarised states), depending on how the coherent ray of photons from a laser source is prepared.
We confirmed that the expectation values of the spin components are equivalent to the Stokes parameters, $\langle {\bf \hat{S}} \rangle = {\bf S}$. 
Therefore, the Stokes parameters are vectorial order parameters in Poincar\'e sphere to describe the coherent nature of photons.
It is also important to recognise that we obtained the proportionality constant of the Stokes parameter as $\hbar N$ for the spin expectation value.
This means that $\hbar$ becomes effectively to  $\hbar N$ for coherent monochromatic photons from a laser source due to the Bose-Einstein condensation.
Therefore, we conclude that the polarisation is a macroscopic manifestation of a quantum-mechanical feature of photons.

\section*{Acknowledgements}
This work is supported by JSPS KAKENHI Grant Number JP 18K19958.
The author would like to express sincere thanks to Prof I. Tomita for continuous discussions and encouragements.

\appendix

\renewcommand{\figurename}{Supplementary Fig}
\setcounter{figure}{0}

\section{Definition of polarisation states}
We use a standard right-hand Cartesian coordinate, ${\bf r}=(x,y,z)$, to describe a space.
In the most parts of this paper, the direction of the propagation is taken along $z$, and we will describe the phase evolution over $t$ and $z$ as ${\rm e}^{i(kz-\omega t)}$, which propagates along $+z$ direction over $t$.
In order to describe the polarisation state, we assume that we will observe from a detector side for the ray not from the laser source side.
This means that we are observing the evolution of the phase front, seen from the $+z$ towards the origin.
This naturally sets the description of the phase front, seen from $+z$ and projected in the $(x,y)$ plane, as shown in Supplementary Fig. 1.
In particular, this definition has an advantage, that left- and right-circular-polarised states are describing counter-clock-wise and clock-wise rotations, respectively \cite{Jackson99}.
On the other hand, the sign of the chirality is usually defined based on the direction of movement of a right-handed screw, for which the way to define rotation is unfortunately seen from the person who rotates the screw, which is in our case a laser source side.
In other words, our left-circular-polarised state is similar to the right-screw, while the right-circular-polarised state is similar to the left-screw.
Consequently, a photon in the left-circular-polarised state should point its spin towards the $+z$ direction, corresponding to the projected spin of $\sigma=+1$ along $z$, and a photon in the right-circular-polarised state points its spin towards the opposite $-z$ direction, corresponding to $\sigma=-1$ \cite{Jackson99}.
It turns out that this is actually true, as shown in the main text.

\begin{figure}[h]
\begin{center}
\includegraphics[width=6cm]{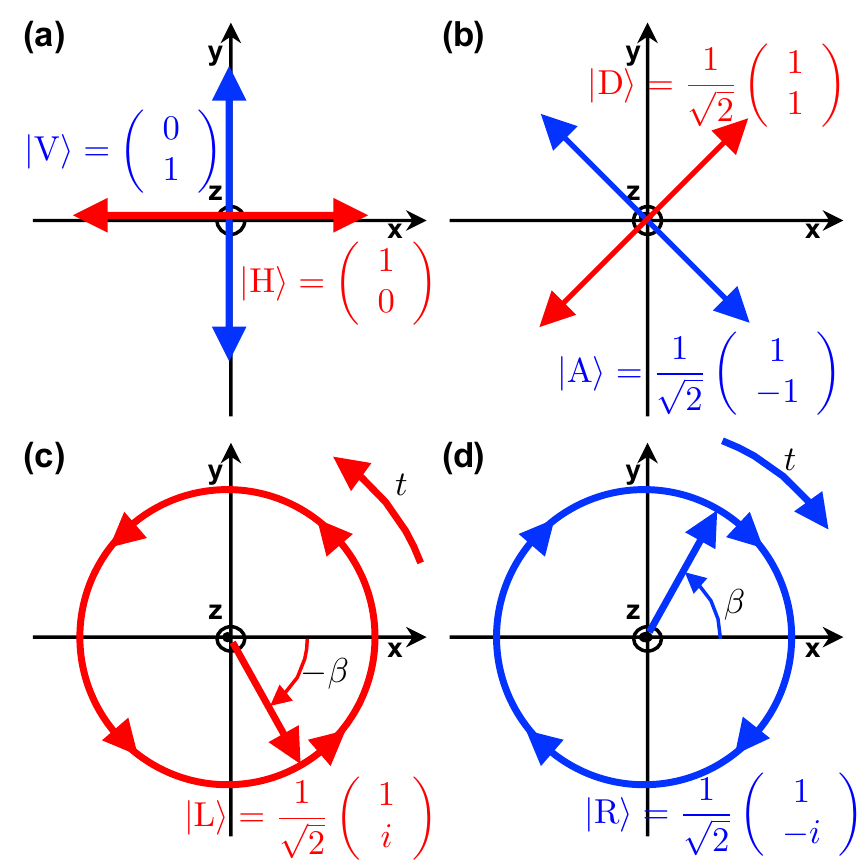}
\caption{
Polarisation states with Jones vectors.
(a) Horizontal (H) and Vertical (V) states.
(b) Diagonal (D) and Anti-Diagonal states.
(c) Left (L) circularly polarised state. The phase front rotate rotates counter-clock-wise. Spin of this state is $\sigma=+1$, since the right screw is pointing to the $+z$ direction.
(d) Right (R) circularly polarised state. The phase front rotates clock-wise. Spin of this state is $\sigma=-1$, since the right screw is pointing to the $-z$ direction.
}
\end{center}
\end{figure}

Here, we explain in more detail about the classical description of the polarisation states, in order to clarify our convention among many other ways of definitions \cite{Max99,Jackson99,Baym69,Sakurai14,Yariv97,Fox06,Grynberg10,Goldstein11,Gil16,Pedrotti07,Hecht17}.
It is quite common to describe the electro-magnetic waves, including photons, by a complex electric field, $\bm{\mathcal{E}}(z,t)$, whose $(x,y)$ components are
\begin{eqnarray}
\left (
  \begin{array}{c}
    \mathcal{E}_{x}\\
    \mathcal{E}_{y}
  \end{array}
\right)
&=&
{\rm e}^{i(kz-\omega t)}
\left (
  \begin{array}{c}
   {E}_{x}^{0} {\rm e}^{i \delta_x} \\
   {E}_{y}^{0} {\rm e}^{i \delta_y}
  \end{array}
\right), \\
&=&
{\rm e}^{i(kz-\omega t+\delta_x)}
\left (
  \begin{array}{c}
   {E}_{x}^{0}  \\
   {E}_{y}^{0}{\rm e}^{i \delta_y}
  \end{array}
\right), \\
&=&
E_{0}{\rm e}^{i\beta}
\left (
  \begin{array}{c}
    \cos \alpha \ \ \ \  \\
    \sin \alpha \ {\rm e}^{i\delta}
  \end{array}
\right),
 \end{eqnarray}
where ${\bf E}_{0}=({E}_{x}^{0},{E}_{y}^{0})=E_0 (\cos \alpha, \sin \alpha)$ is the amplitude of the field, ${\bm \delta}=(\delta_x,\delta_y)$ is the phase for $x$ and $y$ components, $\delta=\delta_y-\delta_x$ is the relative phase, $\beta=kz-\omega t +\delta_x$ is the phase responsible for $t$ and $z$ evolution, and $\alpha$ is the auxiliary angle to define the splitting between $x$ and $y$.
This expression naturally split $\bm{\mathcal{E}}(z,t)$ into its orbital part $\Psi(z,t)=E_{0}{\rm e}^{i\beta}$ and the polarisation part
\begin{eqnarray}
|{\rm Jones} \rangle
&=&
\left (
  \begin{array}{c}
    \cos \alpha \ \ \ \  \\
    \sin \alpha \ {\rm e}^{i\delta}
  \end{array}
\right),
 \end{eqnarray}
which is the Jones vector \cite{Jones41,Poincare92,Yariv97,Goldstein11,Gil16}.

The description using a complex electric field is very useful, while all observables including the electric field must be real, such that the complex electric field should be understood to take the real part at the end of the calculations \cite{Max99,Jackson99,Yariv97,Goldstein11,Gil16}.
Therefore, the real electric field, $\bm E$, becomes 
\begin{eqnarray}
\left (
  \begin{array}{c}
    E_{x} \\
    E_{y}
  \end{array}
\right)
=
\left (
  \begin{array}{c}
    E_{x}^{0} \cos(kz-\omega t+\delta_x) \\
    E_{y}^{0} \cos(kz-\omega t+\delta_y)
  \end{array}
\right).
 \end{eqnarray}

The linearly polarised state corresponds to the cases for $\delta=0$ or $\delta=\pi$.
For $\delta=0$, we see that the slope of $E_{x}/E_{y}$ is constant over $t$, and therefore, the locus of the phase front is linearly oscillating.
These states include the Horizontally (H) polarised state ($\alpha=0$, Supplementary Fig. 1 (a))
\begin{eqnarray}
|{\rm H} \rangle
=|\leftrightarrow \ \rangle
&=&
\left (
  \begin{array}{c}
    1  \\
    0
  \end{array}
\right),
 \end{eqnarray}
and the Diagonally (D) polarised state ($\alpha=\pi/4$, Supplementary Fig. 1 (b))
\begin{eqnarray}
|{\rm D} \rangle
=
|\rotatebox[origin=c]{45}{$\leftrightarrow$}\rangle
&=&
\frac{1}{\sqrt{2}}
\left (
  \begin{array}{c}
    1  \\
    1
  \end{array}
\right).
 \end{eqnarray}
For $\delta=\pi$, we see that the slope of $E_{x}/E_{y}$ is constant over $t$, but the sign is opposite to that for $\delta=0$.
These states include the Vertically (V) polarised state ($\alpha=\pi/2$, Supplementary Fig. 1 (a))
\begin{eqnarray}
|{\rm V} \rangle
=|\updownarrow \ \rangle 
&=&
\left (
  \begin{array}{c}
    0 \\
    1
  \end{array}
\right),
 \end{eqnarray}
and the Anti-diagonally (A) polarised state ($\alpha=\pi/4$, Supplementary Fig. 1 (b))
\begin{eqnarray}
|{\rm A} \rangle
=
|\rotatebox[origin=c]{-45}{$\leftrightarrow$}\rangle
&=&
\frac{1}{\sqrt{2}}
\left (
  \begin{array}{c}
     \ \   1 \\
    -1
  \end{array}
\right).
 \end{eqnarray}

The left-circular-polarised state correspond to $\delta=\pi/2$ and $\alpha=\pi/4$ (Supplementary Fig. 1 (c)), and the Jones vector becomes
\begin{eqnarray}
|{\rm L} \rangle
=|\circlearrowleft \ \rangle
&=&
\frac{1}{\sqrt{2}}
\left (
  \begin{array}{c}
     1 \\
     i
  \end{array}
\right).
 \end{eqnarray}
In this case, the real electric field becomes
\begin{eqnarray}
\left (
  \begin{array}{c}
    E_{x} \\
    E_{y}
  \end{array}
\right)
=
E_{0}
\left (
  \begin{array}{c}
    \cos(-\beta) \\
    \sin(-\beta)
  \end{array}
\right),
 \end{eqnarray}
which shows the counter-clock-wise (left) rotation of the phase front over $t$ (Supplementary Fig. 2 (b)).
This state corresponds to $\sigma=+1$, since the helical wave is pointing towards the $+z$ direction.

Similarly, the right-circular-polarised state corresponds to $\delta=-\pi/2$ and $\alpha=\pi/4$
(Supplementary Fig. 1 (d)), and the Jones vector becomes
\begin{eqnarray}
|{\rm R} \rangle
=|\circlearrowright \ \rangle
&=&
\frac{1}{\sqrt{2}}
\left (
  \begin{array}{c}
     \ \ 1 \\
     -i
  \end{array}
\right).
 \end{eqnarray}
In this case, the real electric field becomes
\begin{eqnarray}
\left (
  \begin{array}{c}
    E_{x} \\
    E_{y}
  \end{array}
\right)
=
E_{0}
\left (
  \begin{array}{c}
    \cos \beta  \\
    \sin \beta 
  \end{array}
\right),
 \end{eqnarray}
which shows the clock-wise (right) rotation of the phase front over $t$ (Supplementary Fig. 2 (b)).
This state corresponds to $\sigma=-1$, since the helical wave is pointing towards the $-z$ direction.

\section{Elliptically polarised state}

\begin{figure}[h]
\begin{center}
\includegraphics[width=5.5cm]{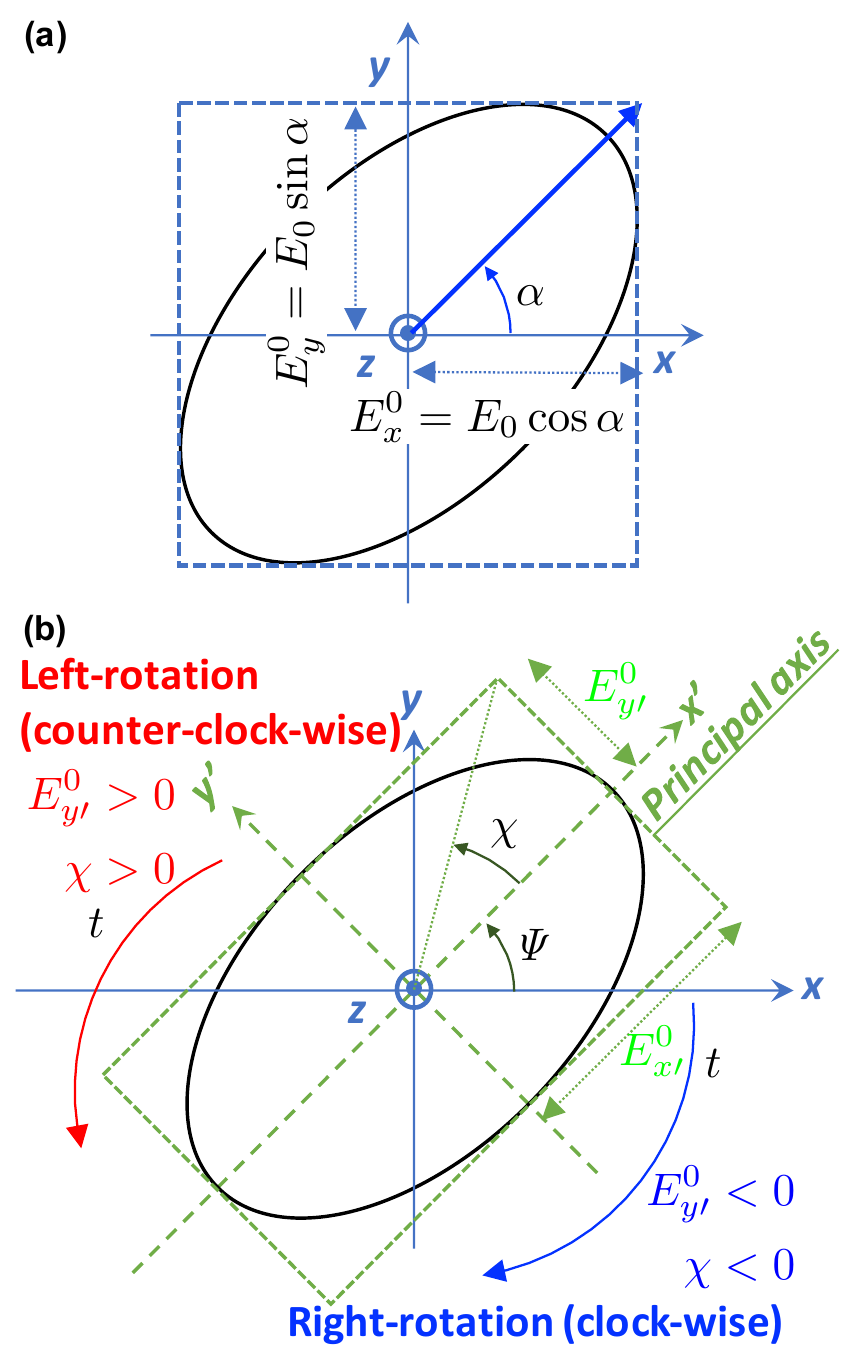}
\caption{
Elliptically polarised state.
(a) Locus of the electric field, described by the auxiliary angle $\alpha$.
(b) Rotation of the axes $(x,y)$ to $(x^{\prime},y^{\prime})$ for aligning the principal axis $x^{\prime}$ to the ellipse.
The inclination angle ($\Psi$) is defined by the angle between $x^{\prime}$ and $x$, while the ellipticity angle ($\chi$) is defined by the ratio of the electric field components along $x^{\prime}$ and $y^{\prime}$ as $\chi=\tan^{-1}(E_{y\prime}^{0}/E_{x\prime}^{0})$.
}
\end{center}
\end{figure}

In general, the locus of the phase front is described by a polarisation ellipse (Supplementary Fig. 2) \cite{Max99,Jackson99,Yariv97,Goldstein11,Gil16}.
This could be confirmed by rewriting
\begin{eqnarray}
\left (
  \begin{array}{c}
    E_{x}/E_{x}^{0} \\
    E_{y}/E_{y}^{0}
  \end{array}
\right)
=
\left (
  \begin{array}{cc}
    1           & 0 \\
    \cos \delta & -\sin \delta \\
  \end{array}
\right)
\left (
  \begin{array}{c}
    \cos \beta \\
    \sin \beta 
  \end{array}
\right),
 \end{eqnarray}
whose inverse equation is
\begin{eqnarray}
\left (
  \begin{array}{c}
    \cos \beta \\
    \sin \beta 
  \end{array}
\right)
=
\left (
  \begin{array}{c}
    \frac{E_{x}}{A_{x}} \\
    \frac{\cos \delta}{\sin \delta}\frac{E_{x}}{E_{x}^{0}} -\frac{1}{\sin \delta}\frac{E_{y}}{E_{y}^{0}}
  \end{array}
\right).
 \end{eqnarray}
We can eliminate $t$ and $z$ dependences by inserting into the identity, $\cos^2 \beta +\sin^2 \beta =1$, and we obtain the polarisation ellipse as 
\begin{eqnarray}
\left (
    \frac{E_{x}}{E_{x}^{0}} 
\right)^2
-2
\left (
    \frac{E_{x}}{E_{x}^{0}} 
\right)
\left (
    \frac{E_{y}}{E_{y}^{0}} 
\right)
\cos \delta
+
\left (
    \frac{E_{y}}{E_{y}^{0}} 
\right)^2
=\sin^2 \delta. \nonumber \\
\end{eqnarray}

We can diagonalise the ellipse by rotating the axes $(x,y)$ to $(x^{\prime},y^{\prime})$, which makes the rotated electric field ${\bf E}^{\prime}$ described as 
\begin{eqnarray}
\left (
  \begin{array}{c}
    E_{x}^{\prime} \\
    E_{y}^{\prime} 
  \end{array}
\right)
=
\left (
  \begin{array}{cc}
    \cos {\it \Psi} & -\sin {\it \Psi}\\
    \sin {\it \Psi} &  \ \ \cos {\it \Psi} 
  \end{array}
\right)
\left (
  \begin{array}{c}
    E_{x} \\
    E_{y} 
  \end{array}
\right),
\end{eqnarray}
where $\Psi$ is the inclination angle.
By imposing a condition to eliminate the cross term, we obtain
\begin{eqnarray}
\tan(2 {\it \Psi})
=
\tan(2 \alpha)
\cos\delta.
\end{eqnarray}
We also obtain the rotated amplitudes as 
\begin{eqnarray}
\frac{1}{(E_{x\prime}^{0})^2}
=
\frac{1}{\sin^2 \delta}
\left(
 \frac{\cos^2 {\it \Psi}}{(E_{x}^{0})^2}
+\frac{\sin^2 {\it \Psi}}{(E_{y}^{0})^2}
-\frac{\sin(2{\it \Psi})\cos \delta}{E_{x}^{0}E_{y}^{0}}
\right), \nonumber \\
\end{eqnarray}
and
\begin{eqnarray}
\frac{1}{(E_{y\prime}^{0})^2}
=
\frac{1}{\sin^2 \delta}
\left(
 \frac{\sin^2 {\it \Psi}}{(E_{x}^{0})^2}
+\frac{\cos^2 {\it \Psi}}{(E_{y}^{0})^2}
+\frac{\sin(2{\it \Psi})\cos \delta}{E_{x}^{0}E_{y}^{0}}
\right).
\nonumber \\
\end{eqnarray}
After the rotation, the polarisation ellipse is simply described as 
\begin{eqnarray}
\frac{
  E_{x}^{\prime 2}
  }
  {(E_{x\prime}^{0})^2}
+
\frac{
  E_{y}^{\prime 2}
  }
  {(E_{y\prime}^{0})^2}
=1
\end{eqnarray}
in the rotated frame.

Combining with the energy conservation law
\begin{eqnarray}
(E_0)^2=
(E_{x}^{0})^2+(E_{y}^{0})^2=
(E_{x\prime}^{0})^2+(E_{y\prime}^{0})^2
\end{eqnarray}
we obtain 
\begin{eqnarray}
\sin (2 \chi)
=
\sin (2 \alpha) \sin \delta . 
\end{eqnarray}

Above calculations are based on classical Maxwell equations, and it was straightforward but lengthy.
On the other hands, if we recognise the $SU(2)$ symmetry of the polarisation state, together with the equivalence between the Poincar\'e sphere and the Bloch sphere, the relationship among angles can be obtained simplify from the geometrical considerations in these spheres, as shown in the main text and the next appendix.

\section{Stokes parameters}
The Stokes parameters \cite{Jones41,Poincare92,Yariv97,Goldstein11,Gil16} are defined as
\begin{eqnarray}
S_0&=&
\mathcal{E}_x \mathcal{E}_x^{*}
+
\mathcal{E}_y \mathcal{E}_y^{*}\\
S_1&=&
\mathcal{E}_x \mathcal{E}_x^{*}
-
\mathcal{E}_y \mathcal{E}_y^{*}\\
S_2&=&
\mathcal{E}_x \mathcal{E}_y^{*}
+
\mathcal{E}_y \mathcal{E}_x^{*}\\
S_3&=&
i(
\mathcal{E}_x \mathcal{E}_y^{*}
-
\mathcal{E}_y \mathcal{E}_x^{*}
),\\
\end{eqnarray}
which represent, the intensity of the beam, the degree of horizontal/vertical linear polarisation, the degree of diagonal/anti-diagonal polarisation, and the degree of left/right circular polarisation, respectively.

These parameters are summarised as a vector 
\begin{eqnarray}
{\bf S}
=
\left (
  \begin{array}{c}
   S_{0} \\
   S_{1} \\
   S_{2} \\
   S_{3} 
  \end{array}
\right)
=
\left (
  \begin{array}{c}
    (E_{x}^{0})^2+(E_{y}^{0})^2  \\
    (E_{x}^{0})^2-(E_{y}^{0})^2  \\
    2E_{x}^{0}E_{y}^{0} \cos \delta  \\
    2E_{x}^{0}E_{y}^{0} \sin \delta  
  \end{array}
\right).
\end{eqnarray}
The time dependence through the phase ${\rm e}^{\i \beta}$ is averaged by taking the complex conjugate for coherent rays.
In particular, the vector, $(S_{1},S_{2},S_{3})$, is described in the Poincar\'e sphere to illustrate the degrees of polarisation \cite{Jones41,Poincare92,Yariv97,Goldstein11,Gil16} .

Using the parameters $(\alpha, \delta)$ in the Jones vector, ${\bf S}$ becomes
\begin{eqnarray}
{\bf S}
=
|E_{0}|^2
\left (
  \begin{array}{c}
    1 \\
    \cos (2\alpha)  \\
    \sin (2\alpha) \cos \delta  \\
    \sin (2\alpha) \sin \delta  
  \end{array}
\right).
 \end{eqnarray}

By using the other chiral parameters $(\chi, \Psi)$, this becomes
\begin{eqnarray}
{\bf S}
&=
|E_{0}|^2
\left (
  \begin{array}{c}
    1 \\
    \cos (2 \chi) \cos (2{\it \Psi}) \\
    \cos (2 \chi) \sin (2{\it \Psi}) \\
    \sin (2 \chi) 
  \end{array}
\right).
 \end{eqnarray}
We have shown that this change is simply coming from the change of the basis states to describe the polarisation states in $SU(2)$ from Jones vectors bases to chiral states.
By comparing the formulas for ${\bf S}$ in Jones and chiral descriptions, we confirm the identities derived in the previous appendix.

\bibliography{QSAM}

\end{document}